\documentclass[a4paper,11pt]{article}
\usepackage[utf8]{inputenc}
\usepackage{authblk}
\usepackage{chngcntr}
\usepackage{graphicx}
\usepackage{booktabs}
\usepackage{ulem}
\usepackage{amsmath}
\usepackage{amssymb}
\usepackage{xcolor}
\usepackage{framed}
\usepackage{pifont}
\usepackage{upgreek}
\usepackage{times}
\usepackage{bm}
\usepackage{mathbbol}
\usepackage{mathrsfs}
\usepackage{tensor}
\usepackage{empheq}
\usepackage[makeroom]{cancel}
\usepackage{needspace}
\usepackage{xspace}
\usepackage{relsize}
\usepackage{nicefrac}
\usepackage{verbatim}
\newcommand{\RB}{\mathbb{R}}
\newcommand{\rmi}{\mathrm{i}}
\newcommand{\HCd}{\mathcal{H}}
\newcommand{\FCd}{\mathcal{F}}
\newcommand{\LCd}{\mathcal{L}}
\newcommand{\KCd}{\mathcal{K}}
\newcommand{\KCdbar}{\bar{\KCd}}
\renewcommand{\d}{\mathrm{d}}
\newcommand{\onehalf}{{\textstyle\frac{1}{2}}}

\newcommand{\quarter}{{\textstyle\frac{1}{4}}}

\newcommand{\oneeighth}{{\textstyle\frac{1}{8}}}

\newcommand{\ihalf}{{\textstyle\frac{\rmi}{2}}}

\newcommand{\iquarter}{{\textstyle\frac{\rmi}{4}}}
\newcommand{\must}{\stackrel{!}{=}}
\newcommand{\partialr}{\overrightarrow{\partial}}
\newcommand{\partiall}{\overleftarrow{\partial}}
\newcommand{\pfracr}[2]{\frac{\partialr{#1}}{\partial{#2}}}
\newcommand{\pfracl}[2]{\frac{\partiall{#1}}{\partial{#2}}}
\newcommand{\Dderr}{\overrightarrow{\mathcal{D}}}
\newcommand{\Dderl}{\overleftarrow{\mathcal{D}}}
\newcommand{\pfrac}[2]{\frac{\partial{#1}}{\partial{#2}}}
\newcommand{\ppfrac}[3]{\frac{\partial^{2}{#1}}{\partial{#2}\partial{#3}}}

\newcommand{\detpartial}[2]{\left|\pfrac{#1}{#2}\right|}
\newcommand{\eref}[1]{Eq.~(\ref{#1})}
\newcommand{\sref}[1]{Section~\ref{#1}}

\newcommand{\aref}[1]{Appendix~\ref{#1}}
\newcommand{\dete}{\varepsilon}
\newcommand{\deteinv}{\frac{1}{\dete}}
\newcommand{\psibar}{\bar{\psi}}
\newcommand{\Psibar}{\bar{\Psi}}

\newcommand{\kappabar}{\bar{\kappa}}
\newcommand{\HO}{\Omega}
\newcommand{\ho}{\omega}
\newcommand{\HOc}{\tilde{Q}}
\newcommand{\hoc}{\tilde{q}}
\definecolor{dunkelgrau}{rgb}{0.33,0.33,0.33}
\definecolor{shadecolor}{rgb}{0.93,0.93,0.93}

\begin{document}
\title{Covariant Canonical Gauge Theory of Classical Gravitation for Scalar, Vector, and Spin-1/2 Particle Fields}
\author[1]{David~Vasak\thanks{vasak@fias.uni-frankfurt.de}} 
\author[1,2]{Jürgen~Struckmeier} 

\affil[1]{Frankfurt Institute for Advanced Studies (FIAS), Ruth-Moufang-Strasse~1, 60438 Frankfurt am Main, Germany}
\affil[2]{Goethe Universit\"at, Max-von-Laue-Strasse~1, 60438~Frankfurt am Main, Germany}
\maketitle
\begin{abstract}
The framework of the Covariant Canonical Gauge theory of Gravity---abbreviated as CCGG---is described in detail,
complementing a series of past publications focusing on many theoretical and phenomenological aspects of the theory.
CCGG is a classical relativistic field theory that emerges naturally in the Palatini formulation,
where the spacetime metric---expressed here by vierbein fields---and the spin connection are independent fundamental
(gauge) fields and where neither torsion nor non-metricity are excluded \emph{a priori}.

The manifestly covariant gauge process is worked out by means of canonical transformations in the realm of the De~Donder-Weyl Hamiltonian
formalism of classical field theories, starting from a small number of basic postulates.
Thereby, the original system of matter fields in flat spacetime, represented by non-degenerate Hamiltonian densities,
is amended by spacetime fields representing the dynamics of a curved geometry.
The coupling of the matter fields with the spacetime fields emerges in a way that
the action integral of the combined system remains invariant under active local Lorentz transformations and passive diffeomorphisms,
and thus implements the Principle of General Relativity.

We consider the real Klein-Gordon, the real Maxwell-Proca, and the Dirac fields in great detail, and derive the corresponding equations of motion.
Albeit the coupling of the given matter fields to the gauge fields are unambiguously determined by CCGG,
the dynamics of the free gauge fields must in addition be postulated based on physical reasoning.
Our choice allows to derive source equations of motion, similar to the Gauss law of electrodynamics, also for curvature and torsion.
The latter is proven to be totally anti-symmetric.
The affine connection is found to be a function of the spin connection and vierbein fields.
Requesting the spin connection to be anti-symmetric gives naturally metric compatibility.
The derivatives of the canonical equations combine to an extension of the Einstein-Hilbert action with a quadratic Riemann-Cartan concomitant that endows spacetime with inertia.
Moreover, the non-degenerate, quadratic (Gasiorowicz) version of the free Dirac Lagrangian is deployed.
When coupled to gravity, the Dirac equation is endowed with an emergent mass (length) parameter,
a curvature dependent correction of the fermion mass, and novel anomalous interactions
between particle spin and spacetime torsion.
\end{abstract}
\newpage
\sloppy
\tableofcontents
\fussy
\newpage
\allowdisplaybreaks[0]
\section{Introduction}
The theory of gravitation and matter as laid down by Einstein and Hilbert more than hundred years ago has since been facing a dilemma.
On the one hand, it is a fairly accurate description of the macroscopic world, be it star formation, black holes and
gravitational waves, or practical daily applications like GPS and satellite navigation.
On the other hand, the mathematics of the Einstein-Hilbert theory fails to provide a pathway to the microscopic --- quantum --- world, nor does it satisfactorily account for many cosmological observations.
Even though many alternative approaches have been investigated in the meantime, they all suffer from ambiguities
and conceptual inconsistencies, and none has so far delivered any convincing and consistent theory.

The early accounts of gauge theories of classical (c-number) fields describing spacetime and matter have been carried out in the
Lagrangian picture~\cite{weyl19,einstein55,YM54,sciama62,kibble61,utiyama56,hehl2023lectures,hayashi80,hayashi80II,hayashi80III,hayashi80IV,hayashi81V}.
In contrast, our approach is based on the framework of covariant canonical transformation theory
in the De Donder-Weyl-Hamiltonian picture, pioneered by Struckmeier and Redelbach~\cite{struckmeier08}.
Its generalization to a dynamic spacetime background is referred to as Canonical Covariant Gauge theory of Gravitation (CCGG).
It is based just on four postulates:
\begin{itemize}
  \item \emph{Hamilton's Principle} or the Principle of Least Action holds, hence, the dynamics, i.e.\ the equations of motion
  of a system of classical physical fields must be derived by variation of an action integral.
  \item \emph{Regularity} (aka non-degeneracy) of the Lagrangian is mandatory for the Legendre transform from the Hamiltonian
  to the Lagrangian picture (and vice versa) to exist, i.e.\ the determinant of the associated Hesse matrix must not vanish.
  This ensures the applicability of the Hamiltonian canonical transformation theory.
  \item Einstein's \emph{Principle of General Relativity} must hold, i.e.\ the action integral
  and the equations of motion must be independent of the chart that maps the base manifold on $\RB^4$.
  Hence the Hamiltonian must be covariant under arbitrary coordinate transformations (diffeomorphisms).
  \item The \emph{Equivalence Principle} must apply, which means that \emph{locally} the system's description in a
  spacetime frame must be equivalent to its description in an inertial frame and be invariant under Lorentz transformations.
\end{itemize}
This restriction to just four fundamental underlying assumptions is possible as the Hamiltonian framework
enforces by construction the form-invariance of the action principle under canonical transformations.
It thus provides a strong formal guidance to ensuring consistency of the emerging equations of motion.
Moreover, many of the ambiguities inherent to the Lagrangian formulation \cite{hehl2023lectures} 
are resolved in the Hamiltonian formulation \cite{Chen:2015vya}.
The validity of this approach was proven on ordinary gauge theories and shown to deliver from first principles the correct
Hamiltonian for any $\mathrm{SU}(N)$ gauge theory~\cite{StrRei12,struckmeier17}.
Its generalization to scalar and vector fields in a dynamic spacetime background was published in Ref.~\cite{struckmeier17a}.

This paper is an extension of that previous work and its application to gravitation, aiming at the inclusion of spin-$\nicefrac{1}{2}$ fields.
Therefore, we introduce an additional structural element to the description of spacetime, namely a global orthonormal
(so called vierbein) basis attached to every point of the tangent space on the base manifold.
Hence, according to the fourth postulate in the above list, we request the frame of any observer to be the
inertial space of \emph{Special Relativity}, where the metric tensor is globally the Minkowski metric.
This ``Lorentzian spacetime'' is represented by a so-called frame bundle, and the vierbein field
is a global section that pulls the Minkowski metric to the curved base manifold.
With the inclusion of that frame we have to deal with the additional Lorentz symmetry and combine the
diffeomorphism-covariance requirement for the base manifold with the Lorentz covariance of the attached inertial frame.
The resulting symmetry group, $\mathrm{SO}(1,3)\times\mathrm{Diff}(M)$, generalizes the ``affine space'' of the Poincar\'{e} gauge theory~\cite{hehl2023lectures,hehl2023lectures}.
Here, though, the Lorentz symmetry is restricted to the \emph{active} orthochronous subgroup, connected to the unit transformation,
while the diffeomorphism is a \emph{passive} transformation.

Throughout this paper we retain the tensor calculus which is more complex
to handle than the formulas of differential geometry but easier to understand for physicists.
In the following, we give a brief overview of the contents of the paper's sections.

Section~\ref{sec:CCTF} starts with a comprehensive account of the formalism of canonical transformations
in the realm of the covariant Hamiltonian description of classical field theories, the latter being founded by De Donder~\cite{dedonder30} and Weyl~\cite{weyl35}.
The real scalar field and its coupling to curved spacetime will be used as an example for working out a canonical transformation under a dynamic spacetime background.
Our approach does not stipulate any \emph{a priori} constraints beyond the requirement that initially a matter Lagrangian exists that is form-invariant
under the global group~$\mathrm{SO}(1,3)$, hence possesses the Minkowski symmetry.
The request for the system's local invariance with respect to both, (passive) diffeomorphism group of chart transitions on the base manifold, and active Lorentz transformations of fields that are sections on the frame bundle with fibres (inertial frames) attached to any point of that base manifold, is implemented via the construction of a generating function.
That generating function fixes the transformation properties of the involved fields under that selected symmetry group.
The induced modifications of the original matter action yield, in the spirit of Weyl~\cite{weyl19},
Yang and Mills~\cite{YM54}, Utiyama~\cite{utiyama56}, Sciama~\cite{sciama62}, Kibble~\cite{kibble61} and Hayashi~\cite{hayashi80},
a gauge theory of dynamic spacetime on an orthonormal frame bundle with the vierbein field representing the global metric.
The Lorentz connection is then identified as the associated gauge field.
The affine connection is found to be a dependent function of the vierbein and spin connection.
Torsion of spacetime and non-metricity are formally not excluded.
The resulting spacetime dynamics is thus driven by the existence of two independent fields, namely the
vierbein and the Lorentz resp.\ the affine connection.
Admitting independent dynamics for the metric, represented by the vierbein, and the connection
is referred to as Palatini formalism~\cite{kibble61,hehl76}.

In Sect.~\ref{sec:cfeq} the coupled canonical field equations are derived from the action that is now invariant
with respect to the $\mathrm{SO}(1,3)\times\mathrm{Diff}(M)$ symmetry.
The formulae derived are generic as they apply for any Hamiltonians resp.\ Lagrangians that describe the dynamics of
the free (uncoupled) scalar, vector, and spinor fields, as well as the free gravitational field.
The ``free'' Hamiltonians cannot be derived from first principles but has to be chosen such that it is consistent
with the basic postulates and the experimentally verified phenomenology.
We observe that our theory that starts off from a few basic physical principles and adheres to the rigorous formalism
of the covariant canonical transformation theory, fixes in a straightforward manner the coupling of spacetime to the various matter fields.

Up to this point, the generic form of a gauge theory of gravity could be derived without any need to refer to particular Lagrangians
resp.\ Hamiltonians that describe the dynamics of free (uncoupled) scalar, vector, and spinor particle fields.
Also, no particular Hamiltonian for the gravity dynamics had to be specified.
In order to set up the \emph{physical} equations for the coupling of matter fields and gravity,
we catch up on discussing the respective Hamiltonians for said fields in Sect.~\ref{sec:matter-fields}.
The Klein-Gordon, Maxwell-Proca, and the regularized Dirac Lagrangians are selected, and
their corresponding De~Donder-Weyl Hamiltonians derived by means of \emph{complete} (covariant) Legendre transformations.
In addition, a particular choice for the Hamiltonian describing the free gravitational field --- going beyond the version advocated
in Ref.~\cite{struckmeier17a} --- is addressed in Sect.~\ref{sec:Hgr}.
The discussion of the pertinent energy-momentum tensors that play a crucial role in the field dynamics is bundled in Sec.~\ref{sec:EMT}.

The coupled set of field equations of matter and dynamic spacetime are then discussed in Sect.~\ref{sec:coupledeqs}.
The particular choice of the kinetic spacetime Hamiltonian, discussed in Sect.~\ref{sec:Hgr}, is coupled to the scalar, vector, and spinor fields.
While for the scalar and vector fields the resulting equations of motion correspond to the ``minimal coupling'' recipe with
torsion included, the dynamics of the spacetime field strengths tensors, curvature and torsion,
in form of conservation equations.
The energy and momentum of matter and spacetime are shown to locally add up to zero, confirming the so called ``Zero-Energy Universe'' conjecture, see e.g. Ref.~\cite{Tryon:1973xi}.
That conservation law that is shown to be a generalization of Einstein's field equation, makes the discussion of the correct energy-momentum of spacetime obsolete.
It is supplemented by Maxwell-like equations for the field strengths 
with fermion spin expressions as source currents.

Regularity of the Dirac Hamiltonian invokes a new length parameter $\ell=1/M$ that, while spurious in the case of
non-interacting spinors, becomes a physical parameter once interaction with spacetime or other gauge fields is turned on.
We observe then that both, the momentum and the mass of the Dirac field, acquire anomalous curvature dependent mass,
and novel spin-dependent contributions that couple to the torsion of spacetime.
The curvature dependent mass term may have a considerable impact on the physics of dense matter in neutron stars
and around black holes, and also on cosmology~\cite{struckvasak18c,Benisty:2019jqz,vasak19a,vasak19b}.


We conclude the paper in Sect.~\ref{sec:conclusion}. In Sect.~\ref{sec:summary} a summary of the paper and a list of the key equations is given.
A brief discussion follows in Sect.~\ref{sec:discussion} on the ambiguities of our ansatz and on the consequences of
including internal fermion-boson interactions stipulated by $\mathrm{U}(1)$ and $\mathrm{SU}(N)$ gauge invariance.

\section{Gauge theory of gravitation in the covariant canonical transformation framework\label{sec:CCTF}}

\subsection{Conventions and notations}
Our conventions are those of Misner, Thorne and Wheeler~\cite{misner} with natural units $\hbar=c=1$ and metric signature $(+,-,-,-)$.
In order to accommodate spinors in the formulation we have to introduce local inertial frames on the tangent space of the base manifold spanned by the unit vectors $e\indices{_i}$. The Latin indices count apply to that Lorentz basis and are called Lorentz indices in contrast to the Greek indices that denote the coordinates on the base manifold. The metric tensor $g_{\mu\nu}$ is then expressed as
\begin{equation}\label{eq:vierbein-def}
g^{\mu\nu}(x)=e\indices{_i^\mu}(x)\,e\indices{_j^\nu}(x)\,\eta^{ij},\qquad
g_{\mu\nu}(x)=e\indices{^i_\mu}(x)\,e\indices{^j_\nu}(x)\,\eta_{ij},
\end{equation}
where the \emph{vierbein field} (vierbeins or tetrads) $e\indices{_i^\mu}(x)$ and their duals, $e\indices{^i_\mu}(x)$, represent the coordinates
of those unit vectors of a global orthonormal basis, and $\eta_{ij}$ denotes the globally constant Minkowski metric. Hence
\begin{equation}\label{def:emuienui}
e\indices{^i_\mu}\,e\indices{_j^\mu}=\delta_j^i,\qquad e\indices{^j_\nu}\,e\indices{_j^\mu}=\delta_\nu^\mu
\end{equation}
holds in a local (intertial) frame attached at each point $p$ with coordinates $x$.
With $\Lambda\indices{^I_i}(x)$ defining a transformation in the local Lorentz frame with
\begin{equation}\label{eq:lam-lam}
\Lambda\indices{^I_i}\Lambda\indices{^i_J}=\delta_J^I,\qquad\det\Lambda\indices{^I_i}=1,
\end{equation}
the vierbeins transform under a combined diffeomorphism and orthochronous Lo\-rentz transformation, $\mathrm{SO}(1,3)\times\mathrm{Diff}(M)$, as
\begin{equation}\label{eq:vierbein-trans}
E\indices{^I_\alpha}(X)=\Lambda\indices{^I_i}(x)\,e\indices{^i_\beta}(x)\,\pfrac{x^\beta}{X^\alpha}.
\end{equation}
This means that the determinants transform as scalar densities:
\begin{equation}\label{eq:vierbeindet-trans}
\left(\det E\indices{^I_\alpha}\right)(X)=\left(\det e\indices{^i_\beta}\right)(x)\,\detpartial{x}{X}=\dete\detpartial{x}{X}.
\end{equation}

The volume form $\d^{4}x$ transforms as a relative scalar of weight $w=-1$
\begin{equation}\label{eq:trans-volumeform}
\d^{4}X=\pfrac{\left(X^{0},\ldots,X^{3}\right)}{\left(x^{0},\ldots,x^{3}\right)}\d^{4}x=
\detpartial{X}{x}\d^{4}x={\detpartial{x}{X}}^{-1}\d^{4}x.
\end{equation}
In conjunction with Eqs.~(\ref{eq:detg-trans}) and~(\ref{eq:vierbeindet-trans}), one concludes that the products $\sqrt{-g}\,\d^{4}x$
as well as $\dete\,\d^{4}x$ transform as scalars of weight $w=0$, hence as absolute scalars
\begin{equation}\label{eq:trans-volform}
\sqrt{-G}\,\d^{4}X=\sqrt{-g}\,\d^{4}x,\qquad\left(\det E\indices{^I_\alpha}\right)\d^{4}X=\left(\det e\indices{^i_\beta}\right)\d^{4}x.
\end{equation}
$\sqrt{-g}\,\d^{4}x\equiv\dete\,\d^{4}x$ is thus referred to as the \emph{invariant volume form}.
In the vierbein formalism, the chart invariant volume form is obtained by multiplying $\d^4x$ with the determinant
$\dete:=\det(e\indices{^i_\alpha})$ of the dual vierbein $e\indices{^i_\alpha}$:
\begin{equation*}
\sqrt{-\det(g_{\alpha\beta})}=\det(e\indices{^i_\alpha})=\dete.
\end{equation*}
For the derivative of the determinant of a matrix we find
\begin{align}
\pfrac{\dete}{e\indices{^i_\xi}}e\indices{^i_\alpha}&=\dete\,\delta_\alpha^\xi
=\dete\,e\indices{_i^\xi}\,e\indices{^i_\alpha}
\qquad\Rightarrow\qquad\pfrac{\dete}{e\indices{^i_\xi}}=\dete\,e\indices{_i^\xi},
\qquad\pfrac{}{e\indices{^i_\xi}}\left(\frac{1}{\dete}\right)=-\frac{e\indices{_i^\xi}}{\dete}\nonumber\\
\pfrac{\dete}{x^\alpha}&=\pfrac{\dete}{e\indices{^i_\xi}}\pfrac{e\indices{^i_\xi}}{x^\alpha}
=\dete\,e\indices{_i^\xi}\pfrac{e\indices{^i_\xi}}{x^\alpha}=
-\dete\,e\indices{^i_\xi}\pfrac{e\indices{_i^\xi}}{x^\alpha}.
\label{eq:dmudet}
\end{align}

\subsection{Relative tensors and their transformation rules}
In addition to quantifying the dynamics of fields, the extended formalism of field theories
describes how dynamic quantities transform under a general coordinate transformation $x\mapsto X$.
$x^\mu$ and $X^\mu$ are coordinates of a point $p$ of a manifold expressed in different charts.
Thus $x\mapsto X$ means here a \emph{passive} transformation (diffeomorphism) of coordinates with the point $p$ unchanged.
For application in physics this requires to generalize the transformation rules of \emph{absolute tensors}
to those of \emph{relative tensors}.
If a tensor $t\indices{^{\alpha_{1}\ldots\alpha_{n}}_{\beta_{1}\ldots\beta_{m}}}$
transforms via chart transition $x\mapsto X$ according to
\begin{equation}\label{eq:def-tensor-denity}
T\indices{^{\xi_{1}\ldots\xi_{n}}_{\eta_{1}\ldots\eta_{m}}}(X)=
t\indices{^{\alpha_{1}\ldots\alpha_{n}}_{\beta_{1}\ldots\beta_{m}}}(x)
\pfrac{X^{\xi_{1}}}{x^{\alpha_{1}}}\ldots\pfrac{X^{\xi_{n}}}{x^{\alpha_{n}}}
\pfrac{x^{\beta_{1}}}{X^{\eta_{1}}}\ldots\pfrac{x^{\beta_{m}}}{X^{\eta_{m}}}\detpartial{x}{X}^{w},
\end{equation}
then $t$ is referred to as a \emph{relative tensor of weight} $w$.
For $w=0$, this definition includes the transformation rule for usual tensors,
which are also called \emph{absolute} tensors if the distinction is to be stressed.
The particular case of a relative tensor of weight $w=1$ is referred to briefly as a \emph{tensor density}.
With \emph{scalars} denoting the particular class of tensors of rank zero,
Eq.~(\ref{eq:def-tensor-denity}) also defines the transformation rule for scalars of weight $w$.
One concludes directly that the tensor product of a relative $(i,n)$ tensor $t$
of weight $w_{1}$ with a relative $(j,m)$ tensor $s$ of weight $w_{2}$ yields a relative $(i+j,n+m)$ tensor $t\otimes s$
of weight $w_{1}+w_{2}$.

Accordingly, a scalar of weight $w=1$ is called a \emph{scalar density}.
As an example, the square root of the determinant of a $(0,2)$ tensor transforms as a scalar density.
The covariant tensor transforms as
\begin{equation*}
T_{\mu\nu}(X)=t_{\alpha\beta}(x)\pfrac{x^{\alpha}}{X^{\mu}}\pfrac{x^{\beta}}{X^{\nu}},
\end{equation*}
hence
\begin{equation*}
(\det{T_{\mu\nu}})(X)=(\det{t_{\alpha\beta}})(x)\detpartial{x}{X}^{2}
\end{equation*}
is a relative scalar of weight $w=2$.
If $t_{\mu\nu}$ stands for the covariant form $g_{\mu\nu}$ of the metric tensor with $g\equiv\det g_{\alpha\beta}<0$, then
\begin{equation}\label{eq:detg-trans}
\sqrt{-G}=\sqrt{-g}\detpartial{x}{X}
\end{equation}
and $\sqrt{-g}$ represents a relative scalar of weight $w=1$, i.e.\ a scalar density.
Correspondingly, the determinant of the contravariant representation of the metric tensor
is a relative scalar of weight $w=-2$
\begin{equation*}
G^{\mu\nu}(X)=g^{\alpha\beta}(x)\pfrac{X^{\mu}}{x^{\alpha}}\pfrac{X^{\nu}}{x^{\beta}}\quad\Rightarrow\quad
(\det{G^{\mu\nu}})(X)=(\det{g^{\alpha\beta}})(x)\detpartial{x}{X}^{-2}.
\end{equation*}

\medskip
In a flat spacetime background, the volume form $\d^4x$ is invariant.
The Lagrangians $\LCd$ of field theories must thus be \emph{absolute scalars} in order for the action
functional to maintain its form under Lorentz transformations, hence $\LCd^\prime(X)=\LCd(x)$ and $\d^4X=d^4x$:
\begin{equation*}
\delta S=\delta\int_V\LCd^\prime\,\d^4x=\delta\int_V\LCd\,\d^4x.
\end{equation*}
In contrast, if spacetime is curved, its volume form $\d^4x$ is no longer invariant, but
transforms as a relative scalar of weight $w=-1$ according to Eq.~(\ref{eq:trans-volumeform}).
With $\LCd$ a Lorentz scalar, the system's Lagrangian $\tilde{\LCd}=\LCd\sqrt{-g}$ must then be a relative scalar of weight $w=+1$, hence a scalar density:
\begin{equation*}
\tilde{\LCd}^{\prime}(X)=\LCd^{\prime}(X)\sqrt{-G}=\LCd(x)\sqrt{-G}=
\LCd(x)\sqrt{-g}\detpartial{x}{X}=\tilde{\LCd}(x)\detpartial{x}{X},
\end{equation*}
such that the product $\tilde{\LCd}\,\d^4x$ transforms as an \emph{absolute} scalar.
As a consequence, the \emph{action integral} (not necessarily the Lagrangian itself!)
maintains its form under chart transition $x\mapsto X$, provided that $\tilde{\LCd}$ is a scalar density:
\begin{equation}\label{eq:inv-Le}
\int_{V^\prime}\tilde{\LCd}^{\prime}(X)\,\d^{4}X=\int_V\tilde{\LCd}^{\prime}(X)\detpartial{X}{x}\d^{4}x=\int_V\tilde{\LCd}(x)\,\d^{4}x.
\end{equation}

\subsection{De Donder-Weyl Hamiltonian formalism}
In the DeDonder-Weyl Hamiltonian formalism\cite{dedonder30,weyl35} the momentum fields $\pi^{\mu}(x)$
are the dual quantities of the derivatives $\partial\varphi/\partial x^{\mu}$
of a set of scalar fields $\varphi$, defined on the basis of a \emph{conventional} Lagrangian $\LCd$ as
\begin{equation*}
\pi^{\mu}(x)=\pfrac{\LCd}{\left(\pfrac{\varphi}{x^{\mu}}\right)}.
\end{equation*}
The corresponding definition in terms of a \emph{scalar density} Lagrangian $\tilde{\LCd}=\LCd\sqrt{-g}$ is then
\begin{equation}\label{eq:pi-tilde-def}
\tilde{\pi}^{\mu}(x)=\pi^{\mu}(x)\sqrt{-g}=
\pfrac{\tilde{\LCd}}{\left(\pfrac{\varphi}{x^{\mu}}\right)},\qquad
\tilde{\Pi}^{\mu}(X)=\Pi^{\mu}(X)\sqrt{-G}=
\pfrac{\tilde{\LCd}^{\prime}}{\left(\pfrac{\Phi}{X^{\mu}}\right)}.
\end{equation}
Therefore, $\tilde{\pi}^{\mu}$ can be regarded as the dual of the derivative $\partial\varphi/\partial x^{\mu}$
with regard to the extended Lagrangian $\tilde{\LCd}$.
While $\pi^{\mu}$ transforms as an absolute tensor, the related $\tilde{\pi}^{\mu}=\pi^{\mu}\sqrt{-g}$
transforms as a relative vector of weight $w=1$, hence as a vector density
\begin{equation*}
\Pi^{\mu}(X)=\pi^{\alpha}(x)\pfrac{X^{\mu}}{x^{\alpha}}=\frac{\tilde{\Pi}^{\mu}(X)}{\sqrt{-G}}=
\frac{\tilde{\pi}^{\alpha}(x)}{\sqrt{-g}}\pfrac{X^{\mu}}{x^{\alpha}},
\end{equation*}
hence by virtue of Eq.~(\ref{eq:detg-trans})
\begin{equation}\label{eq:gen-coord-syst-trans}
\tilde{\Pi}^{\mu}(X)=\tilde{\pi}^{\alpha}(x)\pfrac{X^{\mu}}{x^{\alpha}}\frac{\sqrt{-G}}{\sqrt{-g}}=
\tilde{\pi}^{\alpha}(x)\pfrac{X^{\mu}}{x^{\alpha}}\detpartial{x}{X}.
\end{equation}
Equation~(\ref{eq:gen-coord-syst-trans}) is the general transformation rule for the
vector density $\tilde{\pi}^{\mu}(x)$ under a chart transition.

The transition into the covariant Hamiltonian picture is achieved by applying the covariant Legendre transformation.
Taking the scalar fields as example, this gives
\begin{equation}\label{eq:simpleHamil}
\tilde{\HCd}:=\tilde{\pi}^\alpha\,\pfrac{\varphi}{x^{\alpha}}-\tilde{\LCd}.
\end{equation}
Notice that the DW Hamiltonian is not the energy density as it is familiar from non-relativistic physics.
The energy density is rather given by the ${T}\indices{^0^0}$ component of the canonical energy-momentum tensor
\begin{equation}
\tilde{T}\indices{_\mu^\nu}:=\pfrac{\tilde{\LCd}}{\phi_{,\nu}}\,\phi_{,\mu}-\delta_\mu^\nu\,\tilde{\LCd}.
\end{equation}
However, its trace gives
\begin{equation*}
\tilde{T}\indices{_\mu^\mu} \equiv \tilde{T}=\pfrac{\tilde{\LCd}}{\phi_{,\nu}}\,\phi_{,\nu}
-4\tilde{\LCd}=\tilde{\HCd} - 3\tilde{\LCd},
\end{equation*}
which is the covariant version of the relation $T=\rho-3p$ for a homogeneous fluid in its co-moving frame.

\subsection{The role of the vierbeins } \label{sec:vierbeins}
In the realm of the Poincar\'{e} Gauge Theory the vierbeins, as transformation matrices between the inertial frames and the coordinate base of the tangent space,
are considered to be gauge fields~\cite{kibble61}, or ``translation potentials''~\cite{hehl2023lectures, hehl2023lectures}.

In the covariant De~Donder-Weyl Hamiltonian approach that line of thought can be sketched
using the example of a scalar field.
In a Hamiltonian 
${\HCd}(\phi,\pi^i)$ the momentum field $\pi^i$ is a vector at some point $p$, and its component index $i$
is related to an inertial basis (endowed with the Minkowski metric $\eta_{ij}$) in the ``field space''.
In that frame the field components transform under local (Lorentz) transformations according to
\begin{equation} \label{eq:LTfieldindex}
\pi^I(x)=\Lambda\indices{^I_i}(x)\,\pi^i(x).
\end{equation}
Here ``local'' refers to a chart $p\in U$ in which a local coordinate basis is defined such that the point $p$ has the coordinates $x^\mu$.
Re-expressing or ``gauging'' the field components in that coordinate basis is an invertible linear transformation,
\begin{equation} \label{eq:Lortrafoas gauge}
\pi^\mu(x)=e\indices{_i^\mu}(x)\,\pi^i(x),
\end{equation}
with the vierbein $e\indices{_i^\mu}(x)$ as gauge field, and its dual $e\indices{^i_\mu}(x)$ defined such that
\begin{equation*}
e\indices{^i_\mu}(x)\,e\indices{_j^\mu}(x) = \delta^i_j
\end{equation*}
holds everywhere.
Now under arbitrary chart transitions  $x \mapsto X$ (diffeomorphisms), the newly defined components $\pi^\mu(x)$ must transform as tensors:
\begin{equation} \label{eq:vectortransform}
\pi^\mu(x)\mapsto\Pi^\nu(X) := \pfrac{X^\nu}{x^\mu}\,\pi^\mu(x).
\end{equation}
Obviously, since the gauge transformation \eqref{eq:Lortrafoas gauge} must be independent of the choice of both, the field basis and the coordinate chart,
\begin{equation} \label{eq:Lortrafoas gauge2}
\Pi^\mu(x)=E\indices{_I^\mu}(X)\,\Pi^I(X)
\end{equation}
must hold as well.
Inserting Eqs.~\eqref{eq:vectortransform}, \eqref{eq:Lortrafoas gauge}, and \eqref{eq:LTfieldindex} into \eref{eq:Lortrafoas gauge2} gives
\begin{equation*}
\pfrac{X^\nu}{x^\mu} \,e\indices{_i^\mu}(x)\,\pi^i(x)=E\indices{_I^\nu}(X)\,\Pi^I(X)=E\indices{_I^\nu}(X)\,\Lambda\indices{^I_i}(X)\,\Pi^i(X).
\end{equation*}
Now the components of the field vector depend on the inertial frame only, i.e. $\Pi^i(X)\equiv\pi^i(x)$.
The gauge field vierbein must thus transform according to
\begin{equation*}
e\indices{^i_\mu}(x)\mapsto E\indices{^I_\nu}(X) = \Lambda\indices{^I_j}(x) \, e\indices{^j_\mu}(x)\, \pfrac{x^\mu}{X^\nu},
\end{equation*}
and the Hamiltonian can be expressed in that gauge as
\begin{equation*}
\HCd(\phi,\pi^i)\mapsto\HCd(\phi,\pi^\mu)\equiv\HCd(\phi,e\indices{_i^\mu}\,\pi^i).
\end{equation*}
For ensuring in addition the invariance of the integration in the action, the
Jacobian of that gauge transformation must be taken into account.
Since $\det\Lambda\indices{^I_j}(x) \equiv 1$ by definition, and
\begin{equation*}
\detpartial{x}{X} \equiv \frac{\det E\indices{^I_\nu}(X)}{\det e\indices{^i_\mu}(x)},
\end{equation*}
the Hamiltonian scalar becomes a scalar density via
\begin{equation*}
\HCd(\phi,\pi^\mu)\mapsto\tilde{\HCd}(\phi,\pi^\mu)\equiv\HCd(\phi,e\indices{_i^\mu}\,\pi^i)\det e\indices{^i_\mu}(x) = \dete(x)\,\HCd(\phi,e\indices{_i^\mu}\,\pi^i).
\end{equation*}
Putting in this way the vierbein and the connection under the same ``gauge field'' umbrella might appear quite appealing,
but its physical significance is questionable, especially as it does not have any impact on neither Kibble's results nor on the following analysis.
A gauge field in the sense of Weyl, Fock, Yang and Mills compensates for broken local symmetries and is intimately linked to (covariant) field derivatives.
While the connection qualifies for this interpretation, it is not the case for the vierbeins.
Therefore they are in the following treated as a given structure element of a frame bundle, the Lorentzian manifold.
This is not equivalent to enforcing from the onset a curved spacetime, i.e.\ to plug in upfront any elements of gravity.
It rather provides the ``playground'' for any kind of kinematics.
In that philosophy, similar to Utiyama~\cite{utiyama56}, the essence of gravity is provided by the connection that,
via the autoparallel equation, determines the trajectories of bodies under the influence of gravity.
Whether spacetime is curved or not, i.e.\ whether gravity is present or not,
is determined solely by a non-vanishing or vanishing connection in a holonomous coordinate frame.%
\footnote{It is worthwhile to mention here that without the autoparallel transport equation the notion of a ``straight line'' remains
unclear -- even in flat space -- and a proper definition is thus missing in Newton's Philosophiae naturalis principia mathematica!}

\subsection{Canonical transformation formalism for a scalar field in a curved spacetime} \label{sec:cantransfKG}
Based on the set of general transformations $\varphi(x)\mapsto\Phi(X)$ of a scalar field with simultaneous
transformations of the vierbeins $e\indices{^i_\alpha}(x)\mapsto E\indices{^I_\beta}(X)$ under arbitrary diffemorphisms $x\mapsto X(x)$,
we now consider the particular subset of those transformations that maintain the form of the action functional~(\ref{eq:inv-Le}).
This means, explicitly:
\begin{equation*}
\delta\!\int_{V^\prime}\tilde{\LCd}^\prime\left(\Phi,\pfrac{\Phi}{X^{\nu}},E\indices{^I_\mu},\pfrac{E\indices{^I_\mu}}{X^{\nu}},X\right)\!\d^4X\,\must\,
\delta\!\int_V\tilde{\LCd}\left(\varphi,\pfrac{\varphi}{x^{\nu}},e\indices{^i_\mu},\pfrac{e\indices{^i_\mu}}{x^{\nu}},x\right)\!\d^4x.
\end{equation*}
The postulated variational invariance of the action integrals allows the \emph{integrands}
to differ by the divergence of an arbitrary vector function $\tilde{\FCd}_1^\nu(x)$ of the dynamic fields
\begin{equation} \label{actionintegrand-lag}
\tilde{\LCd}^\prime\left(\Phi,\pfrac{\Phi}{X^\nu},E\indices{^I_\mu},\pfrac{E\indices{^I_\mu}}{X^\nu},X\right)\detpartial{X}{x}\,\must\,
\tilde{\LCd}\left(\varphi,\pfrac{\varphi}{x^\nu},e\indices{^i_\mu},\pfrac{e\indices{^i_\mu}}{x^\nu},x\right)-\pfrac{\tilde{\FCd}_1^\nu}{x^\nu},
\end{equation}
provided that
\begin{equation*}
\delta\int_{V}\pfrac{\tilde{\FCd}_1^\nu}{x^\nu}\,\d^4x=\delta\oint_{\partial V}\tilde{\FCd}_1^\nu\,\d S_\nu\,=\,0.
\end{equation*}

The integrand condition~(\ref{actionintegrand-lag}) can equivalently be expressed in the
De~Donder-Weyl Hamiltonian formulation by means of \emph{complete} Legendre transformations as:
\begin{align}
&\quad\tilde{\pi}^\nu\pfrac{\varphi}{x^\nu}+\tilde{k}\indices{_i^{\mu\nu}}\pfrac{e\indices{^i_\mu}}{x^\nu}-
\tilde{\HCd}\left(\varphi,\tilde{\pi}^\nu,e\indices{^i_\mu},\tilde{k}\indices{_i^{\mu\nu}},x\right)\label{actionintegrand-ham}\\
&\quad-\left[\tilde{\Pi}^\nu\pfrac{\Phi}{X^\nu}+\tilde{K}\indices{_I^{\mu\nu}}\pfrac{E\indices{^I_\mu}}{X^\nu}-
\tilde{\HCd}^\prime \left(\Phi,\tilde{\Pi}^\nu,E\indices{^I_\mu},\tilde{K}\indices{_I^{\mu\nu}},X\right)\right]\detpartial{X}{x}
=\pfrac{\tilde{\FCd}_1^\nu}{x^\nu}\nonumber\\
&=\pfrac{\tilde{\FCd}_1^\nu}{\varphi}\pfrac{\varphi}{x^\nu}+\pfrac{\tilde{\FCd}_1^\beta}{\Phi}\pfrac{X^\nu}{x^\beta}\pfrac{\Phi}{X^\nu}+
\pfrac{\tilde{\FCd}_1^\nu}{e\indices{^i_\mu}}\pfrac{e\indices{^i_\mu}}{x^\nu}+
\pfrac{\tilde{\FCd}_1^\beta}{E\indices{^I_\mu}}\pfrac{X^\nu}{x^\beta}\pfrac{E\indices{^I_\mu}}{X^\nu}+
\left.\pfrac{\tilde{\FCd}_1^\nu}{x^\nu}\right|_{\text{expl}}\hspace*{-3mm}.\nonumber
\end{align}
Herein, $\tilde{k}\indices{_i^{\mu\nu}}$ and $\tilde{K}\indices{_I^{\mu\nu}}$ denote the canonical conjugates
of the vierbeins $e\indices{^i_\mu}$ and $E\indices{^I_\mu}$, respectively.
\eref{actionintegrand-ham} is satisfied iff any matching partial derivatives cancel each other.
For $\tilde{\FCd}_1^\nu=\tilde{\FCd}_1^\nu\left(\varphi,\Phi,e\indices{^i_\mu},E\indices{^I_\mu},x\right)$ this gives
the following transformation rules for the fields:
\begin{subequations}
\begin{align}
\tilde{\pi}^\nu &= \pfrac{\tilde{\FCd}_1^\nu}{\varphi}, &
\tilde{\Pi}^\nu &= -\pfrac{\tilde{\FCd}_1^\beta}{\Phi}\pfrac{X^\nu}{x^\beta}\detpartial{x}{X}\\
\tilde{k}\indices{_i^{\mu\nu}} &= \pfrac{\tilde{\FCd}_1^\nu}{e\indices{^i_\mu}}, &
\tilde{K}\indices{_I^{\mu\nu}} &=-\pfrac{\tilde{\FCd}_1^\beta}{E\indices{^I_\mu}}\pfrac{X^\nu}{x^\beta}\detpartial{x}{X}.
\end{align}
\end{subequations}
The transformation rule for the Hamiltonians involves a possible \emph{explicit} dependence of $\tilde{\FCd}_1^\nu(x)$ on $x$:
\begin{equation}
\tilde{\HCd}^\prime \left(\Phi,\tilde{\Pi}^\nu,E\indices{^I_\mu},\tilde{K}\indices{_I^{\mu\nu}},X\right)\detpartial{X}{x}=
\tilde{\HCd}\left(\varphi,\tilde{\pi}^\nu,e\indices{^i_\mu},\tilde{k}\indices{_i^{\mu\nu}},x\right) +
\left.\pfrac{\tilde{\FCd}_1^\nu}{x^\nu}\right|_{\text{expl}}.
\end{equation}
As the vector density $\tilde{\FCd}_1^\nu$ defines the transformation rules
it is called the \emph{generating function} of type $1$.
There are four possible types of generating functions corresponding to four possible combinations of the original and transformed fields and their conjugate momenta.
A canonical transformation 
$\tilde{\FCd}_3^\nu$, referred to a \emph{generating function} of type $3$,
 is defined as a function of the momenta $\tilde{\pi}^\nu$ and $\tilde{k}\indices{_i^{\mu\nu}}$ in place of the fields $\varphi$
and $e\indices{^i_\mu}$ in $\tilde{\FCd}_1^\nu$. It is related to $\tilde{\FCd}_1^\nu$ by means of the Legendre transformation
\begin{align*}
\tilde{\FCd}_3^\nu\!\left(\tilde{\pi}^\nu(x),\Phi(X),\tilde{k}\indices{_i^{\mu\nu}}(x),E\indices{^I_\mu}(X),x\right)
&=\tilde{\FCd}_1^\nu\!\left(\varphi(x),\Phi(X),e\indices{^i_\mu}(x),E\indices{^I_\mu}(X),x\vphantom{\tilde{K}_I}\right)\\
&\quad-\tilde{\pi}^{\nu}(x)\,\varphi(x)-\tilde{k}\indices{_i^{\mu\nu}}(x)\,e\indices{^i_\mu}(x).
\end{align*}
Hence, in terms of $\tilde{\FCd}_3^\nu$, the integrand condition~(\ref{actionintegrand-ham}) is equivalently expressed as:
\begin{align*}
\pfrac{\tilde{\FCd}_1^\nu}{x^\nu}&=
\pfrac{\tilde{\FCd}_3^\nu}{x^\nu}+\pfrac{\tilde{\pi}^\nu}{x^{\nu}}\,\varphi+\cancel{\tilde{\pi}^\nu\pfrac{\varphi}{x^\nu}}
+\pfrac{\tilde{k}\indices{_i^{\mu\nu}}}{x^{\nu}}\,e\indices{^i_\mu}
+\bcancel{\tilde{k}\indices{_i^{\mu\nu}}\,\pfrac{e\indices{^i_\mu}}{x^{\nu}}}\nonumber\\
&=\cancel{\tilde{\pi}^\nu\pfrac{\varphi}{x^{\nu}}}+\bcancel{\tilde{k}\indices{_i^{\mu\nu}}\pfrac{e\indices{^i_\mu}}{x^\nu}}-
\tilde{\HCd}\left(\varphi,\tilde{\pi}^\nu,e\indices{^i_\mu},\tilde{k}\indices{_i^{\mu\nu}},x\right)\nonumber\\
&\qquad-\left[\tilde{\Pi}^\nu\pfrac{\Phi}{X^{\nu}}+\tilde{K}\indices{_I^{\mu\nu}}\pfrac{E\indices{^I_\mu}}{X^\nu}-
\tilde{\HCd}^\prime \left(\Phi,\tilde{\Pi}^\nu,E\indices{^I_\mu},\tilde{K}\indices{_I^{\mu\nu}},X\right)\right]\detpartial{X}{x}.
\end{align*}
This gives
\begin{align}
&\qquad-\pfrac{\tilde{\pi}^\alpha}{x^{\nu}}\,\delta_\alpha^\nu\,\varphi
-\pfrac{\tilde{k}\indices{_i^{\mu\alpha}}}{x^{\nu}}\,\delta_\alpha^\nu\,e\indices{^i_\mu}
-\tilde{\HCd}\left(\varphi,\tilde{\pi}^\nu,e\indices{^i_\mu},\tilde{k}\indices{_i^{\mu\nu}},x\right)\nonumber\\
&\qquad-\left[\tilde{\Pi}^\nu\pfrac{\Phi}{X^{\nu}}+\tilde{K}\indices{_I^{\mu\nu}}\pfrac{E\indices{^I_\mu}}{X^\nu}-
\tilde{\HCd}^\prime \left(\Phi,\tilde{\Pi}^\nu,E\indices{^I_\mu},\tilde{K}\indices{_I^{\mu\nu}},X\right)\right]\detpartial{X}{x}\label{F3derivative2}\\
&=\!\pfrac{\tilde{\FCd}_3^\alpha}{\Phi}\pfrac{X^\nu}{x^\alpha}\pfrac{\Phi}{X^{\nu}}
+\pfrac{\tilde{\FCd}_3^\alpha}{E\indices{^I_\mu}}\pfrac{X^\nu}{x^\alpha}\pfrac{E\indices{^I_\mu}}{X^\nu}
+\pfrac{\tilde{\FCd}_3^\nu}{\tilde{\pi}^\alpha}\pfrac{\tilde{\pi}^\alpha}{x^{\nu}}
+\pfrac{\tilde{\FCd}_3^\nu}{\tilde{k}\indices{_i^{\mu\alpha}}}\pfrac{\tilde{k}\indices{_i^{\mu\alpha}}}{x^{\nu}}
\!+\!\left.\pfrac{\tilde{\FCd}_3^\nu}{x^\nu}\right|_{\text{expl}}\!\!.\nonumber
\end{align}
Comparing the coefficients in Eq.~(\ref{F3derivative2}), we obtain the field transformation rules
\begin{subequations} \label{eq:102}
\begin{align}
\delta^\nu_\alpha\,\varphi&=-\pfrac{\tilde{\FCd}_3^\nu}{\tilde{\pi}^\alpha}&
\tilde{\Pi}^\nu&=-\pfrac{\tilde{\FCd}_3^\alpha}{\Phi}\,\pfrac{X^\nu}{x^\alpha}\,\detpartial{x}{X}\label{eq:F3-rules-phi}\\
\delta^\nu_\alpha\,e\indices{^i_\mu}&=-\pfrac{\tilde{\FCd}_3^\nu}{\tilde{k}\indices{_i^{\mu\alpha}}}&
\tilde{K}\indices{_I^{\mu\nu}}&=-\pfrac{\tilde{\FCd}_3^\alpha}{E\indices{^I_\mu}}\,\pfrac{X^\nu}{x^\alpha}\,\detpartial{x}{X}
\end{align}
and the corresponding rule for the Hamiltonians:
\begin{equation} \label{F3derivative12}
\tilde{\HCd}^\prime\left(\Phi,\tilde{\Pi}^\nu,E\indices{^I_\mu},\tilde{K}\indices{_I^{\mu\nu}},X\right)\detpartial{X}{x}=
\tilde{\HCd}\left(\varphi,\tilde{\pi}^\nu,e\indices{^i_\mu},\tilde{k}\indices{_i^{\mu\nu}},x\right)
+\left.\pfrac{\tilde{\FCd}_3^\nu}{x^\nu}\right|_{\text{expl}}.
\end{equation}
\end{subequations}
On the basis of the transformation rules~(\ref{eq:F3-rules-phi}), the dependence of $\varphi(x)$ on $\Phi(X)$ follows as:
\begin{equation*}
\delta_\mu^\nu\,\pfrac{\varphi}{\Phi}=-\ppfrac{\tilde{\FCd}_3^\nu}{\tilde{\pi}^\mu}{\Phi}
=\pfrac{\tilde{\Pi}^\alpha}{\tilde{\pi}^\mu}\pfrac{x^\nu}{X^\alpha}\,\detpartial{X}{x}
=\pfrac{\tilde{\Pi}^\nu(x)}{\tilde{\pi}^\mu(x)}\,\detpartial{X}{x},
\end{equation*}
which generally shows that the transformation of fields---as derived from a generating
function---simultaneously determines the transformation of the pertaining conjugate momenta.
In contrast to arbitrary transformations of fields and momenta, it is this feature of
\emph{canonical} transformations which ensures the invariance of the action functionals.
\subsection{Local Lorentz and diffeomorphism transformation and the associated gauge field\label{sec:gauge-field}}
In the next step the particular generating function is defined for a canonical transformation
that provides combined Lorentz and chart transformations, while leaving the scalar field $\varphi$ unchanged:
\begin{equation} \label{def:F3_Lorentzandchart}
\tilde{\FCd}_3^\nu\left(\Phi,\tilde{\pi}^\nu, E\indices{^I_\mu}, \tilde{k}\indices{_i^{\mu\nu}},x\right)
:=-\tilde{\pi}^\nu\,\Phi-\tilde{k}\indices{_i^{\beta\nu}}\,\Lambda\indices{^i_I}\,E\indices{^I_\alpha}\,\pfrac{X^\alpha}{x^\beta}.
\end{equation}
The particular transformation rules~(\ref{eq:102}) follow as
\begin{subequations}\label{F2derivative21}
\begin{alignat}{3}
\tilde{\Pi}^\nu\,&\equiv\,-\pfrac{\tilde{\FCd}_3^\alpha}{\Phi}\,\pfrac{X^\nu}{x^\alpha}\,
\detpartial{x}{X}&=&\,\tilde{\pi}^\alpha \pfrac{X^\nu}{x^\alpha}\detpartial{x}{X}\label{F2derivative211}\\
\delta^\nu_\alpha\,\varphi\,&\equiv\,-\pfrac{\tilde{\FCd}_3^\nu}{\tilde{\pi}^\alpha}
\,&=&\,\delta^\nu_\alpha\,\Phi\label{F3derivative22}\\
\tilde{K}\indices{_I^{\mu\nu}}\,&\equiv\,-\pfrac{\tilde{\FCd}_3^\alpha}{E\indices{^I_\mu}}\pfrac{X^\nu}{x^\alpha}\,\detpartial{x}{X}
\,&=&\,\tilde{k}\indices{_i^{\beta\alpha}}\Lambda\indices{^i_I}\,
\pfrac{X^\mu}{x^\beta}\,\pfrac{X^\nu}{x^\alpha}\,\detpartial{x}{X}\label{F3derivative213}\\
\delta^\nu_\alpha\,e\indices{^i_\beta}\,&\equiv\,-\pfrac{\tilde{\FCd}_3^\nu}{\tilde{k}\indices{_i^{\beta\alpha}}}
\,&=&\,\delta^\nu_\alpha\,\Lambda\indices{^i_I}\,E\indices{^I_\alpha}\,\pfrac{X^\alpha}{x^\beta},\label{F3derivative214}
\end{alignat}
\end{subequations}
which recover the inputted proper transformation rules for the fields and derives that of their conjugates.
The latter transforms as \emph{relative} tensors of weight $w=1$, i.e., as tensor \emph{densities}.

The set of transformation rules is completed by the rule for the Hamiltonian density from Eq.~(\ref{F3derivative12}),
which follows from the \emph{explicit} spacetime dependence of the generating function~(\ref{def:F3_Lorentzandchart})
\begin{align}
\left.\pfrac{\tilde{\FCd}_3^\nu}{x^\nu} \right|_{\text{expl}}&=
-\tilde{k}\indices{_i^{\beta\nu}}\pfrac{}{x^\nu}\left(\Lambda\indices{^i_I}\pfrac{X^\alpha}{x^\beta}\right)E\indices{^I_\alpha}\nonumber\\
&=-\tilde{k}\indices{_i^{(\beta\nu)}}\pfrac{}{x^\nu}\left(\Lambda\indices{^i_I}\pfrac{X^\alpha}{x^\beta}\right)E\indices{^I_\alpha}
-\tilde{k}\indices{_i^{[\beta\nu]}}\pfrac{\Lambda\indices{^i_I}}{x^\nu}\pfrac{X^\alpha}{x^\beta}E\indices{^I_\alpha}.
\label{HLorentzandchart}
\end{align}
In the last line, the right-hand side is split into the symmetric and the skew-sym\-metric contributions of
$\tilde{k}\indices{_i^{\beta\nu}}$ in $\beta$ and $\nu$, considering that the second derivative term of
$X^\alpha$ does not contribute to the skew-symmetric portion of $\tilde{k}\indices{_i^{\nu\beta}}$.
In order to finally cast the transformation rule~(\ref{HLorentzandchart}) into a symmetric form in the
original and the transformed fields, we set up the $x^\nu$-derivative of Eq.~(\ref{F3derivative214}):
\begin{equation}\label{deri-F3derivative214}
\pfrac{}{x^\nu}\left(\Lambda\indices{^i_I}\pfrac{X^\alpha}{x^\beta}\right)E\indices{^I_\alpha}
+\Lambda\indices{^i_I}\pfrac{X^\alpha}{x^\beta}\pfrac{X^\xi}{x^\nu}\pfrac{E\indices{^I_\alpha}}{X^\xi}
=\pfrac{e\indices{^i_\beta}}{x^\nu}.
\end{equation}
Inserting Eqs.~(\ref{deri-F3derivative214}) and (\ref{F2derivative211}) for the symmetric contribution of $\tilde{k}\indices{_i^{\beta\nu}}$ in $\beta$ and $\nu$ yields:
\begin{align}
\left.\pfrac{\tilde{\FCd}_3^\nu}{x^\nu} \right|_{\text{expl}}
&=-\tilde{k}\indices{_i^{(\beta\nu)}}\left(\pfrac{e\indices{^i_\beta}}{x^\nu}-
\Lambda\indices{^i_I}\pfrac{X^\alpha}{x^\beta}\pfrac{X^\xi}{x^\nu}\pfrac{E\indices{^I_\alpha}}{X^\xi}\right)
-\tilde{k}\indices{_i^{[\beta\nu]}}\pfrac{\Lambda\indices{^i_I}}{x^\nu}\pfrac{X^\alpha}{x^\beta}E\indices{^I_\alpha}\nonumber\\
&=-\tilde{k}\indices{_i^{(\beta\nu)}}\pfrac{e\indices{^i_\beta}}{x^\nu}
+\tilde{K}\indices{_I^{(\beta\nu)}} \pfrac{E\indices{^I_\beta}}{X^\nu}\detpartial{X}{x}
+\tilde{k}\indices{_i^{[\beta\nu]}}\Lambda\indices{^i_I}\pfrac{\Lambda\indices{^I_j}}{x^\nu} e\indices{^j_\beta}\label{F3derivative12a}\\
&=\tilde{\HCd}^\prime\detpartial{X}{x}-\tilde{\HCd}.\nonumber
\end{align}
Clearly, the Hamiltonians and hence the actions are no longer form-invariant for spacetime-dependent Lorentz transformation coefficients $\Lambda\indices{^I_j}(x)$.
The only way to re-establish the form invariance of the actions is to amend them
by \emph{gauge Hamiltonians} whose transformation rule absorbs the symmetry-breaking term proportional
to $\partial\Lambda\indices{^I_j}/\partial x^\nu$ in Eq.~(\ref{F3derivative12a}).
This entails the following transformation requirement for the gauge Hamiltonians:
\begin{equation}\label{eq:inv-cond}
\tilde{\HCd}_{\mathrm{Gau}_1}^{\prime}\,\detpartial{X}{x}-\tilde{\HCd}_{\mathrm{Gau}_1}=
\tilde{k}\indices{_i^{[\mu\nu]}}\Lambda\indices{^i_I}\pfrac{\Lambda\indices{^I_j}}{x^\nu}\,e\indices{^j_\mu}.
\end{equation}
The  gauge Hamiltonian $\tilde{\HCd}_{\mathrm{Gau}_1}$ must be devised in the way that the \emph{external} index structure of
$\Lambda\indices{^i_I}\partial\Lambda\indices{^I_j}/\partial x^\nu$ is matched by some compensating \emph{gauge field} $\ho\indices{^i_{j\nu}}$.
Its simplest form is given by:
\begin{equation}\label{g-ham1}
\tilde{\HCd}_{\mathrm{Gau}_1}=-\tilde{k}\indices{_i^{[\mu\nu]}}\,\ho\indices{^i_{j\nu}}\,e\indices{^j_\mu},
\end{equation}
which is required to be form-invariant in terms of the transformed gauge field $\HO\indices{^I_{J\nu}}$:
\begin{equation}\label{g-ham2}
\tilde{\HCd}_{\mathrm{Gau}_1}^{\prime}=-\tilde{K}\indices{_I^{[\mu\nu]}}\,\HO\indices{^I_{J\nu}}\,E\indices{^J_\mu}.
\end{equation}
The sign of the gauge field was chosen such that $\ho\indices{^i_{j\nu}}$ can later be identified with the \emph{spin connection}.
We derive the ensuing transformation rule for the gauge field $\ho\indices{^{i}_{k\mu}}$,
by inserting the Hamiltonians~(\ref{g-ham1}) and~(\ref{g-ham2}) into Eq.~(\ref{eq:inv-cond}).
Beforehand, the Hamiltonian~(\ref{g-ham2}) is expressed in terms of the original fields
according to the canonical transformation rules~(\ref{F2derivative21}):
\begin{equation*}
\tilde{\HCd}_{\mathrm{Gau}_1}^{\prime}=-\tilde{k}\indices{_i^{[\mu\nu]}}
\Lambda\indices{^i_I}\,\HO\indices{^{I}_{J\alpha}}\,\Lambda\indices{^J_j}
\pfrac{X^\alpha}{x^\nu}\,e\indices{^j_\mu}\detpartial{x}{X}.
\end{equation*}
It follows that the gauge field $\ho\indices{^i_{j\nu}}$ transforms inhomogeneously as:
\begin{equation}\label{omegatransform1}\boxed{%
\ho\indices{^i_{j\nu}}=\Lambda\indices{^i_I}\,\HO\indices{^I_{J\alpha}}\,\Lambda\indices{^J_j}\,\pfrac{X^\alpha}{x^\nu}
+\Lambda\indices{^i_I}\,\pfrac{\Lambda\indices{^I_j}}{x^{\nu}}.}
\end{equation}
The inhomogeneous term in~(\ref{omegatransform1}) is skew-symmetric in $i,j$ according to Eq.~(\ref{eq:lam-lam}):
\begin{equation}\label{eq:skew-inhom}
\Lambda\indices{^i_I}\pfrac{\Lambda\indices{^I_j}}{x^\nu}=-\Lambda\indices{^I_j}\pfrac{\Lambda\indices{^i_I}}{x^\nu}.
\end{equation}
Consequently, for the gauge field $\ho\indices{_n_{j\nu}}=\eta_{ni}\,\ho\indices{^i_{j\nu}}$,
the condition $\ho\indices{_n_{j\nu}}\neq\ho\indices{_j_{n\nu}}$ must hold.

The now form-invariant ``gauged'' action functional writes:
\begin{equation}\label{action-integral2}
S=\int_{V}\left[\tilde{\pi}^\nu\pfrac{\varphi}{x^\nu}+
\tilde{k}\indices{_i^{[\mu\nu]}}\left(\pfrac{e\indices{^i_\mu}}{x^\nu}+\ho\indices{^i_{j\nu}}\,e\indices{^j_\mu}\right)
-\tilde{\HCd}_0\right]\d^4x.
\end{equation}
Herein, the gauge field $\ho\indices{^i_{j\nu}}(x)$ enters as an
\emph{external} field whose dynamics is not described by the action~(\ref{action-integral2}).
The action functional of Eq.~(\ref{action-integral2}) fulfills also the postulate of form-invariance under diffeomorphisms,
historically referred to as the ``Principle of General Relativity''.
We observe that no \emph{direct} coupling of the scalar field $\varphi(x)$ with the vierbein field
$e\indices{^j_\nu}$ and the gauge field $\ho\indices{^i_{j\nu}}$ emerges.
Rather, the coupling occurs merely via the common dependence of $\tilde{\HCd}_0$
and $\tilde{\HCd}_{\mathrm{Gau}_1}$ on the vierbein field $e\indices{^j_\nu}$.
The reason is that $\tilde{\pi}^\nu\,\partial\varphi/\partial x^\nu$ 
constitutes a world scalar density and is, therefore, already form-invariant under diffeomorphisms.
This changes if we include spin particle fields into our analysis, which is worked out below.
Moreover, the system as constructed so far is not yet complete as the gauge field $\ho\indices{^i_{j\nu}}$
introduced to eliminate the symmetry violation of the original field system
is yet an \emph{external} entity that is missing a dynamic closure.
This is the subject of the next step in the transformation theory.

\subsection{Including the canonical transformation of the gauge field $\ho\indices{^i_{j\mu}}$}
In that second step the newly introduced gauge field $\ho\indices{^i_{j\mu}}$,
defined in Eq.~(\ref{g-ham1}), will be treated as an \emph{internal} dynamic object.
Then the system will be closed, whereby the action functional~(\ref{action-integral2}) must
be extended further to define the transformation property and the dynamics of the gauge field, i.e. to
include the conjugate 
momentum field 
$\hoc\indices{_i^{j\mu\nu}}$:
\begin{equation*}
S=\int_{V}\left[\tilde{\pi}^\nu\pfrac{\varphi}{x^\nu}+
\tilde{k}\indices{_i^{[\mu\nu]}}\pfrac{e\indices{^i_\mu}}{x^\nu}
+\hoc\indices{_i^{j\mu\nu}}\pfrac{\ho\indices{^i_{j\mu}}}{x^\nu}-\tilde{\HCd}_{\mathrm{Gau}_1}-\tilde{\HCd}_0\right]\d^4x.
\end{equation*}
If we describe in addition the dynamics of a real vector field $a_\mu$ and a complex spinor field $\psi$
by the given Hamiltonian $\tilde{\HCd}_0$, the corresponding action functional now writes:
\begin{align}
S=\int_{V}\d^4x\bigg(&\tilde{\pi}^\nu\pfrac{\varphi}{x^\nu}+\tilde{p}^{\mu\nu}\pfrac{a_\mu}{x^\nu}
+\tilde{\kappabar}^\nu\pfrac{\psi}{x^\nu}+\pfrac{\psibar}{x^\nu}\tilde{\kappa}^\nu\nonumber\\
&+\tilde{k}\indices{_i^{\mu\nu}}\pfrac{e\indices{^i_\mu}}{x^\nu}
+\hoc\indices{_i^{\,j\mu\nu}}\pfrac{\ho\indices{^i_{j\mu}}}{x^\nu}-\tilde{\HCd}_{\mathrm{Gau}_2}-\tilde{\HCd}_0\bigg).
\label{action-integral3}
\end{align}
The form of the additional terms for the spin-$1$ and spin-$\nicefrac{1}{2}$ fields derives from the canonical definitions of the pertinent momentum fields
in analogy to the steps carried out in Sect.~(\ref{sec:cantransfKG}) for scalar fields.
Accordingly will now the gauging process 
lead to further amendments
of the total Hamiltonian, extending the gauge Hamiltonian $\tilde{\HCd}_{\mathrm{Gau}_1}$  to
$\tilde{\HCd}_{\mathrm{Gau}_2}$ accommodate the additional fields.

Thus, in analogy to the procedure that led to the gauge Hamiltonian~(\ref{g-ham1}),
our task is now to determine the gauge Hamiltonian $\tilde{\HCd}_{\mathrm{Gau}_2}$ that renders the action~(\ref{action-integral3})
invariant for a given Hamiltonian $\tilde{\HCd}_0$ describing the dynamics of scalar, vector, and spinor fields with the given transformation properties.
In other words, $\tilde{\HCd}_{\mathrm{Gau}_2}$ must make the integrand of~(\ref{action-integral3}) into a \emph{world scalar density}.
The generating function of type $\tilde{\FCd}_3$ must then be extended to encompass also the additional conjugate fields,
i.e.\ $\tilde{\FCd}_3 = \tilde{\FCd}_{3}^\nu(\Phi, \tilde{\pi}, A, \tilde{p},%
\Psi, \tilde{\kappabar}, \Psibar, \tilde{\kappa}, E, \tilde{k},\HO, \hoc, x)$.
The general transformation rules for an action of type~(\ref{action-integral3}) are worked out explicitly in \aref{app1}.

Now in order to promote the connection $\ho\indices{^i_{j\alpha}}$ to a dynamic field,
the particular generating function~$\tilde{\FCd}^\nu_3$ from~(\ref{def:F3_Lorentzandchart}) must be amended to
define in addition the transformation law~(\ref{omegatransform1}). 
Suppressing the spinor indices of $\psi,\tilde{\kappabar},\psibar,\tilde{\kappa}$ and of the yet to be specified spinor transformation matrix, $S$, this gives:
\begin{align}
&\quad\,\,\tilde{\FCd}_{3}^\nu\Big(\Phi,\tilde{\pi},A,\tilde{p},\Psi,\tilde{\kappabar},
\Psibar,\tilde{\kappa},E,\tilde{k},\HO,\hoc,x\Big)\nonumber\\
&=-\tilde{\pi}^\nu\,\Phi-\tilde{p}^{\,\mu\nu}\,A_{\alpha}\,\pfrac{X^\alpha}{x^\mu}-
\tilde{\kappabar}\indices{^\nu}\,S^{-1}\,\Psi-\Psibar\,S\,\tilde{\kappa}^{\nu}-
\tilde{k}\indices{_i^{\mu\nu}}\Lambda\indices{^i_I}\,E\indices{^I_\alpha}\pfrac{X^\alpha}{x^\mu}\nonumber\\
&\quad-\hoc\indices{_i^{j\mu\nu}}\left(\Lambda\indices{^i_I}\,\HO\indices{^I_{J\alpha}}\,
\Lambda\indices{^J_j}\,\pfrac{X^\alpha}{x^\mu}+\Lambda\indices{^i_I}\,\pfrac{\Lambda\indices{^I_j}}{x^\mu}\right).
\label{eq:genfu-total}
\end{align}
From the general set of canonical transformation rules~(\ref{canonicalrules}), the complete set of
specific rules for the generating function~(\ref{eq:genfu-total}) are:
\allowdisplaybreaks
\begin{subequations}\label{eq:all-rules}
\begin{alignat}{3}
\delta_\beta^\nu\,\varphi&\equiv-\pfrac{\tilde{\FCd}_{3}^\nu}{\tilde{\pi}^\beta}
&=&\,\delta_\beta^\nu\,\Phi\\
\tilde{\Pi}^\nu&\equiv-\pfrac{\tilde{\FCd}_{3}^\lambda}{\Phi}\,\pfrac{X^\nu}{x^\lambda}\detpartial{x}{X}
&=&\,\tilde{\pi}^\lambda\,\pfrac{X^\nu}{x^\lambda}\detpartial{x}{X}\\
\delta_\beta^\nu\,a_{\mu}&\equiv-\pfrac{\tilde{\FCd}_{3}^\nu}{\tilde{p}^{\mu\beta}}
&=&\,\delta_\beta^\nu\,A_{\alpha}\pfrac{X^\alpha}{x^\mu}\label{eq:a-rule}\\
\tilde{P}^{\mu\nu}&\equiv-\pfrac{\tilde{\FCd}_{3}^\lambda}{A_{\mu}}\,\pfrac{X^\nu}{x^\lambda}\detpartial{x}{X}
&=&\,\tilde{p}^{\xi\lambda}\,\pfrac{X^\mu}{x^\xi}\pfrac{X^\nu}{x^\lambda}\detpartial{x}{X}\\
\delta_\beta^\nu\,\psibar&\equiv-\pfrac{\tilde{\FCd}_{3}^\nu}{\tilde{\kappa}^{\beta}}
&=&\,\delta_\beta^\nu\,\Psibar\,S\\\
\tilde{\KCd}^{\nu}&\equiv-\pfrac{\tilde{\FCd}_{3}^\lambda}{\Psibar}\,\pfrac{X^\nu}{x^\lambda}\detpartial{x}{X}
&=&\,S\,\tilde{\kappa}^{\lambda}\,\pfrac{X^\nu}{x^\lambda}\detpartial{x}{X}\\
\delta_\beta^\nu\,\psi&\equiv-\pfrac{\tilde{\FCd}_{3}^\nu}{\tilde{\kappabar}\indices{^\beta}}
&=&\,\delta_\beta^\nu\,S^{-1}\,\Psi\\
\tilde{\KCdbar}\indices{^\nu}&\equiv-\pfrac{\tilde{\FCd}_{3}^\lambda}{\Psi}\,\pfrac{X^\nu}{x^\lambda}\detpartial{x}{X}
&=&\,\tilde{\kappabar}\indices{^\lambda}\,S^{-1}\,\pfrac{X^\nu}{x^\lambda}\detpartial{x}{X}\\
\delta_\beta^\nu\,e\indices{^i_\mu}&\equiv-\pfrac{\tilde{\FCd}_{3}^\nu}{\tilde{k}\indices{_i^{\mu\beta}}}
&=&\,\delta_\beta^\nu\,\Lambda\indices{^i_I}\,
E\indices{^I_\alpha}\pfrac{X^\alpha}{x^\mu} \label{eq:tet-trans}\\
\tilde{K}\indices{_I^{\mu\nu}}&\equiv-\pfrac{\tilde{\FCd}_{3}^\lambda}{E\indices{^I_\mu}}\,\pfrac{X^\nu}{x^\lambda}\detpartial{x}{X}
&=&\,\tilde{k}\indices{_i^{\xi\lambda}}\,\Lambda\indices{^i_I}\,\pfrac{X^\mu}{x^\xi}\pfrac{X^\nu}{x^\lambda}\detpartial{x}{X}
\end{alignat}
\allowdisplaybreaks[0]
and
\begin{alignat}{3}
\delta^\nu_\beta\,\ho\indices{^i_{j\mu}} &\equiv -\pfrac{\tilde{\FCd}_{3}^\nu}{\hoc\indices{_i^{j\mu\beta}}}
&=&\,\delta^\nu_\beta\left(\Lambda\indices{^i_I}\,\HO\indices{^I_{J\alpha}}\,\Lambda\indices{^J_j}\pfrac{X^\alpha}{x^\mu}+
\Lambda\indices{^i_I}\,\pfrac{\Lambda\indices{^I_j}\,}{x^\mu}\right) \label{eq:conn-trans}\\
\HOc\indices{_I^{J\mu\nu}} &\equiv -\pfrac{\tilde{\FCd}_{3}^\lambda}{\HO\indices{^I_{J\mu}}}
\,\pfrac{X^\nu}{x^\lambda} \detpartial{x}{X}
&=&\,\hoc\indices{_i^{j\xi\lambda}}\,\Lambda\indices{^i_I}\,\Lambda\indices{^J_j}\,\pfrac{X^\mu}{x^\xi}\pfrac{X^\nu}{x^\lambda}
\detpartial{x}{X}.\label{rhorhoLT}
\end{alignat}
\end{subequations}
Rule~(\ref{eq:conn-trans}) calculated in detail in \aref{appS} reproduces the inhomogeneous transformation property of the gauge field,
$\ho\indices{^i_{j\mu}}$, as required by Eq.~(\ref{omegatransform1}), whereas rule~(\ref{rhorhoLT})
determines the transformation property of the pertaining conjugate momentum field, $\hoc\indices{_i^{j\mu\nu}}$.

\subsection{Derivation of the gauge Hamiltonian}
The key benefit of the canonical transformation framework is that it provides
the prescription for gauging the initial Hamiltonian density $\tilde{\HCd}_0$,
hence to derive the gauge Hamiltonian $\tilde{\HCd}_{\mathrm{Gau}_2}$ such that the
combined system $\tilde{\HCd}_0+\tilde{\HCd}_{\mathrm{Gau}_2}$ becomes invariant with respect to the given transformation of the constituent fields.
The gauge Hamiltonian is ultimately determined by the \emph{explicit}
$x^\mu$-dependence of the generating function according to the general rule from Eq.~(\ref{F3derivative12}).
For the actual generating function~(\ref{eq:genfu-total}), the $x^\nu$-derivative
of the parameters in the generating function evaluates to:
\begin{align}
\pfrac{\tilde{\FCd}_{3}^\nu}{x^\nu}\bigg|_{\text{expl}}&=
-\,\tilde{p}^{(\mu\nu)}A_{\alpha}\,\ppfrac{X^\alpha}{x^\mu}{x^\nu}
-\tilde{\kappabar}\indices{^\nu}\,\pfrac{S^{-1}}{x^\nu}\,\Psi-\Psibar\,\pfrac{S}{x^\nu}\,\tilde{\kappa}^{\nu}\nonumber\\
&\quad-\tilde{k}\indices{_i^{\mu\nu}}\pfrac{}{x^\nu}\left(\Lambda\indices{^i_I}\,\pfrac{X^\alpha}{x^\mu}\right)E\indices{^I_\alpha}\nonumber\\
&\quad-\hoc\indices{_i^{j\mu\nu}}\left[\HO\indices{^I_{J\alpha}}\pfrac{}{x^\nu}
\left(\Lambda\indices{^i_I}\Lambda\indices{^J_j}\pfrac{X^\alpha}{x^\mu}\right)
+\pfrac{}{x^\nu}\left(\Lambda\indices{^i_I}\pfrac{\Lambda\indices{^I_j}}{x^\mu}\right)\right].
\label{F21derivative2}
\end{align}
Due to the symmetry of the second derivative term of $X^\alpha(x)$ in $\mu,\nu$, merely the \emph{symmetric}
part of $\tilde{p}^{\mu\nu}$ contributes to the transformation rule for the Hamiltonian.
The gauge Hamiltonian is then obtained from~(\ref{F21derivative2}) by expressing all its parameters, namely
$\Lambda\indices{^i_I}$, $S$, $\partial X^\alpha/\partial x^\mu$ and their respective derivatives, in terms
of the physical fields of the system according to the set of canonical transformation rules~(\ref{eq:all-rules}).
The individual contributions from all involved fields are worked out in the next subsections.
\subsubsection{Contribution of the vector field $A_\alpha$ to Eq.~(\ref{F21derivative2})\label{sec:a-contribution}}
The scalar field $\varphi(x)$ does not contribute to the divergence of the generating function.
Regarding the contribution of the vector field, we have two options.
From the canonical transformation rule~(\ref{eq:tet-trans}) written in the form
\begin{equation*}
\pfrac{X^\alpha}{x^\mu}=E\indices{^\alpha_I}\,\Lambda\indices{^I_i}\,e\indices{^i_\mu},
\end{equation*}
we can use its $x^\nu$-derivative and Eq.~(\ref{omegatransform1}) to compute the following transformation
of the vector field term in Eq.~(\ref{F21derivative2}):
\begin{align*}
-\tilde{p}^{\mu\nu}A_\alpha\,\ppfrac{X^\alpha}{x^\mu}{x^\nu}
&=-\tilde{p}^{\mu\nu}\,e\indices{_i^\beta}\,a_\beta\left(\pfrac{e\indices{^i_\mu}}{x^\nu}+\ho\indices{^i_{j\nu}}\,e\indices{^j_\mu}\right)\\
&\quad\,+\tilde{P}^{\mu\nu}\,E\indices{_I^\beta}\,A_\beta\left(\pfrac{E\indices{^I_\mu}}{X^\nu}+\HO\indices{^I_{J\nu}}E\indices{^J_\mu}\right)\detpartial{X}{x}.
\end{align*}
This provides a \emph{direct} coupling of the vector field $e\indices{_i^\beta}\,a_\beta$ with the connection $\ho\indices{^i_{j\nu}}$
and leads to a \emph{covariant} derivative term of $a_\beta$ in the final, diffeomorphism-invariant action functional, as laid out in Ref.~\cite{struckmeier17a}.

Alternatively, we can express the second derivative of $X^\alpha(x)$ through the deri\-vative of the canonical transformation rule~(\ref{eq:a-rule}) of the vector field
\begin{equation*}
a_\mu=A_\alpha\pfrac{X^\alpha}{x^\mu},
\end{equation*}
which yields
\begin{equation*}
A_\alpha\ppfrac{X^\alpha}{x^\mu}{x^\nu}=\pfrac{a_\mu}{x^\nu}-\pfrac{A_\alpha}{X^\tau}\pfrac{X^\tau}{x^\nu}\pfrac{X^\alpha}{x^\mu}.
\end{equation*}
Consequently, the term related to the vector field $A_\alpha$ in Eq.~(\ref{F21derivative2}) transforms as:
\begin{equation*}
-\tilde{p}^{\mu\nu}A_{\alpha}\,\ppfrac{X^\alpha}{x^\mu}{x^\nu}
=-\tilde{p}^{(\mu\nu)}\pfrac{a_\mu}{x^\nu}+\tilde{P}^{(\mu\nu)}\pfrac{A_\mu}{X^\nu}\detpartial{X}{x}.
\end{equation*}
As will be shown in the following, this leads to the conventional skew-symmetric field tensor $\tilde{p}^{\mu\nu}$
and hence to the \emph{exterior} derivative term of $a_\beta$ in the final, diffeo\-morphism-invariant action functional.
We will pursue here the second option.
Which one is correct depends on the particle, which one wishes to describe. Since all known vector particles are gauge bosons, which means, that their Lagrangian resp. Hamiltonian must obey an $\mathrm{SU}(N)$ symmetry, the first type of coupling is not allowed in that case. Due to that, we will stick with the case involving the exterior derivative. We still keep the potential mass term to possibly also describe theories with effectively massive gauge bosons like electroweak theory.

In both cases, one encounters the required form of a transformation rule for the Hamiltonian: it no longer depends on
parameters but instead on the respective physical fields in a symmetric representation in the
original and---with the opposite sign---the transformed fields.
This scheme will be be shown in the following to emerge for all terms of the sum of Eq.~(\ref{F21derivative2})
and thereby to produce a valid gauge Hamiltonian.

\subsubsection{Contribution of the spinor fields $\Psi,\Psibar$ to Eq.~(\ref{F21derivative2})}
By means of the transformation rule~(\ref{eq:conn-trans-spinor}) for $S$, the replacement
of the coefficients associated with the spinor terms in the first line of Eq.~(\ref{F21derivative2}) follows as:
\begin{align*}
-\tilde{\kappabar}^\nu\pfrac{S^{-1}}{x^\nu}\,\Psi
&=\tilde{\kappabar}^\nu S^{-1}\,\pfrac{S}{x^\nu}\,\psi\\
&=\iquarter\tilde{\kappabar}^\nu\left(S^{-1}\,\Omega\indices{_{IJ\alpha}}\,
\Sigma^{IJ}\,S\,\pfrac{X^\alpha}{x^\nu}-\ho\indices{_{ij\nu}}\,\sigma^{ij}\right)\psi\\
&=\tilde{\KCdbar}^\nu\,\iquarter\Omega\indices{_{IJ\nu}}\,\Sigma^{IJ}\,\Psi\detpartial{X}{x}
-\tilde{\kappabar}^\nu\,\iquarter\ho\indices{_{ij\nu}}\,\sigma^{ij}\,\psi.
\end{align*}
Similarly
\begin{align*}
-\Psibar\,\pfrac{S}{x^\nu}\,\tilde{\kappa}^{\nu}
&=-\psibar\,S^{-1}\,\pfrac{S}{x^\nu}\,\tilde{\kappa}^{\nu}\\
&=\psibar\,\iquarter\ho\indices{_{ij\nu}}\,\sigma^{ij}\,\tilde{\kappa}^{\nu}
-\Psibar\,\iquarter\Omega\indices{_{IJ\nu}}\,\Sigma^{IJ}\tilde{\KCd}^{\nu}\detpartial{X}{x}.
\end{align*}
Hence, the parameters of the spinor-related terms in Eq.~(\ref{F21derivative2})
are replaced by the connection fields according to
\begin{align*}
-\tilde{\kappabar}\indices{^\nu}\,\pfrac{S^{-1}}{x^\nu}\,\Psi-\Psibar\,\pfrac{S}{x^\nu}\,\tilde{\kappa}^{\nu}
&=\iquarter\left(\psibar\,\ho\indices{_{ij\nu}}\,\sigma^{ij}\,\tilde{\kappa}^{\nu}-
\tilde{\kappabar}\indices{^\nu}\,\ho\indices{_{ij\nu}}\,\sigma^{ij}\,\psi\right)\\
&\quad-\iquarter\left(\Psibar\,\Omega\indices{_{IJ\nu}}\,\Sigma^{IJ}\,\tilde{\KCd}^{\nu}-
\tilde{\KCdbar}\indices{^\nu}\,\Omega\indices{_{IJ\nu}}\,\Sigma^{IJ}\,\Psi\right)\detpartial{X}{x}.
\end{align*}

\subsubsection{Contribution of the vierbein field $E\indices{^I_\alpha}$ to Eq.~(\ref{F21derivative2})}
The coefficients in the term proportional to $\tilde{k}\indices{_i^{\mu\nu}}$ can similarly
be expressed in terms of the physical fields $e\indices{^i_\mu}$ and $\ho\indices{^i_{n\mu}}$ as follows:
\allowdisplaybreaks[0]
\begin{align*}
-\tilde{k}\indices{_i^{\mu\nu}}\,\pfrac{}{x^\nu}&\left(\Lambda\indices{^i_J}\pfrac{X^\alpha}{x^\mu}\right)E\indices{^J_\alpha}
=-\onehalf\tilde{k}\indices{_i^{\mu\nu}}\left(\pfrac{e\indices{^i_\mu}}{x^\nu}+\pfrac{e\indices{^i_\nu}}{x^\mu}-
\ho\indices{^i_{j\nu}}e\indices{^j_\mu}+\ho\indices{^i_{j\mu}}e\indices{^j_\nu}\right)\nonumber\\
&\qquad+\onehalf\tilde{K}\indices{_I^{\mu\nu}}\left(\pfrac{E\indices{^I_\mu}}{X^\nu}+\pfrac{E\indices{^I_\nu}}{X^\mu}-
\Omega\indices{^I_{J\nu}}E\indices{^J_\mu}+\Omega\indices{^I_{J\mu}}E\indices{^J_\nu}\right)\detpartial{X}{x}.
\end{align*}
The explicit calculation is worked out in \aref{app2}.
\subsubsection{Contribution of the gauge field $\HO\indices{^I_{J\alpha}}$ to Eq.~(\ref{F21derivative2})}
As the last step, we express the coefficients in the term proportional to $\hoc\indices{_i^{j\mu\nu}}$ of Eq.~(\ref{F21derivative2})
in terms of the physical fields $\ho\indices{^i_{j\mu}}$ according to the canonical transformation rules~(\ref{eq:all-rules}).
The explicit calculation is also presented in \aref{app2}, so that we can here restrict ourselves again to presenting the final result:
\begin{align}
&\quad-\hoc\indices{_i^{j\mu\nu}}\left[\Omega\indices{^I_{J\alpha}}\,\pfrac{}{x^\nu}
\left(\Lambda\indices{^i_I}\,\Lambda\indices{^J_j}\,\pfrac{X^\alpha}{x^\mu}\right)
+\pfrac{}{x^\nu}\left(\Lambda\indices{^i_I}\,\pfrac{\Lambda\indices{^I_j}}{x^\mu}\right)\right]\nonumber\\
&=-\onehalf\hoc\indices{_i^{j\mu\nu}}\left(
\pfrac{\ho\indices{^i_{j\mu}}}{x^\nu}+\pfrac{\ho\indices{^i_{j\nu}}}{x^\mu}
+\ho\indices{^i_{n\mu}}\,\ho\indices{^n_{j\nu}}-\ho\indices{^i_{n\nu}}\,\ho\indices{^n_{j\mu}}\right)\nonumber\\
&\quad\,+\onehalf\HOc\indices{_I^{J\mu\nu}}\left(\pfrac{\Omega\indices{^I_{J\mu}}}{X^\nu}+\pfrac{\Omega\indices{^I_{J\nu}}}{X^\mu}
+\Omega\indices{^I_{N\mu}}\,\Omega\indices{^N_{J\nu}}-\Omega\indices{^I_{N\nu}}\,\Omega\indices{^N_{J\mu}}\right)\detpartial{X}{x}.
\label{F21derivative3}
\end{align}
\subsubsection{Final gauge Hamiltonian, gauge-invariant action}
The divergence of the \emph{explicit} $x^\mu$-dependence of the generating
function according to Eq.~(\ref{eq:ham-rule}) sums up to
\begin{align}
\pfrac{\tilde{\FCd}_{3}^\nu}{x^\nu} \bigg|_{\text{expl}}
=&\,-\onehalf\tilde{p}^{\mu\nu}\left(\pfrac{a_\mu}{x^\nu}+\pfrac{a_\nu}{x^\mu}\right)
-\iquarter\!\left(\tilde{\kappabar}^\nu\,\ho\indices{_{ij\nu}}\,\sigma^{ij}\,\psi
-\psibar\,\ho\indices{_{ij\nu}}\,\sigma^{ij}\,\tilde{\kappa}^\nu\right)\nonumber\\
&-\onehalf\tilde{k}\indices{_i^{\mu\nu}}\left(\pfrac{e\indices{^i_\mu}}{x^\nu}+\pfrac{e\indices{^i_\nu}}{x^\mu}
+\ho\indices{^i_{j\mu}}\,e\indices{^j_\nu}-\ho\indices{^i_{j\nu}}\,e\indices{^j_\mu}\right)\nonumber\\
&-\onehalf\hoc\indices{_i^{j\mu\nu}}\left(
\pfrac{\ho\indices{^i_{j\mu}}}{x^\nu}+\pfrac{\ho\indices{^i_{j\nu}}}{x^\mu}
+\ho\indices{^i_{n\mu}}\,\ho\indices{^n_{j\nu}}-\ho\indices{^i_{n\nu}}\,\ho\indices{^n_{j\mu}}\right)\nonumber\\
&+\detpartial{X}{x}\left[\onehalf\tilde{P}^{\mu\nu}\left(\pfrac{A_\mu}{X^\nu}+\pfrac{A_\nu}{X^\mu}\right)\right.\nonumber\\
&+\iquarter\left(\tilde{\KCdbar}\indices{^\nu}\,\Omega\indices{_{IJ\nu}}\,\Sigma^{IJ}\,\Psi
-\Psibar\,\Omega\indices{_{IJ\nu}}\,\Sigma^{IJ}\tilde{\KCd}^{\nu}\right)\nonumber\\
&+\onehalf\tilde{K}\indices{_I^{\mu\nu}}\!\left(\pfrac{E\indices{^I_\mu}}{X^\nu}+\pfrac{E\indices{^I_\nu}}{X^\mu}
+\Omega\indices{^I_{J\mu}}\,E\indices{^J_\nu}-\Omega\indices{^I_{J\nu}}\,E\indices{^J_\mu}\right)\nonumber\\
&\left.+\,\onehalf\HOc\indices{_I^{J\mu\nu}}\left(\pfrac{\Omega\indices{^I_{J\mu}}}{X^\nu}+\pfrac{\Omega\indices{^I_{J\nu}}}{X^\mu}
+\Omega\indices{^I_{N\mu}}\,\Omega\indices{^N_{J\nu}}-\Omega\indices{^I_{N\nu}}\,\Omega\indices{^N_{J\mu}}\right)\right].
\label{eq:f3-div-total}
\end{align}
The final gauge Hamiltonian $\tilde{\HCd}_{\mathrm{Gau}_2}$ follows from Eq.~(\ref{eq:f3-div-total}) as:
\begin{align}
\tilde{\HCd}_{\mathrm{Gau}_2}
&=\iquarter\tilde{\kappabar}\indices{^\beta}\,\ho\indices{^i_{j\beta}}\,\sigma\indices{_i^j}\,\psi
-\iquarter\psibar\,\ho\indices{^i_{j\beta}}\,\sigma\indices{_i^j}\,\tilde{\kappa}^{\beta}\label{eq:HGgammaomega}\\
&\quad+\onehalf\tilde{k}\indices{_i^{\alpha\beta}}\left(
\ho\indices{^{i}_{j\alpha}}\,e\indices{^j_\beta}-\ho\indices{^{i}_{j\beta}}\,e\indices{^j_\alpha}\right)
+\onehalf\hoc\indices{_{i}^{j\alpha\beta}}\left(
\ho\indices{^{i}_{n\alpha}}\,\ho\indices{^{n}_{j\beta}}
-\ho\indices{^{i}_{n\beta}}\,\ho\indices{^{n}_{j\alpha}}\right).\nonumber
\end{align}
The partial derivatives associated with $\tilde{p}^{\mu\nu}$, $\tilde{k}\indices{_i^{\mu\nu}}$, and
$\hoc\indices{_i^{j\mu\nu}}$ in~(\ref{eq:f3-div-total}) are merged with
the corresponding derivatives contained in the initial action functional~(\ref{action-integral3})
to yield the following modified action functional
\begin{align}
S&=\int_{V}\Bigg[\tilde{\pi}^\nu\pfrac{\varphi}{x^{\nu}}+\onehalf\tilde{p}^{\mu\nu}\left(\pfrac{a_\mu}{x^\nu}-\pfrac{a_\nu}{x^\mu}\right)
+\tilde{\kappabar}^\nu\pfrac{\psi}{x^{\nu}}+\pfrac{\psibar}{x^{\nu}}\tilde{\kappa}^\nu\nonumber\\
&+\onehalf\tilde{k}\indices{_i^{\mu\nu}}\left(\pfrac{e\indices{^i_\mu}}{x^{\nu}}-\pfrac{e\indices{^i_\nu}}{x^\mu}\right)
+\onehalf\hoc\indices{_i^{\,k\mu\nu}}\left(\pfrac{\ho\indices{^i_{k\mu}}}{x^{\nu}}-\pfrac{\ho\indices{^i_{k\nu}}}{x^\mu}\right)
-\tilde{\HCd}_0-\tilde{\HCd}_{\mathrm{Gau}_2}\Bigg]\d^4x.
\label{action-integral4}
\end{align}
Finally the ``free gravity'' Hamiltonian $\tilde{\HCd}_{\mathrm{Gr}}(\tilde{q},\tilde{k},e)$ must be added to $\tilde{\HCd}_0$,
which is supposed to describe the dynamics of the free gravitational field, hence its dynamics in the absence of any sources of gravitation.
Such a Hamiltonian will be discussed in Sect.~\ref{sec:Hgr}.
The total system Hamiltonian
\begin{equation}\label{H2-def}
\tilde{\HCd}_2=\tilde{\HCd}_0+\tilde{\HCd}_{\mathrm{Gau}_2}+\tilde{\HCd}_{\mathrm{Gr}}
\end{equation}
is thus given as the sum of the Hamiltonian of the scalar, vector, and spinor source fields, $\tilde{\HCd}_0$,
the free gravity Hamiltonian $\tilde{\HCd}_{\mathrm{Gr}}$, and the gauge Hamiltonian $\tilde{\HCd}_{\mathrm{Gau}_2}$.

The final form-invariant action functional is now given for the Hamiltonians
of the initially uncoupled system of real scalar, real vector, and complex spinor fields
$\tilde{\HCd}_{0}\big(\tilde{\pi},\varphi,\tilde{p},a,\tilde{\kappa},\psibar,\tilde{\kappabar},\psi,e\big)$
and the system of the source-free gravitational field $\tilde{\HCd}_{\mathrm{Gr}}\big(\tilde{k},e,\hoc\,\big)$:
\begin{align}
S=\int_{V}&\d^4x\Bigg\{\tilde{\pi}^\nu\pfrac{\varphi}{x^\nu}
+\onehalf\tilde{p}^{\mu\nu}\left(\pfrac{a_\mu}{x^\nu}-\pfrac{a_\nu}{x^\mu}\right)\nonumber\\
&+\tilde{\kappabar}^\nu\left(\pfrac{\psi}{x^\nu}-\iquarter\ho\indices{^i_{j\nu}}\,\sigma\indices{_i^j}\,\psi\right)
+\left(\pfrac{\psibar}{x^\nu}+\iquarter\psibar\,\ho\indices{^i_{j\nu}}\,\sigma\indices{_i^j}\right)\tilde{\kappa}^\nu\nonumber\\
&+\onehalf\tilde{k}\indices{_i^{\mu\nu}}\left(\pfrac{e\indices{^i_\mu}}{x^\nu}-\pfrac{e\indices{^i_\nu}}{x^\mu}
+\ho\indices{^i_{j\nu}}\,e\indices{^j_\mu}-\ho\indices{^i_{j\mu}}\,e\indices{^j_\nu}\right)\nonumber\\
&+\onehalf\hoc\indices{_i^{j\mu\nu}}\left(\pfrac{\ho\indices{^i_{j\mu}}}{x^\nu}
-\pfrac{\ho\indices{^i_{j\nu}}}{x^\mu}+\ho\indices{^i_{n\nu}}\,\ho\indices{^n_{j\mu}}
-\ho\indices{^i_{n\mu}}\,\ho\indices{^n_{j\nu}}\right)-\tilde{\HCd}_0-\tilde{\HCd}_{\mathrm{Gr}}\Bigg\},\label{action-integral5}
\end{align}
which is closed as it does not contain external dependencies.
It includes torsion of spacetime and is \emph{not} restricted to metric compatibility, hence to
vanishing covariant derivatives of the metric.
Obviously, only the skew-symmetric portions of $\tilde{k}\indices{_i^{\mu\nu}}$ and
$\hoc\indices{_i^{j\mu\nu}}$ in $\mu$ and $\nu$ contribute to the action $S$.
We may thus generally assume these tensors to be skew-symmetric in those indices.
Also, only the skew-symmetric portion of $\ho_{ij\nu}$ contributes to the action $S$
in the spinor-related terms due to the skew-symmetry of $\sigma^{ij}$.

\subsection{Restriction to metric compatibility}
In this section the complex expressions encountered in the final action integral~(\ref{action-integral5}) will be linked to physical entities which will allow to introduce a well defined simplified notations.
Metric compatibility then pops up naturally by restricting the symmetry of the spin connection.

To start off we define the quantity $\gamma\indices{^\xi_{\mu\nu}}$ via
\begin{equation}
 \gamma\indices{^\xi_{\mu\nu}}:=e\indices{_i^\xi}\left(\pfrac{e\indices{^i_\mu}}{x^\nu}+\ho\indices{^i_{j\nu}}\,e\indices{^j_\mu}\right)
\label{eq:def-affconn}
\end{equation}
in terms of the vierbein and the spin connection. This relation can easily be reversed to
\begin{equation}
\ho\indices{^j_{i\nu}}
=e\indices{_i^\alpha}\left(\gamma\indices{^\xi_{\alpha\nu}}e\indices{^j_\xi}-\pfrac{e\indices{^j_\alpha}}{x^\nu}\right)
\label{eq:def-spinconn}
\end{equation}
or re-written as
\begin{equation}\label{eq:partial-deri-tetr}
\pfrac{e\indices{^i_\mu}}{x^\nu}=e\indices{^i_\xi}\gamma\indices{^\xi_{\mu\nu}}-\ho\indices{^i_{j\nu}}\,e\indices{^j_\mu},
\end{equation}
and, similarly, for the inverse vierbein:
\begin{equation}\label{eq:partial-deri-tetr-inv}
\pfrac{e\indices{_i^\mu}}{x^\nu}=\ho\indices{^j_{i\nu}}\,e\indices{_j^\mu}-e\indices{_i^\xi}\gamma\indices{^\mu_{\xi\nu}}.
\end{equation}
Applying now the transformation rule for the spin connections $\ho\indices{^i_{j\nu}}$ from Eq.~(\ref{omegatransform1}),
\begin{equation}\label{eq:omegatransform2}
\ho\indices{^i_{j\nu}}(x)=\Lambda\indices{^i_I}\,\HO\indices{^I_{J\alpha}}(X)\,\Lambda\indices{^J_j}\,\pfrac{X^\alpha}{x^\nu}
+\Lambda\indices{^i_I}\,\pfrac{\Lambda\indices{^I_j}}{x^{\nu}},
\end{equation}
translates into the transformation rule for the $\gamma\indices{^\xi_{\mu\nu}}$ as follows:
\begin{equation}\label{eq:affinetransform}
\gamma\indices{^\xi_{\mu\nu}}(x)=\Gamma\indices{^\tau_\beta_\alpha}(X)\pfrac{x^\xi}{X^\tau}\pfrac{X^\beta}{x^\mu}\pfrac{X^\alpha}{x^\nu}
+\ppfrac{X^\alpha}{x^\mu}{x^\nu}\pfrac{x^\xi}{X^\alpha}.
\end{equation}
This obviously represents the usual transformation law for the \emph{affine connection},
and the quantity $\gamma\indices{^\xi_{\mu\nu}}$ can be \emph{identified} with it. 
It is common to call Eq.~(\ref{eq:partial-deri-tetr}) \emph{Vierbein Postulate} and introduce
therein a covariant derivative that acts on both kinds of indices, on the Lorentz indices using the spin connection,
and on the coordinate indices using the affine connection:
\begin{equation}\label{def:vierbeinpostulate}
e\indices{^i_{\mu;\nu}}:=\pfrac{e\indices{^i_\mu}}{x^\nu}-e\indices{^i_\xi}\gamma\indices{^\xi_{\mu\nu}}+\ho\indices{^i_{k\nu}}\,e\indices{^k_\mu}=0.
\end{equation}
We then find for the covariant derivative of the metric
\begin{align}
g_{\mu\nu;\alpha}&=e\indices{^i_{\mu;\alpha}}\,e\indices{^j_{\nu}}\,\eta_{ij}+e\indices{^j_{\nu;\alpha}}\,e\indices{^i_{\mu}}\,\eta_{ij}
+e\indices{^i_{\mu}}\,e\indices{^j_{\nu}}\,\eta_{ij;\alpha}\nonumber\\
&=e\indices{^i_{\mu}}\,e\indices{^j_{\nu}}\left(\eta_{kj}\,\ho\indices{^k_i_\alpha}+\eta_{ik}\,\ho\indices{^k_j_\alpha}\right)\nonumber\\
&=2e\indices{^i_{\mu}}\,e\indices{^j_{\nu}}\,\ho\indices{_{(ij)}_\alpha}.
\label{eq:metricity-condition}
\end{align}
With the above definition of the affine connection, hence with the validity of the Vierbein Postulate,
we conclude that for \emph{metric compatibility} the spin connection must be skew-symmetric in its Lorentz indices.
This is consistent with the transformation rule for the spin connection as becomes evident when lowering the index $i$ in Eq.~(\ref{eq:omegatransform2}).
We find that $\ho\indices{_i_{j\nu}}$ must indeed have a skew-symmetric portion due to
the skew-symmetry of the inhomogeneous term, as shown in Eq.~(\ref{eq:skew-inhom}).
On the other hand, $\ho\indices{_i_{j\nu}}$ may also have a symmetric portion in $i,j$,
that would lead to a non-zero metricity tensor $Q_{\mu\nu\alpha}:=g_{\mu\nu;\alpha}$,
as a symmetric $\HO\indices{^I^J_\alpha}$ transforms into a symmetric $\ho\indices{^i^j_\alpha}$ without affecting the inhomogeneous term:
\begin{align*}
\Lambda\indices{_i_I}\,\HO\indices{^{(IJ)}_\alpha}\,\Lambda\indices{_J_j}
&=\Lambda\indices{_I_i}\,\HO\indices{^{(IJ)}_\alpha}\,\Lambda\indices{_j_J}
=\Lambda\indices{_j_J}\,\HO\indices{^{(IJ)}_\alpha}\,\Lambda\indices{_I_i}
=\Lambda\indices{_j_I}\,\HO\indices{^{(JI)}_\alpha}\,\Lambda\indices{_J_i}\\
&=\Lambda\indices{_j_I}\,\HO\indices{^{(IJ)}_\alpha}\,\Lambda\indices{_J_i}.
\end{align*}
Conversely, the affine connection $\gamma\indices{^\xi_{\mu\nu}}$ must have a symmetric portion due to the symmetry of the inhomogeneous term in $\mu,\nu$.
Yet, it may as well have a skew-symmetric portion (aka torsion) if both, $\gamma\indices{^\xi_{\mu\nu}}$ and
$\Gamma\indices{^\tau_\beta_\alpha}$ are not symmetric in their lower indices.
The transformation laws~(\ref{eq:omegatransform2}) and~(\ref{eq:affinetransform}) are thus equivalent
as both do not require any symmetries of the connections $\ho\indices{^i_{j\nu}}$ and $\gamma\indices{^\xi_{\mu\nu}}$,
hence both comprise the full range of $4^3=64$ independent coefficients in a four-dimensional spacetime.

\medskip
Henceforth we request that the spin connection is anti-symmetric and allow the affine connection to be asymmetric, i.e. we stipulate metric compatibility and allow torsion as a structural element of spacetime.

\medskip
The definitions
\begin{subequations}\label{eq:T-R-def}
\begin{align}
S\indices{^i_{\mu\nu}}&:=\onehalf \left(\pfrac{e\indices{^i_\mu}}{x^\nu}-\pfrac{e\indices{^i_\nu}}{x^\mu}
+\ho\indices{^i_{n\nu}}\,e\indices{^n_\mu}-\ho\indices{^i_{n\mu}}\,e\indices{^n_\nu} \right)
\label{eq:T-def2} \\
R\indices{^i_{j\nu\mu}}&:=\pfrac{\ho\indices{^i_{j\mu}}}{x^\nu}-\pfrac{\ho\indices{^i_{j\nu}}}{x^\mu}
+\ho\indices{^i_{n\nu}}\,\ho\indices{^n_{j\mu}}-\ho\indices{^i_{n\mu}}\,\ho\indices{^n_{j\nu}}
\label{eq:omega-deri3}
\end{align}
\end{subequations}
acquire a physical meaning once we plug in the definition~(\ref{eq:def-affconn}) of the affine connection:
\begin{subequations}
\begin{align}
S\indices{^i_{\mu\nu}}
&\equiv \onehalf e\indices{^i_\xi}\left(\gamma\indices{^\xi_\mu_\nu}-\gamma\indices{^\xi_\nu_\mu}\right)
= e\indices{^i_\xi}\,S\indices{^\xi_{\mu\nu}} \label{eq:T-def3} \\
R\indices{^i_{j\nu\mu}}&\equiv e\indices{^i_\xi}\,e\indices{_j^\eta}R\indices{^\xi_{\eta\nu\mu}}.
\label{eq:R-def}
\end{align}

Obviously $S\indices{^\xi_{\mu\nu}}$ thus denotes the Cartan torsion tensor, and $R\indices{^\xi_{\eta\nu\mu}}$ is the Rie\-mann-Cartan curvature tensor.
The proof of the correlation of $R\indices{^\xi_{\eta\nu\mu}}$ and $R\indices{^i_{j\nu\mu}}$
is given in \aref{app:riemanntensor}.

For further simplifying the notation in Eq.~(\ref{action-integral5}), we define the Proca field strength tensor
\begin{align}
 f_{\nu\mu} &:= \pfrac{a_\mu}{x^\nu}-\pfrac{a_\nu}{x^\mu}. \label{def:fmunu}
\end{align}
\end{subequations}
Notice in addition that a new \emph{spinor covariant derivatives} has emerged for the spinor fields that we abbreviate as
\begin{subequations}\label{Def:Spincovderivative}
\begin{alignat}{4}
\Dderr_\mu&:=\pfracr{}{x^\mu}-\iquarter\sigma^{lj}\,\omega\indices{_{lj\mu}}
\label{Def:Spincovderivative1}\\
\Dderl_\mu&:=\pfracl{}{x^\mu}+\iquarter\sigma^{lj}\,\omega\indices{_{lj\mu}}
\equiv\gamma^0\,\Dderr_\mu^\dagger\,\gamma^0.\label{Def:Spincovderivative2}
\end{alignat}
\end{subequations}
The Lorentz connection corrects the partial derivative as it is common for gauge fields.
Notice that while the spinor and coordinate indices in these tensor operators are open, the Lorentz indices are contracted out.
With the spinor indices $a,b$ made explicit this reads
\begin{equation*} 
\left(\Dderr_\eta\right)^a_b := \delta^a_b\,\pfracr{}{x^\eta} - \iquarter\left(\sigma^{lj}\right)^a_b\,\omega\indices{_{lj\eta}}.
\end{equation*}

Then the action integral~(\ref{action-integral5}), invariant w.r.t. the transformation group SO(1,3) $\times$ Diff(M), can be re-written in the compact form
\begin{align} \label{action-integral6}
S=\int_{V}\d^4x\Bigg\{&\tilde{\pi}^\nu\pfrac{\varphi}{x^\nu}
+\onehalf\tilde{p}^{\mu\nu}\,f_{\nu\mu}+\tilde{\kappabar}^\nu\,\Dderr_\nu\,\psi+\psibar\,\Dderl_\nu\,\tilde{\kappa}^\nu \nonumber\\
&+\tilde{k}\indices{_i^{\mu\nu}}\,S\indices{^i_{\mu\nu}}
+\onehalf\hoc\indices{_i^{j\mu\nu}}\,R\indices{^i_{j\nu\mu}}-\tilde{\HCd}_0-\tilde{\HCd}_{\mathrm{Gr}}\Bigg\}.
\end{align}

\section{Canonical field equations\label{sec:cfeq}}
The variation of the action integral with respect to the conjugate pairs of fields is now based on the amended version \eref{action-integral6}, and is carried out in the next subsections.
\subsection{Canonical equations for the scalar field}
As the scalar field $\varphi$ does not couple to the gauge field $\ho\indices{^i_{j\nu}}$
and hence does not contribute to $\HCd_{\mathrm{Gau}_2}$, the field equation take their usual form:
\begin{equation}\label{eq:phi-feqs}
\pfrac{\varphi}{x^\nu}=\pfrac{\tilde{\HCd}_0}{\tilde{\pi}^\nu},\qquad\pfrac{\tilde{\pi}^\nu}{x^\nu}=-\pfrac{\tilde{\HCd}_0}{\varphi}.
\end{equation}
The coupling of scalar fields to a dynamic spacetime thus merely occurs via the common dependence of
$\tilde{\HCd}_0$ and $\tilde{\HCd}_{\mathrm{Gr}}$ on the vierbein field $e\indices{^i_\nu}$.
In contrast, vector and spinor fields couple directly to the gauge field $\ho\indices{^i_{j\nu}}$,
as will be shown in the following.
\subsection{Canonical equations for the vector field}
The equation for the derivative of the vector field $a_\mu$ emerges as:
\begin{equation}
\pfrac{a_\mu}{x^\nu}-\pfrac{a_\nu}{x^\mu} \equiv f_{\nu\mu} = 2\pfrac{\tilde{\HCd}_2}{\tilde{p}^{\,\mu\nu}}=2\pfrac{\tilde{\HCd}_0}{\tilde{p}^{\,\mu\nu}}.
\label{eq:a-deri}
\end{equation}
This shows that $\tilde{p}^{\,\mu\nu}$ must be skew-symmetric.
The conjugate canonical equation for the divergence of the momentum field $\tilde{p}^{\mu\nu}$ is then:
\begin{equation}
\pfrac{\tilde{p}^{[\mu\nu]}}{x^\nu}=-\pfrac{\tilde{\HCd}_2}{a_\mu}=-\pfrac{\tilde{\HCd}_0}{a_\mu},
\label{eq:p-div}
\end{equation}
which is actually a tensor equation as will be demonstrated explicitly below.
\subsection{Canonical equations for the spinor fields}
For the spinors $\psi$ and $\psibar$ the set of canonical equations is:
\begin{subequations}\label{eq:spinor-feqs}
\begin{align}
\pfrac{\psi}{x^\nu}&=\hphantom{-}\pfrac{\tilde{\HCd}_2}{\tilde{\kappabar}^\nu}
=\hphantom{-}\pfrac{\tilde{\HCd}_0}{\tilde{\kappabar}^\nu}+\iquarter\,\ho\indices{^i_{j\nu}}\,\sigma\indices{_i^j}\,\psi\\
\pfrac{\tilde{\kappabar}^\nu}{x^\nu}&=-\pfrac{\tilde{\HCd}_2}{\psi}=-\pfrac{\tilde{\HCd}_0}{\psi}-
\iquarter\,\tilde{\kappabar}^\nu\,\ho\indices{^i_{j\nu}}\,\sigma\indices{_i^j}\\
\pfrac{\psibar}{x^\nu}&=\hphantom{-}\pfrac{\tilde{\HCd}_2}{\tilde{\kappa}^\nu}
=\hphantom{-}\pfrac{\tilde{\HCd}_0}{\tilde{\kappa}^\nu}-\iquarter\,\psibar\,\ho\indices{^i_{j\nu}}\,\sigma\indices{_i^j}\\
\pfrac{\tilde{\kappa}^\nu}{x^\nu}&=-\pfrac{\tilde{\HCd}_2}{\psibar}=-\pfrac{\tilde{\HCd}_0}{\psibar}+
\iquarter\,\ho\indices{^i_{j\nu}}\,\sigma\indices{_i^j}\,\tilde{\kappa}^\nu.
\end{align}
\end{subequations}

\subsection{Canonical equations for the vierbein field}

For the canonical equation for the vierbein field $e\indices{^i_{\mu}}$ we get
\begin{equation*}
\pfrac{e\indices{^i_\mu}}{x^\nu}-\pfrac{e\indices{^i_\nu}}{x^\mu}=2\pfrac{\tilde{\HCd}_2}{\tilde{k}\indices{_i^{\mu\nu}}}=
2\pfrac{\tilde{\HCd}_\mathrm{Gr}}{\tilde{k}\indices{_i^{\mu\nu}}}+
\ho\indices{^i_{j\mu}}\,e\indices{^j_\nu}-\ho\indices{^i_{j\nu}}\,e\indices{^j_\mu}.
\end{equation*}
Regrouping gives then the simple relation
\begin{align}
\pfrac{\tilde{\HCd}_\mathrm{Gr}}{\tilde{k}\indices{_i^{\mu\nu}}}&=\onehalf\left(\pfrac{e\indices{^i_{\mu}}}{x^\nu}
-\pfrac{e\indices{^i_\nu}}{x^\mu}+\ho\indices{^i_{j\nu}}\,e\indices{^j_\mu}-\ho\indices{^i_{j\mu}}\,e\indices{^j_\nu}\right)
=S\indices{^i_\mu_\nu}.
\label{eq:e-deri}
\end{align}
From the action integral \eref{action-integral6} we conclude that $\tilde{k}\indices{_i^{\mu\nu}}$ is skew-symmetric in its last index pair.
Then the canonical equation for its divergence becomes
\begin{align}
\pfrac{\tilde{k}\indices{_i^{[\mu\alpha]}}}{x^\alpha}&=-\pfrac{\tilde{\HCd}_2}{e\indices{^i_{\mu}}}=
-\pfrac{\tilde{\HCd}_0}{e\indices{^i_{\mu}}}-\pfrac{\tilde{\HCd}_\mathrm{Gr}}{e\indices{^i_{\mu}}}-\pfrac{\tilde{\HCd}_\mathrm{Gau_2}}{e\indices{^i_{\mu}}}\nonumber\\
&=-\pfrac{\tilde{\HCd}_0}{e\indices{^i_{\mu}}}-\pfrac{\tilde{\HCd}_{\mathrm{Gr}}}{e\indices{^i_{\mu}}}
+\onehalf\left(\tilde{k}\indices{_j^{\mu\alpha}}-\tilde{k}\indices{_j^{\alpha\mu}}\right)\ho\indices{^{j}_{i\alpha}}\nonumber\\
&=-\pfrac{\tilde{\HCd}_0}{e\indices{^i_{\mu}}}-\pfrac{\tilde{\HCd}_{\mathrm{Gr}}}{e\indices{^i_{\mu}}}
+\tilde{k}\indices{_j^{[\mu\alpha]}}\,\ho\indices{^{j}_{i\alpha}}.
\label{eq:k-div}
\end{align}
Regrouping the terms of Eq.~(\ref{eq:k-div}) yields:
\begin{equation}\label{eq:k-div3}
\left(\pfrac{\tilde{k}\indices{_i^{[\mu\alpha]}}}{x^\alpha}-\tilde{k}\indices{_j^{[\mu\alpha]}}\,\ho\indices{^{j}_{i\alpha}}\right)e\indices{^i_{\nu}}
=-\pfrac{\tilde{\HCd}_{\mathrm{Gr}}}{e\indices{^i_{\mu}}}e\indices{^i_{\nu}}-\pfrac{\tilde{\HCd}_0}{e\indices{^i_{\mu}}}e\indices{^i_{\nu}}.
\end{equation}
The terms on the right-hand side of Eq.~(\ref{eq:k-div3}) are minus the Hamiltonian representation of the \emph{metric}
energy-momentum tensors of the free gravitational system $\tilde{\HCd}_{\mathrm{Gr}}$ and the source system $\tilde{\HCd}_0$. In terms of these, the field equations reduce to
\begin{align}
\left(\pfrac{\tilde{k}\indices{_i^{[\mu\alpha]}}}{x^\alpha}-\tilde{k}\indices{_j^{[\mu\alpha]}}\,\ho\indices{^{j}_{i\alpha}}\right)e\indices{^i_{\nu}}
=-\tilde{T}_{\text{Gr\,}\nu}{}^\mu-\tilde{T}_{0\,\nu}{}^\mu.
\end{align}
With
\begin{align}
\deteinv \, \pfrac{\tilde{k}_i{}^{[\mu\alpha]}}{x^\alpha} = {k}_i{}^{[\mu\alpha]}{}_{;\alpha}
+\omega^j{}_{i\alpha}{k}_j{}^{[\mu\alpha]}-\gamma^\mu{}_{\beta\alpha}{k}_i{}^{[\beta\alpha]}-\gamma^\alpha{}_{\beta\alpha}{k}_i{}^{[\mu\beta]} + \gamma^\alpha{}_{\alpha\beta}{k}_i{}^{[\mu\beta]}
\end{align}
this gives  a
tensor equation of motion for the momentum field ${k}_i{}^{[\mu\alpha]}{}$ the source of which is the total energy momentum of matter and gravity:
\begin{align}
{k}_i{}^{[\mu\alpha]}{}_{;\alpha} - S^\mu{}_{\beta\alpha}{k}_i{}^{[\beta\alpha]}
- 2S^\alpha{}_{\beta\alpha}{k}_i{}^{[\mu\beta]}
=-{T}_{\text{Gr\,}\nu}{}^\mu-{T}_{0\,\nu}{}^\mu.
\end{align}
\subsection{Canonical equations for the connection field}
The canonical equation for the derivative of the gauge field $\ho\indices{^i_{j\mu}}$ follows as
\begin{equation*}
\pfrac{\ho\indices{^i_{j\mu}}}{x^\nu}-\pfrac{\ho\indices{^i_{j\nu}}}{x^\mu}
=2\pfrac{\tilde{\HCd}_2}{\hoc\indices{_i^{j\mu\nu}}}=
2\pfrac{\tilde{\HCd}_\mathrm{Gr}}{\hoc\indices{_i^{j\mu\nu}}}+
\ho\indices{^i_{n\mu}}\,\ho\indices{^n_{j\nu}}-\ho\indices{^i_{n\nu}}\,\ho\indices{^n_{j\mu}},
\end{equation*}
which shows that $\hoc\indices{_i^{j\mu\nu}}$ must be skew-symmetric in its last index pair.
With the definition \eqref{eq:R-def} this gives:
\begin{equation}\label{eq:omega-deri}
\pfrac{\tilde{\HCd}_\mathrm{Gr}}{\hoc\indices{_i^{j\mu\nu}}}=
\onehalf\left(\pfrac{\ho\indices{^i_{j\mu}}}{x^\nu}-\pfrac{\ho\indices{^i_{j\nu}}}{x^\mu}
+\ho\indices{^i_{n\nu}}\,\ho\indices{^n_{j\mu}}-\ho\indices{^i_{n\mu}}\,\ho\indices{^n_{j\nu}}\right)
=\onehalf R\indices{^i_{j\nu\mu}}.
\end{equation}
The canonical equation for the divergence of $\hoc\indices{_i^{j\mu\nu}}$ follows as
\begin{align}
\pfrac{\hoc\indices{_i^{j\mu\nu}}}{x^\nu}&=-\pfrac{\tilde{\HCd}_2}{\ho\indices{^i_{j\mu}}}
=-\pfrac{\tilde{\HCd}_{\mathrm{Gau}_2}}{\ho\indices{^i_{j\mu}}}\nonumber\\
&=-\iquarter\tilde{\kappabar}^\mu\,\sigma\indices{_i^j}\,\psi+\iquarter\psibar\,\sigma\indices{_i^j}\,\tilde{\kappa}^\mu
+\tilde{k}\indices{_i^{\beta\mu}}e\indices{^j_\beta}
+\hoc\indices{_i^{n\beta\mu}}\ho\indices{^j_{n\beta}}-\hoc\indices{_n^{j\beta\mu}}\ho\indices{^n_{i\beta}}\label{eq:t-div}.
\end{align}
Again, realizing that
\begin{align*}
\deteinv \, \pfrac{\tilde{q}_i{}^{j\mu\nu}}{x^\nu} =&\; {q}_i{}^{j\mu\nu}{}_{;\nu}
-\omega^n{}_{i\nu}{q}_n{}^{j\nu\mu}+\omega^j{}_{n\nu}{q}_i{}^{n\nu\mu}\\
&-\gamma^\mu{}_{\lambda\nu}{q}_i{}^{n\lambda \nu}-\gamma^\nu{}_{\lambda\nu}{q}_i{}^{n\mu\lambda}-\gamma^\nu{}_{\nu\lambda}{q}_i{}^{n\mu\lambda},
\end{align*}
we can re-write \eref{eq:t-div} as
\begin{equation} \label{eq:t-divv}
{q}_i{}^{j\mu\nu}{}_{;\nu} -S^\mu{}_{\lambda\nu}{q}_i{}^{j\lambda \nu} - 2S^\nu{}_{\lambda\nu}{q}_i{}^{j\mu\lambda} = {k}\indices{_i^{\beta\mu}}e\indices{^j_\beta} + \Sigma _i{}^{j\mu}.
\end{equation}
\subsection{Summary of the coupled set of field equations}
The complete closed set of twelve coupled field equations for a system of scalar, vector, and spinor fields
in a dynamic spacetime---which can also directly be read off the variation of the action functional~(\ref{action-integral5})---is summarized in the following.
The field equations related to the given system Hamiltonian $\tilde{\HCd}_0$
describe the dynamics of the scalar, vector, and spinor fields as well as their
coupling to the dynamic spacetime, as expressed by the dynamic vierbeins $e\indices{^j_\mu}$ and the connection field~$\ho\indices{^i_{j\nu}}$:
\allowdisplaybreaks
\begin{subequations}\label{eq:feqs-all}
\begin{align}
\pfrac{\varphi}{x^\nu}&=\hphantom{-}\pfrac{\tilde{\HCd}_0}{\tilde{\pi}^\nu}\label{eq:phi_deri}\\
\pfrac{\tilde{\pi}^\nu}{x^\nu}&=-\pfrac{\tilde{\HCd}_0}{\varphi}\label{eq:pi_div}\\
f_{\nu\mu}&=2\,\pfrac{\tilde{\HCd}_0}{\tilde{p}^{\,\mu\nu}}\label{eq:a_deri}\\
\pfrac{\tilde{p}^{[\mu\nu]}}{x^\nu}&=-\pfrac{\tilde{\HCd}_0}{a_\mu}\label{eq:p_div}\\
\Dderr_\nu\,\psi &=\hphantom{-}\pfrac{\tilde{\HCd}_0}{\tilde{\kappabar}^\nu}
\label{eq:psi-deri2} \\
\tilde{\kappabar}^\nu\,\Dderl_\nu &=-\pfrac{\tilde{\HCd}_0}{\psi}
\label{eq:kappabar-div2}\\
\psibar\,\Dderl_\nu&=\hphantom{-}\pfrac{\tilde{\HCd}_0}{\tilde{\kappa}^\nu}
\label{eq:psibar-deri2}\\
\Dderr_\nu\,\tilde{\kappa}^\nu&=-\pfrac{\tilde{\HCd}_0}{\psibar}
.\label{eq:kappa-div2}
\end{align}
\allowdisplaybreaks[0]
The dynamics of the vierbeins, the connection fields, and their respective conjugates are described by:
\begin{align}
S\indices{^i_{\mu\nu}}&=\pfrac{\tilde{\HCd}_\mathrm{Gr}}{\tilde{k}\indices{_i^{\mu\nu}}}
\label{eq:e-deri2}\\
R\indices{^i_{j\nu\mu}}&=2\pfrac{\tilde{\HCd}_\mathrm{Gr}}{\hoc\indices{_i^{j\mu\nu}}}
\label{eq:omega-deri2}\\
{k}_i{}^{[\mu\nu]}{}_{;\nu} - S^\mu{}_{\beta\nu}{k}_i{}^{[\beta\nu]}
- 2S^\nu{}_{\beta\nu}{k}_i{}^{[\mu\beta]}
&=-{T}_{\text{Gr\,}\nu}{}^\mu-{T}_{0\,\nu}{}^\mu \label{eq:k-div2}\\
{q}_i{}^{j\mu\nu}{}_{;\nu} -S^\mu{}_{\beta\nu}{q}_i{}^{j\beta \nu} - 2S^\nu{}_{\beta\nu}{q}_i{}^{j\mu\beta}
&= {k}\indices{_i^{\beta\mu}}e\indices{^j_\beta} - \Sigma _i{}^{j\mu}
\label{eq:t-div-2}.
\end{align}
\end{subequations}
\subsection{Consistency equation\label{sec:consistency-general}}
The canonical equations as derived in the Hamiltonian formalism are first-order differential equations that determine the dynamics
of the coupled conjugate pairs of matter and spacetime fields.
In contrast, the field equation that emerge in the Lagrangian picture are second order differential equations where the momentum fields
are substituted by the derivatives of the fields, in analogy to momenta in point mechanics being replaced by particle velocities.
Hence, in order to get an equation that is comparable to Einstein's field equation, we need to carry out a similar substitution.
The resulting equation is also called \emph{consistency equation} as it relates matter and spacetime expressions.

We start by taking a derivative of equation~(\ref{eq:t-div}) with respect to $x^{\beta}$.
The left-hand side vanishes due to the skew-symmetry of $\tilde{q}\indices{_i^{j[\beta\alpha]}}$ in its last index pair:
\allowdisplaybreaks[0]
\begin{align}
0=&\,\pfrac{}{x^\beta}\left(
\iquarter\psibar\,\sigma\indices{_i^j}\,\tilde{\kappa}^\beta-\iquarter\tilde{\kappabar}^\beta\,\sigma\indices{_i^j}\,\psi
+\tilde{k}\indices{_i^{[\alpha\beta]}}e\indices{^j_\alpha}\right)\nonumber\\
&+\hoc\indices{_i^{n[\alpha\beta]}}\,\pfrac{\ho\indices{^j_{n\alpha}}}{x^\beta}
-\pfrac{\ho\indices{^n_{i\alpha}}}{x^\beta}\,\hoc\indices{_n^{\,j[\alpha\beta]}}
+\pfrac{\hoc\indices{_i^{n\beta\alpha}}}{x^\alpha}\ho\indices{^j_{n\beta}}
-\ho\indices{^n_{i\beta}}\pfrac{\hoc\indices{_n^{j\beta\alpha}}}{x^\alpha}.\label{eq:consistency1}
\end{align}
Inserting the above cited field equation~(\ref{eq:t-div}) for the first derivatives of $\tilde{q}$ gives:
\begin{align}
0&=\pfrac{}{x^\beta}\left(
\iquarter\psibar\sigma\indices{_i^j}\tilde{\kappa}^\beta-\iquarter\tilde{\kappabar}^\beta\sigma\indices{_i^j}\psi
+\tilde{k}\indices{_i^{[\alpha\beta]}}e\indices{^j_\alpha}\right)
+\hoc\indices{_i^{n[\alpha\beta]}}\pfrac{\ho\indices{^j_{n\alpha}}}{x^\beta}
-\pfrac{\ho\indices{^n_{i\alpha}}}{x^\beta}\hoc\indices{_n^{\,j[\alpha\beta]}}\nonumber\\
&+\left(\iquarter\psibar\sigma\indices{_i^n}\tilde{\kappa}^\beta
-\iquarter\tilde{\kappabar}^\beta\sigma\indices{_i^n}\psi
+\tilde{k}\indices{_i^{[\alpha\beta]}}e\indices{^n_\alpha}
+\hoc\indices{_i^{m[\alpha\beta]}}\ho\indices{^n_{m\alpha}}
-\cancel{\hoc\indices{_m^{\,n[\alpha\beta]}}\ho\indices{^m_{i\alpha}}}\,\right)\ho\indices{^j_{n\beta}}\nonumber\\
&-\ho\indices{^n_{i\beta}}\!\left(\iquarter\psibar\sigma\indices{_n^j}\tilde{\kappa}^\beta
-\iquarter\tilde{\kappabar}^\beta\sigma\indices{_n^j}\psi
+\tilde{k}\indices{_n^{[\alpha\beta]}}e\indices{^j_\alpha}
+\cancel{\hoc\indices{_n^{\,m[\alpha\beta]}}\ho\indices{^j_{m\alpha}}}
-\hoc\indices{_m^{\,j[\alpha\beta]}}\ho\indices{^m_{n\alpha}}\right)\!.
\label{eq:consistency0}
\end{align}
A lengthy calculation laid out in \aref{sec:gen-consistapp} yields finally:
\begin{align}
0&=\frac{\rmi}{4}\left(\pfrac{\tilde{\HCd}_0}{\psi}\sigma\indices{_i^j}\psi
-\tilde{\kappabar}^\alpha\sigma\indices{_i^j}\pfrac{\tilde{\HCd}_0}{\tilde{\kappabar}^\alpha}
+\pfrac{\tilde{\HCd}_0}{\tilde{\kappa}^\alpha}\sigma\indices{_i^j}\tilde{\kappa}^\alpha
-\psibar\,\sigma\indices{_i^j}\pfrac{\tilde{\HCd}_0}{\psibar}\right)
-\pfrac{\tilde{\HCd}_0}{e\indices{^i_\alpha}}e\indices{^j_\alpha}\nonumber\\
&\quad-\pfrac{\tilde{\HCd}_{\mathrm{Gr}}}{e\indices{^i_\alpha}}e\indices{^j_\alpha}
+\tilde{k}\indices{_i^{[\alpha\beta]}}\pfrac{\tilde{\HCd}_\mathrm{Gr}}{\tilde{k}\indices{_j^{\alpha\beta}}}
+\hoc\indices{_i^{m[\alpha\beta]}}\pfrac{\tilde{\HCd}_\mathrm{Gr}}{\hoc\indices{_j^{m\alpha\beta}}}
-\hoc\indices{_m^{j[\alpha\beta]}}\pfrac{\tilde{\HCd}_\mathrm{Gr}}{\hoc\indices{_m^{i\alpha\beta}}}.
\label{eq:consistency2}
\end{align}
This Einstein-type equation relates the derivatives of the Hamiltonian $\tilde{\HCd}_0$ of the matter fields
with the derivatives emerging from the model Hamiltonian $\tilde{\HCd}_{\mathrm{Gr}}$ for the free gravitational field.
It holds for any given Hamiltonian densities $\tilde{\HCd}_0$ and $\tilde{\HCd}_{\mathrm{Gr}}$.
The particular equation for the scalar, vector, spinor, and gravitational fields can be set up only after
specifying those Hamiltonians which will be done in the following sections.

\section{Free field Hamiltonians in curved spacetime\label{sec:matter-fields}}
The derivations presented in the previous sections were based solely on the transformation properties of the spin-0, spin-1, and spin-\nicefrac{1}{2}
fields and the derived transformation properties of the gauge fields.
While the transformation framework fixed the interaction between the matter and spacetime fields,
the specific form of the pertinent ``kinetic'' terms, the free, non-interacting Lagrangians $\tilde{\LCd}$  or equivalently Hamiltonians $\tilde{\HCd}$ have so far been kept open.
In this section that gap shall be closed and the details of the matter fields, namely the free Klein-Gordon, Maxwell-Proca and Dirac systems specified.
We thereby transfer the well known Lorentz covariant field equations of those matter fields with help of the vierbeins from the
inertial frames with the standard Minkowski metric into a curvilinear geometry with an arbitrary metric.
Then the particular Hamiltonians of mutually non-interacting scalar, vector, and spinor matter fields are derived, constituting the total free matter Hamiltonian density:
\begin{equation*}
 \tilde{\HCd}_{\text{0}} = \tilde{\HCd}_{\text{KG}} + \tilde{\HCd}_{\text{P}} + \tilde{\HCd}_{\text{D}}.
\end{equation*}
(Introducing mutual matter-matter interactions, e.g. Yang-Mills or Higgs models, are delegated to upcoming work. We are also aware that the Proca field is not a fundamental field but rather an element of effective field theories. Nevertheless, since vector fields will be key to abelian and non-abelian gauge theories we have included it in its Maxwell form in our analyzes.)
In addition, we present a particular, phenomenologically justified model Hamiltonian for the free gravitational field.
\subsection{Scalar fields\label{sec:KGfields}}
As the first term in the yet unspecified matter Hamiltonian $\tilde{\HCd}_{\text{0}}$ we select for the scalar field its real massive Klein-Gordon version. That gives a simple first impression of how the curvilinear nature of a a manifold is transferred to the dynamics of physical fields that live in local inertial frames.
The standard Lorentz invariant free Klein-Gordon Lagrangian in the inertial
frame is given by
\begin{equation}
\mathcal{L}^0_{\text{KG}}(\phi,\partial\phi)=\onehalf
\pfrac{\varphi}{x^i}\eta^{ij}\pfrac{\varphi}{x^j}-\onehalf m^2 \varphi^2.
\end{equation}
It is a function of fields and their derivatives, analogous to generalized coordinates and velocities in point dynamics. The Lagrangian must have the dimension $[\mathcal{L}_{\text{KG}}]=L^{-4}$, so that we find for the scalar field $[\varphi]=L^{-1}$. As $\partial\varphi/\partial x^i$ is a vector in the tangent space of the frame bundle written in the inertial frame base, it can be re-written in the coordinate base as
\begin{equation*}
\pfrac{\varphi}{x^i}=\pfrac{\varphi}{x^\alpha}\,e\indices{_i^\alpha},\qquad
\pfrac{\varphi}{x^\alpha}=e\indices{^{i}_{\alpha}}\,\pfrac{\varphi}{x^i}.
\end{equation*}
In order to express this Klein-Gordon Lagrangian in the coordinate frame, we use the relations~(\ref{eq:vierbein-def}) and the orthonormality identities~(\ref{def:emuienui}) of the vierbein fields. The coordinate frame Klein-Gordon Lagrangian is then
\begin{align}\label{KGLagrangian}
\tilde{\LCd}^0_{\mathrm{KG}}\left(\varphi,\partial_\alpha\varphi,e\indices{_k^\alpha}\right)
&=\onehalf\dete\left(\pfrac{\varphi}{x^\alpha}\,e\indices{_k^\alpha
}\,\eta^{kj}\,e\indices{_j^\beta}\,\pfrac{\varphi}{x^\beta}-m^2\,\varphi^2\right)\nonumber\\
&=\onehalf\dete\left(\pfrac{\varphi}{x^\alpha}\,g^{\alpha\beta}\,\pfrac{\varphi}{x^\beta}-m^2\,\varphi^2\right)
\end{align}
The factor $\dete$ makes the absolute scalar Lagrangian $\LCd^0_{\mathrm{KG}}$
from Eq.~(\ref{KGLagrangian}) into a relative scalar of weight $w=1$ and thus
maintains the correct integral measure.\\
To derive the corresponding Hamiltonian, we apply in the following the generic definition given in \sref{sec:cantransfKG} for the canonical conjugate ``momentum'' field:
\begin{equation}
\tilde{\pi}^\nu=\pfrac{\tilde{\mathcal{L}}^0_{\text{KG}}}{\left(\pfrac{\varphi}{
x ^\nu}\right)}=\varepsilon
e\indices{_k^\nu}\eta^{kj}e\indices{_j^\beta}\pfrac{\varphi}{x^\beta},
\end{equation}
which has the dimensions $[\tilde{\pi}^\nu]=L^{-2}$. The Hamiltonian can now be obtained by inverting the above relation and inserting it into the definition of the Hamiltonian:
\begin{align}\label{HKG1}
\tilde{\HCd}_{\text{KG}}(\phi,\tilde{\pi}^\nu,e\indices{^i_\mu})&=\tilde{\pi}^
\nu\pfrac{\varphi}{x^\nu}-\tilde{\LCd}^0_\text{KG}(\phi,\partial\phi,
e\indices{^i_\mu})\nonumber\\
&=\frac{1}{2\varepsilon}\tilde{\pi}^\alpha e\indices{^n_\alpha}\eta_{nm}e\indices{^m_\beta}\tilde{\pi}^\beta+\frac{\varepsilon}{2}m^2\varphi^2.
\end{align}
This Hamiltonian can now simply be inserted into the equations (\ref{eq:phi_deri}) and (\ref{eq:pi_div}) to obtain the field equations for a scalar field in curved spacetime. This step will be performed in Section \ref{sec:coupledeqs}.

\medskip
Notice in passing that the above calculations and those following below
also hold if we replace the mass term $\onehalf m^2 \varphi^2$ by a more general,
possibly explicitly spacetime-dependent potential term $V(\varphi,x)$.

\subsection{Vector fields}
Next, we want to deduce the Hamiltonian for the vector field. As our model, we employ the Maxwell-Proca system for a massive spin-1 field. In terms of the field strength tensor, which is the exterior derivative of the vector potential and thus a 2-form,
\begin{align*}
f_{ij}=\pfrac{a_j}{x^i}-\pfrac{a_i}{x^j},
\end{align*}
the Maxwell-Proca Lagrangian in the local inertial frame is given by
\begin{equation}
{\mathcal{L}}^0_\text{P}=-\quarter f_{ij}\eta^{ik}\eta^{jl}f_{kl}+\onehalf m^2
a_i \eta^{ij}a_j
\end{equation}
The field strength tensor can be re-written in the general coordinate base with help of the proper vierbeins,
\begin{equation}
f_{ij}=e\indices{_i^\alpha}e\indices{_j^\beta}f_{\alpha\beta},\qquad a_i=e\indices{_i^\alpha}a_\alpha.
\end{equation}
Multiplicating then ${\mathcal{L}}_\text{P}$ by the volume factor $\varepsilon$ gives the coordinate frame Lagrangian density
\begin{align}
\tilde{\mathcal{L}}^0_\text{P}=\left(-\quarter\eta^{nk}\eta^{jl}\,
e\indices{_n^\tau}\,e\indices{_k^\sigma}\,
e\indices{_j^\lambda}\,e\indices{_l^\beta}\,f_{\tau\lambda}\,f_{\sigma\beta}+
\onehalf m^2\eta^{jl}\,e\indices{_j^\alpha}\,e\indices{_l^\beta}\,\,a_\alpha\,a_\beta\right)\dete.
\end{align}
For swapping into the Hamiltonian picture we need to compute the pertinent conjugate momentum field:
\begin{align}\label{eq:pro_mom}
\tilde{p}^{\mu\nu}=\dfrac{\tilde{\mathcal{L}}_\text{P}}{\left(\pfrac{a_\mu}{x^\nu}\right)}=\varepsilon \eta^{nk}\eta^{jl}e\indices{_n^\mu}e\indices{_k^\sigma}e\indices{_j^\nu}e\indices{_l^\beta}f_{\sigma\beta}\equiv \tilde{f}^{\mu\nu}
\end{align}
The skew-symmetry of $\tilde{f}^{\mu\nu}$ thus induces the skew-symmetry of $\tilde{p}^{\mu\nu}$. The Maxwell-Proca Hamiltonian now follows as
\begin{align}
\tilde{\HCd}_\text{P}(a_\mu,\tilde{p}^{\mu\nu},e\indices{^i_\mu})
&=
\tilde{p}^{\mu\nu}\,\pfrac{a_\mu}{x^\nu}-\tilde{\LCd}^0_{\mathrm{P}}
\left(a_\mu,\partial_\nu a_\mu,e\indices{^i_\mu}\right)\nonumber\\
&=-\frac{1}{4\dete}\eta_{ik}\,\eta_{jl}\,e\indices{^i_\xi}\,e\indices{^k_\alpha}
\,e\indices{^j_\lambda}\,e\indices{^l_\beta}\,
\tilde{p}^{\,\xi\lambda}\,\tilde{p}^{\,\alpha\beta}-\onehalf
m^2\eta^{ij}\,e\indices{_i^\alpha}\,e\indices{_j^\beta}\,a_\alpha\,a_\beta\,
\dete\nonumber\\
&=-\frac{1}{4\dete}g_{\xi\alpha}\,g_{\lambda\beta}\tilde{p}^{\,\xi\lambda}\,\tilde{p}^{\,\alpha\beta}-\onehalf m^2g^{\alpha\beta}\,a_\alpha\,a_\beta\,\dete.
\end{align}

\subsection{Spinor fields}
The dynamics of uncoupled particles and antiparticles with spin-$\nicefrac{1}{2}$ and mass $m$
is in the inertial frame described by the Dirac equation and its conjugate,
\begin{subequations}\label{eq:Diracinlorentz}
\begin{align}
\rmi\gamma^{k}\pfrac{\psi}{x^k}-m\,\psi&=0\label{eq:Diracinlorentz1}\\
\pfrac{\psibar}{x^k}\,\rmi\gamma^{k}+m\psibar&=0\label{eq:Diracinlorentz2}
\end{align}
\end{subequations}
with  $\gamma^k$, $k=0,\ldots,3$ denoting the $4\times 4$ Dirac matrices in the Minkowski spacetime.
$\psi$ and  $\psibar$ are the independent fields, where $\psi$ is a four component Dirac spinor and
$\psibar\equiv\psi^{\dagger}\gamma^{0}$ the adjoint spinor of $\psi$.

The symmetric form of the usual free Dirac Lagrangian $\LCd^0_{\mathrm{D}}\left(\psi,\partial_k\psi,\psibar,\partial_k\psibar\right)$
which entails~Eqs.~(\ref{eq:Diracinlorentz}) by means of the Euler-Lagrange equations
\begin{equation}\label{eq:elgl}
\pfrac{}{x^k}\pfrac{\LCd^0_{\mathrm{D}}}{(\partial_k\psi)}-\pfrac{\LCd^0_{\mathrm{D}}}{\psi}=0,\qquad
\pfrac{}{x^k}\pfrac{\LCd^0_{\mathrm{D}}}{(\partial_k\psibar)}-\pfrac{\LCd^0_{\mathrm{D}}}{\psibar}=0
\end{equation}
is given by
\begin{equation}\label{eq:ld-dirac-nonregular}
\LCd^0_{\mathrm{D}}=\frac{\rmi}{2}\left(\psibar\,\gamma^{k}\pfrac{\psi}{x^k}-
\pfrac{\psibar}{x^k}\,\gamma^{k}\,\psi\right)-m\,\psibar\psi.
\end{equation}
As this Lagrangian is merely \emph{linear} in the derivatives of the fields, it cannot directly
be Legendre-transformed into an equivalent Dirac Hamiltonian $\HCd_{\mathrm{D}}$.
In order for the Legendre transformation to exist, this Lagrangian
must be amended by adding the divergence of the vector
\begin{equation*}
F^j=\frac{\rmi}{6M}\left(\psibar\,\sigma^{jk}\pfrac{\psi}{x^k}+
\pfrac{\psibar}{x^k}\sigma^{kj}\psi\right)\quad\Rightarrow\quad
\pfrac{F^j}{x^j}=\frac{\rmi}{3M}\pfrac{\psibar}{x^k}\sigma^{kj}\pfrac{\psi}{x^j}.
\end{equation*}
Herein, $\sigma^{jk}$ denotes the commutator of the Dirac matrices $\gamma^{j}$
as the generator of the spinor representation SL$(2,\mathbb{C})$ of the Lorentz group,
\begin{equation}\label{eq:sigma-def0}
\sigma^{jk}=\frac{\rmi}{2}\left(\gamma^j\gamma^k-\gamma^k\gamma^j\right),\qquad
\eta^{jk}\Eins=\frac{1}{2}\left(\gamma^j\gamma^k+\gamma^k\gamma^j\right),
\end{equation}
and $\eta^{jk}$ the Minkowski metric of the Lorentz frame, with $\Eins$ denoting
the unit matrix in spinor space, and $M$ a coupling constant with dimension of mass.

The regularized Gasiorowicz-Dirac Lagrangian~\cite{gasiorowicz66} is thus given by:
\begin{equation}\label{eq:ld-dirac-regular}
\LCd^0_{\mathrm{D}}=\frac{\rmi}{2}\left(\psibar\,\gamma^{k}\pfrac{\psi}{x^k}-
\pfrac{\psibar}{x^k}\,\gamma^{k}\,\psi\right)-m\,\psibar\psi+\frac{\rmi}{3M}\pfrac{\psibar}{x^k}\sigma^{kj}\pfrac{\psi}{x^j}.
\end{equation}
Making use of the identities
\begin{equation}\label{eq:sigma-identity}
\gamma_k\,\sigma^{kj}\equiv\sigma^{jk}\,\gamma_k\equiv 3\rmi\,\gamma^j,\qquad\gamma_k\,\sigma^{kj}\gamma_j\equiv 12\rmi\,\Eins,
\end{equation}
it is equivalently expressed as the product
\begin{equation}\label{eq:ld-dirac-coupling0}
\LCd^0_{\mathrm{D}}=\left(\pfrac{\psibar}{x^k}-
\frac{\rmi M}{2}\psibar\,\gamma_k\right)\frac{\rmi\sigma^{kj}}{3M}
\left(\pfrac{\psi}{x^j}+\frac{\rmi M}{2}\gamma_j\psi\right)-\left(m-M\right)\psibar\psi,
\end{equation}
which remarkably suggests a ``minimal'' coupling of the spinor to the $\gamma$-matrices with coupling constant $M/2$.
The Euler-Lagrange equations for~(\ref{eq:ld-dirac-coupling0}) again yield the Dirac equations~(\ref{eq:Diracinlorentz}),
as amending the Lagrangian~(\ref{eq:ld-dirac-nonregular}) by the divergence term $\partial F^k/\partial x^k$
does not modify the resulting field equations.\\
The inertial frame vectors $\partial \psi/\partial x^k$ and $\partial \bar{\psi}/\partial x^k$ can again be expressed in the coordinate frame by means of the vierbeins, which provides the Dirac Lagrangian density $\tilde{\mathcal{L}}^0_\text{D}(\psi, \partial \psi,\bar{\psi},\partial \bar{\psi},e\indices{_k^\alpha})$ in the form
\begin{equation}
\tilde{\mathcal{L}}^0_\text{D}=\left[\left(\pfrac{\bar{\psi}}{x^\alpha} e\indices{_k^\alpha}-\frac{\rmi M}{2}\bar{\psi}\gamma_k\right)\frac{\rmi\sigma^{kj}}{3M}\left(e_j{}^\beta \pfrac{\psi}{x^\beta}+\frac{\rmi M}{2}\gamma_j\psi\right)-(m-M)\bar{\psi}\psi\right]\varepsilon.
\end{equation}
One finds for the conjugate momenta:
\begin{align}
\tilde{\kappa}^\alpha=\pfrac{\tilde{\LCd}^0_{\mathrm{D}}}{\left(\pfrac{\psibar}{x^\alpha}\right)}&=
\left(-\ihalf\,e\indices{_k^\alpha}\,\gamma^k\,\psi+\frac{\rmi}{3M}e\indices{_k^\alpha}\sigma^{kj}e\indices{_j^\beta}\pfrac{\psi}{x^\beta}\right)\dete,\\
\tilde{\bar{\kappa}}^\alpha=\pfrac{\tilde{\LCd}^0_{\mathrm{D}}}{\left(\pfrac{\psi}{x^\alpha}\right)}&=
\left(\ihalf\psibar\,\gamma^j\,e\indices{_j^\alpha}+\frac{\rmi}{3M}\pfrac{\psibar}{x^\beta}e\indices{_k^\beta}\sigma^{kj}e\indices{_j^\alpha}\right)\dete.
\end{align}
Since the Lagrangian is quadratic in the ``velocities'', the conjugate momenta have a linear ``velocity'' dependence, so that they can be inverted, and the corresponding covariant Hamiltonian~\cite{struckmeier08,StrRei12} is obtained via a regular Legendre transformation.
\begin{align}
\tilde{\HCd}_\text{D}&=\tilde{\bar{\kappa}}^\alpha\pfrac{\psi}{x^\alpha}+\pfrac{
\bar{\psi}}{x^\alpha}\tilde{\kappa}^\alpha-\tilde{\mathcal{L}}^0_\text{D}(\psi,\partial\psi,\bar{\psi},\partial \bar{\psi},e_k{}^\alpha)\nonumber\\
&=\frac{\rmi M}{2}\left(\psibar\,\gamma_k\,e\indices{^k_\alpha}\,\tilde{\kappa}^\alpha-\tilde{\kappabar}^\alpha\,e\indices{^k_\alpha}\frac{6\tau_{kj}}{\dete}e\indices{^j_\beta}\,\tilde{\kappa}^\beta-\tilde{\kappabar}^\alpha\,e\indices{^k_\alpha}\,\gamma_k\,\psi\right)+\left(m-M\right)\psibar\psi\,\dete,\nonumber\\
&=\left(\tilde{\bar{\kappa}}^\alpha-\frac{\rmi}{2}\varepsilon\bar{\psi}\gamma^i e_i{}^\alpha\right)\frac{3M e^k{}_\alpha \tau_{kj}e^j{}_\beta}{\rmi\varepsilon}\left(\tilde{\kappa}^\beta +\frac{\rmi}{2}e_l{}^\beta \gamma^l \psi\varepsilon\right)+m\bar{\psi}\psi\varepsilon,
\end{align}
with $\tau_{kj}$ the inverse of the matrix $\sigma^{jk}$, the latter defined in Eq.~(\ref{eq:sigma-def0}):
\begin{equation*}
\tau_{kj}=\frac{\rmi}{6}\left(\gamma_{k}\gamma_{j}+3\gamma_{j}\gamma_{k}\right),\qquad
\tau_{ik}\,\sigma^{kj}=\delta_{i}^{j}\,\Eins.
\end{equation*}
This definition entails the identities
\begin{equation}\label{eq:tau-identity}
\gamma^{k}\tau_{kj}=\frac{1}{3\rmi}\,\gamma_{j},\qquad
\tau_{kj}\gamma^{j}=\frac{1}{3\rmi}\,\gamma_{k}.
\end{equation}

\subsection{Gravitational field\label{sec:Hgr}}
At last, we have to postulate the dynamics of the free gravitational field. Its Hamiltonian $\tilde{\HCd}_{\mathrm{Gr}}$ is postulated as~\cite{struckmeier17a}:
\begin{align}
\tilde{\HCd}_{\mathrm{Gr}}&=\frac{1}{4g_1\dete}\tilde{q}\indices{_l^{m\alpha\beta}}
\tilde{q}\indices{_m^{l\xi\lambda}}\,\eta_{kn}\,\eta_{ij}\,e\indices{^k_\alpha}\,e\indices{^n_\xi}\,e\indices{^i_\beta}\,e\indices{^j_\lambda}
-g_2\,\tilde{q}\indices{_l^{m\alpha\beta}}e\indices{^l_\alpha}\,e\indices{^n_\beta}\,\eta_{mn}\nonumber\\
&\quad+\frac{1}{2g_3\dete}\tilde{k}\indices{_l^{\alpha\beta}}\tilde{k}\indices{_m^{\xi\lambda}}\,
\eta^{lm}\eta_{kn}\,\eta_{ij}\,e\indices{^k_\alpha}\,e\indices{^n_\xi}\,e\indices{^i_\beta}\,e\indices{^j_\lambda}.
\label{eq:ham-free-grav}
\end{align}
It is quadratic in the conjugate momentum $\tilde{q}\indices{_l^{m\alpha\beta}}$ of the spin connection $\ho\indices{^l_m_\alpha}$
with $g_1$ the pertaining dimensionless coupling constant.
The term \emph{linear} in $\tilde{q}$ is associated with the coupling constant $g_2$ of dimension Length$^{-2}$
and corresponds to the linear dependence on the Riemann-Cartan tensor in the Lagrangian description, as given by the
Hilbert-Einstein Lagrangian of classical General Relativity.
Including this term that in fact breaks conformal invariance of the quadratic term is motivated by phenomenology as it ensures consistency with
observations on the scale of the solar system.
In any Hamiltonian description, at least a \emph{quadratic} momentum dependence of the Hamiltonian is mandatory
in order to encounter a well-defined correlation of the momentum to the respective ``velocity''.
The quadratic term must necessarily be a full momentum tensor concomitant~\cite{Benisty:2018ufz} in order for the Legendre transformation
to retain all information on the gauge field in the equivalent Lagrangian description.
That quadratic $\tilde{q}$-term then describes a ``deformation'' of the actual theory from classical General Relativity represented by the linear term,
and its relative contribution is governed by the coupling constant $g_1$, the \emph{deformation parameter}.

\medskip
The contribution of $\tilde{k}\indices{_l^{\alpha\beta}}$
describes the free (uncoupled) dynamics of the dimensionless vierbein field
$e\indices{^l_\alpha}$ and represents, as usual, the \emph{dual} of the ``velocity'' $\partial e\indices{^l_\alpha}/\partial x^\beta$.
We chose by Occam's razor the simplest, quadratic term in $\tilde{k}\indices{_l^{\alpha\beta}}$.
Since $\tilde{k}$ is of dimension Length$^{-3}$ the pertaining coupling constant $g_3$ in the denominator must have the dimension Length$^{-2}$.
Thus restricting the variety of possible terms to just three terms with three coupling constants is thus the minimum choice.
Analysing more complex gravity models can be pursued in a straightforward but is out of scope of this paper.

\medskip
In order to understand now the physical content of the conjugate momenta $k^j{}_{\mu\nu}$ and $q^j{}_{i\mu\nu}$, we compute the equations \eqref{eq:e-deri2} and \eqref{eq:omega-deri2}, respectively.
From
\begin{equation}
S\indices{^j_{\mu\nu}}=\pfrac{\tilde{\HCd}_\mathrm{Gr}}{\tilde{k}\indices{_j^{\mu\nu}}}
=\frac{1}{g_3}\,k\indices{^j_{\mu\nu}},
\label{eq:T-def}
\end{equation}
we conclude that $k\indices{^i_{\mu\nu}}$ can be identified with the torsion tensor $S\indices{^\xi_{\mu\nu}}=S\indices{^j_{\mu\nu}}\,e\indices{_j^\xi}$
and is thus skew-symmetric in $\mu,\nu$:
\begin{equation}\label{eq:eqm-k}
k\indices{^j_{\mu\nu}}\equiv k\indices{^j_{[\mu\nu]}}=g_3\,e\indices{^j_\xi}\,S\indices{^\xi_{\mu\nu}}.
\end{equation}
Eq.~(\ref{eq:omega-deri2}) gives, on the other hand:
\begin{equation*}
-R\indices{^j_{i\mu\nu}}=2\pfrac{\tilde{\HCd}_\mathrm{Gr}}{\hoc\indices{_j^{i\mu\nu}}}
=\frac{1}{g_1}q\indices{_i^{j\xi\lambda}}\,g_{\mu\xi}\,g_{\nu\lambda}-g_2\,e\indices{^j_\mu}\,e\indices{_i^\lambda}\,g_{\nu\lambda}
+g_2\,e\indices{^j_\nu}\,e\indices{_i^\lambda}\,g_{\mu\lambda}.
\end{equation*}
Resolving for the momentum field gives
\begin{align}
q\indices{^j_{i\mu\nu}}&=g_1\,R\indices{^j_{i\mu\nu}}
+g_1\,g_2\left(e\indices{^j_\nu}\,g_{\mu\lambda}-e\indices{^j_\mu}\,g_{\nu\lambda}\right)e\indices{_i^\lambda}\nonumber\\
&=g_1\left(R\indices{^j_{i\mu\nu}}-\bar{R}\indices{^j_{i\mu\nu}}\right),
\label{eq:R-q}
\end{align}
wherein $\bar{R}\indices{^j_{i\mu\nu}}$ denotes the Riemann tensor of the maximally symmetric spacetime:
\begin{equation}\label{eq:Rbar}
\bar{R}\indices{^j_{i\mu\nu}}=g_2\left(e\indices{^j_\mu}\,g_{\nu\lambda}-e\indices{^j_\nu}\,g_{\mu\lambda}\right)e\indices{_i^\lambda}.
\end{equation}
The momentum tensor $q\indices{^j_{i\mu\nu}}$ thus describes the \emph{deformation} of the actual curvature $R\indices{^j_{i\mu\nu}}$ w.r.t. the de Sitter ($g_2 > 0$) or Anti-de Sitter ($g_2 < 0$) ``ground state''
\cite{Vasak:2022gps,carroll13}, given by $\bar{R}\indices{^j_{i\mu\nu}}$.
Calling the parameter $g_1$ ``deformation parameter'' has thus a twofold meaning --- deformation of curvature, and deformation of the theory w.r.t. the linear Einstein ansatz.
Moreover, being in the denominator in the Hamiltonian, and multiplying the ``velocity'' represented by the curvature tensor, it resembles the role of mass in point mechanics.
Here it thus assigns to geometry a property similar to inertia w.r.t. curvature deformations~\cite{Vasak:2022gps}.
Spacetime is thus not following matter rigidly as in General Relativity, but acquires a dynamic resistance ability on its own.

For the sake of completeness we list here also the pertaining Lagrangian density, now in contrast to the previous sections, as the Legendre-transform of the free gravity Hamiltonian
%
$\tilde{\HCd}_{\mathrm{Gr}}$ in the final action functional~(\ref{action-integral5}):
\begin{equation}\label{eq:LT-Hdyn}
\tilde{\LCd}_{\mathrm{Gr}}=\tilde{k}\indices{_i^{\mu\nu}}S\indices{^i_{\mu\nu}}+
\onehalf\,\tilde{q}\indices{_i^{j\mu\nu}}R\indices{^i_{j\mu\nu}}-\tilde{\HCd}_{\mathrm{Gr}}.
\end{equation}
Recall that $S\indices{^i_{\mu\nu}}$ and $R\indices{^i_{j\mu\nu}}$,
defined by Eqs.~(\ref{eq:T-def2}) and~(\ref{eq:omega-deri3}), are determined by the Hamiltonian $\tilde{\HCd}_{\mathrm{Gr}}$
via the canonical equations~(\ref{eq:e-deri2}) and (\ref{eq:omega-deri2}).
Hence
\begin{equation}\label{eq:LT-Hdyn2}
\tilde{\LCd}_{\mathrm{Gr}}=\tilde{k}\indices{_i^{\mu\nu}}\pfrac{\tilde{\HCd}_{\mathrm{Gr}}}{\tilde{k}\indices{_i^{\mu\nu}}}
+\tilde{q}\indices{_i^{j\mu\nu}}\pfrac{\tilde{\HCd}_{\mathrm{Gr}}}{\tilde{q}\indices{_i^{j\mu\nu}}}-\tilde{\HCd}_{\mathrm{Gr}}.
\end{equation}
Inserting $\tilde{\HCd}_{\mathrm{Gr}}$ from Eq.~(\ref{eq:ham-free-grav}) into Eq.~(\ref{eq:LT-Hdyn2}) gives
\begin{align*}
\tilde{\LCd}_{\mathrm{Gr}}&=\frac{1}{4g_1\dete}\tilde{q}\indices{_l^{m\alpha\beta}}
\tilde{q}\indices{_m^{l\xi\lambda}}\,\eta_{kn}\,\eta_{ij}\,e\indices{^k_\alpha}\,e\indices{^n_\xi}\,e\indices{^i_\beta}\,e\indices{^j_\lambda}\nonumber\\
&\quad+\frac{1}{2g_3\dete}\tilde{k}\indices{_l^{\alpha\beta}}\tilde{k}\indices{_m^{\xi\lambda}}\,
\eta^{lm}\eta_{kn}\,\eta_{ij}\,e\indices{^k_\alpha}\,e\indices{^n_\xi}\,e\indices{^i_\beta}\,e\indices{^j_\lambda}.
\end{align*}
This reproduces the quadratic momentum terms of the Hamiltonian, whereas the term linear in $\tilde{q}$ cancels.
Of course, any Lagrangian density proper must be expressed in terms of the derivatives of the
respective fields rather than by the canonical momenta of the Hamiltonian description.
To this end, the canonical momenta $\tilde{k}\indices{_i^{\mu\nu}}$ and $\tilde{q}\indices{_i^{j\mu\nu}}$
must be expressed in terms of the actual Lagrangian dynamic quantities given by the
specific canonical equations~(\ref{eq:eqm-k}) and~(\ref{eq:R-q}), which leads to
\begin{align*}
\tilde{\LCd}_{\mathrm{Gr}}&=\frac{g_1\dete}{4}\left(R\indices{_l^{m\alpha\beta}}-\bar{R}\indices{_l^{m\alpha\beta}}\right)
\left(R\indices{_m^{l\xi\lambda}}-\bar{R}\indices{_m^{l\xi\lambda}}\right)\eta_{kn}\,\eta_{ij}\,
e\indices{^k_\alpha}\,e\indices{^n_\xi}\,e\indices{^i_\beta}\,e\indices{^j_\lambda}\nonumber\\
&\quad+\frac{g_3\dete}{2}S\indices{_\sigma^{\alpha\beta}}\,S\indices{_\rho^{\xi\lambda}}\,
\eta^{lm}\eta_{kn}\,\eta_{ij}\,e\indices{_l^\sigma}\,e\indices{_m^\rho}\,e\indices{^k_\alpha}\,e\indices{^n_\xi}\,e\indices{^i_\beta}\,e\indices{^j_\lambda}.
\end{align*}
Inserting Eq.~(\ref{eq:Rbar}) for $\bar{R}$ yields
\begin{align*}
\tilde{\LCd}_{\mathrm{Gr}}&=g_1\dete\left(\quarter R\indices{^l_{m\xi\lambda}}R\indices{^m_{l\alpha\beta}}\,\eta^{kn}\,\eta^{ij}\,
e\indices{_k^\alpha}\,e\indices{_n^\xi}\,e\indices{_i^\beta}\,e\indices{_j^\lambda}
+g_2\,R\indices{^l^j_\lambda_\beta}\,e\indices{_j^\beta}\,e\indices{_l^\lambda}-6g_2^2\right)\nonumber\\
&\quad+\frac{g_3\dete}{2}S\indices{_\sigma^{\alpha\beta}}\,S\indices{_\rho^{\xi\lambda}}\,
\eta^{lm}\eta_{kn}\,\eta_{ij}\,e\indices{_l^\sigma}\,e\indices{_m^\rho}\,e\indices{^k_\alpha}\,e\indices{^n_\xi}\,e\indices{^i_\beta}\,e\indices{^j_\lambda}.
\end{align*}
The corresponding Lagrangian ${\LCd}_{\mathrm{Gr}}$ is expressed equivalently in terms of the metric and the Ricci scalar
$R\equiv R\indices{^l^j_\lambda_\beta}\,e\indices{_j^\beta}\,e\indices{_l^\lambda}$ as:
\begin{align}
{\LCd}_{\mathrm{Gr}}&= \frac{g_1}{4} R\indices{^l_{m\xi\lambda}}R\indices{^m_{l\alpha\beta}}\,g^{\alpha\xi}\,g^{\beta\lambda}
+g_1\,g_2\,R-6g_1\,g_2^2\nonumber\\
&\quad+\frac{g_3}{2}S\indices{_\sigma^{\alpha\beta}}\,S\indices{_\rho^{\xi\lambda}}\,
g^{\sigma\rho}\,g_{\alpha\xi}\,g_{\beta\lambda}.
\end{align}
One thus encounters the Hilbert Lagrangian with an additional (cosmological) constant,
plus the quadratic Riemann tensor term that is exactly the one proposed by Einstein
in his personal letter to H.~Weyl~\cite{einstein18}.
(For a discussion of the physical meaning of the constants see \sref{sec:consistency}.)

\subsection{Action integral of the overall system}
At this point, it is instructive to once again write down the action integral \eqref{action-integral6} of the entire now specific system as the basis for the following analyzes.
Splitting up the matter Lagrangians, which now include the gravitational interaction and hence do not carry the superscript $0$, and
the corresponding Hamiltonian expressions by the different matter fields,
\begin{align} \label{eq:actionsplit}
&\int_{V}\d^4x\,\{ \tilde{\LCd}_{\mathrm{matter}}  + \tilde{\LCd}_{\mathrm{Gr}} \}
= \int_{V}\d^4x\,\{ \tilde{\LCd}_{\mathrm{KG}} + \tilde{\LCd}_{\mathrm{P}} + \tilde{\LCd}_{\mathrm{D}} + \tilde{\LCd}_{\mathrm{Gr}} \} \\
=
&\int_{V}\d^4x\,\Bigg\{
( \tilde{\pi}^\nu\pfrac{\varphi}{x^\nu}-\tilde{\HCd}_{\mathrm{KG}} )
+ (\onehalf\tilde{p}^{\mu\nu} f_{\nu\mu} - \tilde{\HCd}_{\mathrm{P}} )
+ ( \tilde{\kappabar}^\nu \Dderr \psi +\psibar \,\Dderl \tilde{\kappa}^\nu - \tilde{\HCd}_{\mathrm{D}} ) \nonumber \\
& \qquad \qquad +( \tilde{k}\indices{_i^{\mu\nu}} S\indices{^i_{\mu\nu}} +\onehalf\hoc\indices{_i^{j\mu\nu}} R\indices{^i_{j\mu\nu}} - \tilde{\HCd}_{\mathrm{Gr}} ) \Bigg\} ,\nonumber
%
\end{align}
we get by collecting all pieces discussed above expressed in terms of the proper dynamic fields:
\begin{align}
S=\int_{V}&\d^4x\Bigg\{\tilde{\pi}^\nu\pfrac{\varphi}{x^\nu}-\frac{1}{2\varepsilon}\tilde{\pi}^\alpha g_{\alpha\beta}\tilde{\pi}^\beta +\frac{\varepsilon}{2}m^2\varphi^2 \\
&+\onehalf\tilde{p}^{\mu\nu}\left(\pfrac{a_\mu}{x^\nu}-\pfrac{a_\nu}{x^\mu}\right)+\frac{1}{4\varepsilon}g_{\xi\alpha}g_{\lambda\beta}\tilde{p}^{\xi\lambda}\tilde{p}^{\alpha\beta}+\onehalf m^2g^{\alpha\beta}a_\alpha a_\beta \varepsilon\nonumber\\
&+\tilde{\kappabar}^\nu\left(\pfrac{\psi}{x^\nu}-\iquarter\ho\indices{^i_{j\nu}}\,\sigma\indices{_i^j}\,\psi\right)
+\left(\pfrac{\psibar}{x^\nu}+\iquarter\psibar\,\ho\indices{^i_{j\nu}}\,\sigma\indices{_i^j}\right)\tilde{\kappa}^\nu\nonumber\\
&\quad-\frac{3M }{\rmi\varepsilon} \left(\tilde{\bar{\kappa}}^\alpha-\frac{\rmi}{2}\varepsilon\bar{\psi}\gamma^i e_i{}^\alpha\right)
 e^k{}_\alpha \tau_{kj}e^j{}_\beta
\left(\tilde{\kappa}^\beta +\frac{\rmi}{2}e_l{}^\beta \gamma^l \psi\varepsilon\right)-m\bar{\psi}\psi\nonumber\\
&+\onehalf\tilde{k}\indices{_i^{\mu\nu}}\left(\pfrac{e\indices{^i_\mu}}{x^\nu}-\pfrac{e\indices{^i_\nu}}{x^\mu}
+\ho\indices{^i_{j\nu}}\,e\indices{^j_\mu}-\ho\indices{^i_{j\mu}}\,e\indices{^j_\nu}\right)\nonumber\\
&\quad-\frac{1}{2g_3\dete}\tilde{k}\indices{_l^{\alpha\beta}}\tilde{k}\indices{_m^{\xi\lambda}}\,
\eta^{lm}g_{\alpha\xi}g_{\beta\lambda}\nonumber\\
&+\onehalf\hoc\indices{_i^{j\mu\nu}}\left(\pfrac{\ho\indices{^i_{j\mu}}}{x^\nu}
-\pfrac{\ho\indices{^i_{j\nu}}}{x^\mu}+\ho\indices{^i_{n\nu}}\,\ho\indices{^n_{j\mu}}
-\ho\indices{^i_{n\mu}}\,\ho\indices{^n_{j\nu}}\right)\nonumber\\
&\quad-\frac{1}{4g_1\dete}\tilde{q}\indices{_l^{m\alpha\beta}}
\tilde{q}\indices{_m^{l\xi\lambda}}g_{\alpha\xi}g_{\beta\lambda}
+g_2\,\tilde{q}\indices{_l^{m\alpha\beta}}e^l{}_\alpha e^n{}_\beta \eta_{mn}\Bigg\}.\nonumber
\end{align}

\section{Energy-momentum tensors of fields} \label{sec:EMT}
The energy-momentum tensor of matter, aka stress-energy tensor, has been the key physical entity in Einstein's derivation of General Relativity since it determines the curvature of spacetime.
In the CCGG framework the energy-momentum tensor is naturally identified with $g^{\nu\alpha}e^i{}_\alpha\partial \tilde{\HCd}/\partial e^i_\mu$. This definition applies though to both, matter and spacetime.
Its dynamic impact depends on the concrete specification of the involved matter fields and on the underlying gravity model.
For the matter portion of the total Lagrangian in the action integral \eqref{eq:actionsplit},
\begin{equation} \label{def:matterlagrangian}
\tilde{\LCd}_{\mathrm{matter}}  \equiv \tilde{\LCd}_{\mathrm{KG}} + \tilde{\LCd}_{\mathrm{P}} + \tilde{\LCd}_{\mathrm{D}},
\end{equation}
the factors facilitating the Legendre transform to the Hamiltonian do not depend on the vierbein field. Then
\begin{equation}
 \tilde{T}_{\mathrm{matter}}^{\mu\nu} := g^{\nu\alpha}e^i{}_\alpha\pfrac{\tilde{\HCd}_{\mathrm{matter}}}{e^i_\mu} \equiv - g^{\nu\alpha}e^i{}_\alpha\pfrac{\tilde{\LCd}_{\mathrm{matter}}}{e^i_\mu},
\end{equation}
which is the standard textbook definition. The formulas for free non-interacting fields derive from the free Hamiltonian (or Lagrangian) densities as follows.

\subsection{Energy-momentum tensor of the scalar field}
For the free Klein-Gordon field we compute the energy-momentum tensor  via \eref{eq:dmudet} from the Hamiltonian (\ref{HKG1}):
\begin{equation}
\tilde{T}_\text{KG}^{\nu\mu}=g^{\nu\alpha} e^i{}_\alpha\pfrac{\tilde{\HCd}_\text{KG}}{e^i{}_\mu}
=\frac{1}{\varepsilon}\tilde{\pi}^\nu\tilde{\pi}^\mu -g^{\nu\mu}\left(\frac{1}{2\varepsilon}\tilde{\pi}^\gamma g_{\gamma\beta}\tilde{\pi}^\beta-\frac{\varepsilon}{2}m^2\varphi^2\right).
\end{equation}
In terms of the field and its derivative we appply Eq.~(\ref{eq:phi_deri}),
\begin{align*}
\tilde{\pi}^\mu=g^{\mu\nu}\partial_\nu \varphi \varepsilon,
\end{align*}
to get a symmetric energy-momentum tensor
\begin{align}\label{eq:emt-KG-local}
\tilde{T}^{\nu\mu}_\text{KG}=\varepsilon\left[g^{\nu\alpha}g^{\mu\beta}\partial_\alpha \varphi\partial_\beta \varphi -g^{\nu\mu}\onehalf\left( g^{\alpha\beta}\partial_\alpha \varphi\partial_\beta \varphi -m^2 \varphi^2\right)\right].
\end{align}

\subsection{Energy-momentum tensor of the vector field}
The energy-momentum tensor of the vector field is obtained in analogous steps. Taking into account the derivative of $e_n{}^\tau$ with respect to the inverse vierbein $e^i{}_\xi$,
\begin{align*}
\pfrac{e_n{}^\tau}{e^i{}_\xi}=-e_i{}^\tau e_n{}^\xi,
\end{align*}
this gives again the symmetric expression
\begin{align}
\tilde{T}_\text{P}^{\nu\mu}&=g^{\nu\alpha}e^i{}_\alpha\pfrac{\tilde{\HCd}_\text{P}}{e^i{}_\mu}\nonumber\\
&=-\frac{1}{\varepsilon}\tilde{p}^{\nu\lambda}g_{\lambda\beta}\tilde{p}^{\mu\beta}+m^2g^{\nu\lambda}a_\lambda g^{\mu\alpha}a_\alpha\varepsilon\nonumber\\
&\quad-g^{\nu\mu}\left(-\frac{1}{4\varepsilon}g_{\xi\alpha}g_{\lambda\beta}\tilde{p}^{\xi\lambda}\tilde{p}^{\alpha\beta}+\frac{1}{2} m^2g^{\alpha\beta}a_\alpha a_\beta \varepsilon\right).
\end{align}
Expressing this in terms of the field and its derivatives means replacing $\tilde{p}^{\mu\nu}$ by $\varepsilon f^{\mu\nu}$:
\begin{align}\label{eq:emt-Proca-local}
\tilde{T}_\text{P}^{\nu\mu}&=\varepsilon\left[-f^{\nu\lambda}g_{\lambda\beta}f^{\mu\beta}+m^2g^{\nu\lambda}a_\lambda g^{\mu\alpha}a_\alpha\nonumber\right.\\
&\left.\quad-g^{\nu\mu}\left(-\quarter g_{\xi\alpha}g_{\lambda\beta}f^{\xi\lambda}f^{\alpha\beta}+\onehalf m^2g^{\alpha\beta}a_\alpha a_\beta \right)\right].
\end{align}

\subsection{Energy-momentum tensor of the spinor field}
Just like before, the metric energy-momentum tensor density of the Dirac field follows from the derivative of $\tilde{\HCd}_\text{D}$ with respect to the vierbein:
\begin{align*}
\tilde{T}^{\nu\mu}_\text{D}&=g^{\nu\alpha}e^i{}_\alpha \pfrac{\tilde{\HCd}_\text{D}}{e^i{}_\mu}\nonumber\\
&=g^{\nu\alpha}e^i{}_\alpha\frac{iM}{2}\left(\bar{\psi}\gamma_i\tilde{\kappa}^\mu-\tilde{\bar{\kappa}}^\mu\gamma_i\psi\right)+g^{\nu\alpha}e^i{}_\alpha (m-M)\bar{\psi}\psi e_i{}^\mu \varepsilon\nonumber\\
&\quad-g^{\nu\alpha}e^i{}_\alpha\frac{3iM}{\varepsilon}\left(\tilde{\bar{\kappa}}^\mu\tau_{ik}e^k{}_\beta\tilde{\kappa}^\beta+\tilde{\bar{\kappa}}^\beta e^k{}_\beta \tau_{ki}\tilde{\kappa}^\mu-\tilde{\bar{\kappa}}^\gamma e^k{}_\gamma \tau_{kn}e^n{}_\beta \tilde{\kappa}^\beta e_i{}^\mu\right).
\end{align*}
For a better readability we introduce the abbreviations
\begin{align*}
\gamma_i e^i{}_\alpha \equiv \gamma_\alpha,
\end{align*}
with analogous definitions for the $\sigma$ and $\tau$ matrices. Then the energy-momentum tensor reads
\begin{align}
\tilde{T}^{\nu\mu}_\text{D}&=g^{\nu\alpha}\frac{iM}{2\varepsilon}\left[(\varepsilon\bar{\psi}\gamma_\alpha-6\tilde{\bar{\kappa}}^\beta \tau_{\beta\alpha})\tilde{\kappa}^\mu-\tilde{\bar{\kappa}}^\mu(\varepsilon\gamma_\alpha\psi+6\tau_{\alpha\beta}\tilde{\kappa}^\beta)\right]\nonumber\\
&\quad-g^{\nu\mu}\left[(m-M)\bar{\psi}\psi\epsilon-\frac{3iM}{\varepsilon}\tilde{\bar{\kappa}}^\gamma \tau_{\gamma\beta}\tilde{\kappa}^\beta \right].
\end{align}
To proceed further, we need to insert the explicit form of the canonical momenta, which can be obtained from Equation (\ref{eq:psi-deri2}). They result as
\begin{subequations}
\begin{align}
\tilde{\kappa}^\alpha&=\left[-\frac{\rmi}{2}\gamma^j\,\psi+\frac{\rmi}{3M}\,\sigma^{ji}\,e\indices{_i^\beta}
\left(\pfrac{\psi}{x^\beta}-\frac{\rmi}{4}\ho_{nm\beta}\,\sigma^{nm}\,\psi\right)\right]e\indices{_j^\alpha}\dete,\\
\tilde{\kappabar}^\alpha&=\left[\hphantom{-}\frac{\rmi}{2}\psibar\,\gamma^j-\frac{\rmi}{3M}
\left(\pfrac{\psibar}{x^\beta}+\frac{\rmi}{4}\ho_{nm\beta}\,\psibar\,\sigma^{nm}\right)\sigma^{ji}\,e\indices{_i^\beta}\right]e\indices{_j^\alpha}\dete.
\end{align}
\end{subequations}
To keep the formulas as brief and readable as possible, we introduce the spinor covariant derivative according to equation (\ref{Def:Spincovderivative1}) and (\ref{Def:Spincovderivative2}). By inserting the conjugate momenta, we receive
\begin{align}\label{eq:emt-dirac-local}
\tilde{T}^{\nu\mu}_\text{D}&=\varepsilon g^{\nu \alpha}\left[\bar{\psi}\Dderl_\alpha\left(-\frac{\rmi}{2}\gamma^\mu\,\psi+\frac{\rmi}{3M}\,\sigma^{\mu\beta}\Dderr_\beta \psi\right)\right.\nonumber\\
&\quad\quad\left.+\left(\frac{\rmi}{2}\psibar\,\gamma^\mu+\frac{\rmi}{3M}\psibar 
\Dderl_\beta \sigma^{\beta\mu}\right) \Dderr_\alpha \psi\right]\nonumber\\
&\quad-g^{\nu\mu}\tilde{\mathcal{L}}'_{\text{D}}.
\end{align}
In the process we identified the Lagrangian of the gravitationally minimal coupled Dirac system
\begin{align}
\tilde{\mathcal{L}}'_D&=\left(\psibar \Dderl_\beta-\frac{\rmi M}{2}\psibar\,\gamma_\beta\right)\frac{\rmi\sigma^{\beta\alpha}}{3M}\left(\Dderr_\alpha \psi +\frac{\rmi M}{2}\gamma_\alpha \psi\right)-\left(m-M\right)\psibar\psi.
\end{align}
One interesting feature is, that the skew symmetric part does not vanish in this case. Instead, it is given by
\begin{align} \label{eq:emt-dirac-local-skew}
\tilde{T}_{\text{D}[\mu\nu]}&=\frac{\rmi\varepsilon}{4}\left(\bar{\psi}\gamma_\nu \Dderr_\mu\psi-\bar{\psi}\gamma_\mu\Dderr_\nu \psi-\bar{\psi}\Dderl_\mu \gamma_\nu \psi +\bar{\psi}\Dderl_\nu \gamma_\mu \psi\right)\\
&\quad+\frac{\rmi\varepsilon}{6M}\bar{\psi}\Dderl_\alpha\left( \sigma_\nu{}^\beta\delta^\alpha_\mu-\sigma_\mu{}^\beta \delta^\alpha_\nu-\sigma_\nu{}^\alpha \delta^\beta_\mu +\sigma_\mu{}^\alpha \delta^\beta_\nu\right)\Dderr_\beta \psi \nonumber \neq 0. \nonumber
\end{align}

For this reason, the Dirac systems does act as a source of torsion of spacetime, as will be laid out in the following.

\subsection{Energy-momentum tensor for the gravitational field}
In analogy to the accepted definition of the metric energy-momentum tensor of matter fields,
the derivative of the Hamiltonian density~(\ref{eq:ham-free-grav}) with respect to the vierbein is defined as
the metric energy-momentum tensor density of spacetime: 
\begin{align}
\tilde{T}_\text{Gr}^{\mu\nu}&=g^{\mu\alpha}\pfrac{\tilde{\HCd}_{\mathrm{Gr}}}{e\indices{^i_\nu}}e\indices{^i_\alpha}\nonumber\\
&=-\frac{1}{g_1\dete}\left(\hoc\indices{_l_m_\tau_\alpha}\,\hoc\indices{^{\,l}^m^\tau^\nu}g^{\mu\alpha}
-\quarter g^{\mu\nu}\,\hoc\indices{_l_m_\tau_\beta}\,\hoc\indices{^{\,l}^m^\tau^\beta}\right)\label{eq:emt-HGr}\\
&\quad+\frac{2}{g_3\dete}\left(\tilde{k}\indices{_l_\tau_\alpha}\,\tilde{k}\indices{^l^\tau^\nu}\, g^{\mu\alpha}
-\quarter g^{\mu\nu}\,\tilde{k}\indices{_l_\tau_\beta}\,\tilde{k}\indices{^l^\tau^\beta}\right)-2g_2\,\hoc\indices{_l_i^\alpha^\nu}e\indices{^l_\alpha}\,\eta^{ij}e\indices{_j^\mu}.
\nonumber
\end{align}
Inserting Eqs.~(\ref{eq:eqm-k}) and~(\ref{eq:R-q}), the energy-momentum tensor $\tilde{T}_\text{Gr}^{\mu\nu}$
is equivalently expressed in terms of the curvature and torsion tensors, $R$ and $S$, as:
\begin{align}
\tilde{T}_\text{Gr}^{\mu\nu}&=-g_1\,\dete\left(R\indices{_l_m_\tau^\nu}\,R\indices{^l^m^\tau_\alpha}g^{\mu\alpha}
-\quarter g^{\mu\nu} R\indices{_l_m_\tau_\beta}\,R\indices{^l^m^\tau^\beta}\right)\nonumber\\
&\quad+2g_1g_2\,\dete\left(R\indices{_l^j^\beta_\alpha}\,e\indices{^l_\beta}\,e\indices{_j^\nu}g^{\alpha\mu}
-\onehalf g^{\mu\nu}\,R\indices{_l^n^\alpha_\beta}\,e\indices{^l_\alpha}\,e\indices{_n^\beta}+3g_2\,g^{\mu\nu}\right)\nonumber\\
&\quad+2g_3\,\dete\left(S\indices{_l_\tau_\alpha}\,S\indices{^l^\tau^\nu}\,g^{\alpha\mu}
-\quarter g^{\mu\nu}\,S\indices{_l_\tau_\beta}\,S\indices{^l^\tau^\beta}\right).
\label{eq:emt-HGr-2}
\end{align}
It is not symmetric.
Its skew-symmetric portion is
\begin{equation*}
\tilde{T}_\text{Gr}^{[\mu\nu]}=\frac{1}{2}\pfrac{\tilde{\HCd}_{\mathrm{Gr}}}{e\indices{^n_\beta}}\left(\delta^\nu_\beta e^n{}_\alpha g^{\alpha\mu}-\delta^\mu_\beta e^n{}_\alpha g^{\alpha\nu}\right)
=g_1g_2\,\dete\left(R\indices{_\alpha^{j\alpha i}}-R\indices{_\alpha^{i\alpha j}}\right)e_i{}^\mu e_j{}^\nu.
\end{equation*}
Re-writing the r.h.s. as the skew-symmetric portion of the Ricci tensor $R^{\mu\nu}:=R\indices{_\alpha^i^\alpha^j}e_i{}^\mu e_j{}^\nu$ gives: 
\begin{equation}\label{eq:HGr-skewpart}
T_\text{Gr}^{[\mu\nu]}=-2g_1g_2\,R^{[\mu\nu]}.
\end{equation}
All other terms of~\eref{eq:emt-HGr-2} establish the symmetric portion of the energy-momen\-tum tensor of $\tilde{\HCd}_{\mathrm{Gr}}$:
\begin{align}
T_\text{Gr}^{(\mu\nu)}&=-g_1\,\left(R\indices{_l_m_\tau^\nu}\,R\indices{^l^m^\tau^\mu}
-\quarter g^{\mu\nu} R\indices{_l_m_\tau_\beta}\,R\indices{^l^m^\tau^\beta}\right)\nonumber\\
&\quad+2g_1g_2\left(R^{(\mu\nu)}-\onehalf g^{\mu\nu}\,R+3g_2\,g^{\mu\nu}\right)\nonumber\\
&\quad+2g_3\left(S\indices{_l_\tau^\mu}\,S\indices{^l^\tau^\nu}\
-\quarter g^{\mu\nu}\,S\indices{_l_\tau_\beta}\,S\indices{^l^\tau^\beta}\right).
\label{eq:TGr-symmpart}
\end{align}
The contraction of the Ricci tensor, $R := R_\alpha{}^\alpha$, is the Ricci scalar.

\section{Coupled field equations of matter and dynamic spacetime\label{sec:coupledeqs}}
The diffeomorphism invariant field equations emerging from the covariant canonical transformation theory
are now discussed for the particular Hamiltonians of scalar, vector, and spinor matter fields in conjunction
with the particular model Hamiltonian for the free gravitational field.

\subsection{Klein-Gordon equation in curved spacetime}
In order to obtain the modified Klein-Gordon equation we simply have to compute the canonical equations (\ref{eq:phi_deri}) and (\ref{eq:pi_div}) for the Klein-Gordon Hamiltonian (\ref{HKG1}). This gives
\begin{subequations}
\begin{align}
\pfrac{\varphi}{x^\nu}&=\hphantom{-}\pfrac{\tilde{\HCd}_\text{KG}}{\tilde{\pi}^\nu}=\frac{1}{\varepsilon}\tilde{\pi}^\alpha e^n{}_\alpha\eta_{nm}e^m{}_\eta=\pi_\nu\label{eq:ceqkg1}\\
\pfrac{\tilde{\pi}^\alpha}{x^\alpha}&=-\pfrac{\tilde{\HCd}_\text{KG}}{\varphi}=-\varepsilon m^2 \varphi
\end{align}
\end{subequations}
Solving Eq.~(\ref{eq:ceqkg1}) for $\tilde{\pi}^\alpha$ is now possible:
\begin{equation*}
\tilde{\pi}^\alpha=e\indices{_i^\alpha}\,\eta^{ij}\,e\indices{_j^\beta}\,\pfrac{\varphi}{x^\beta}\,\dete=g^{\alpha\beta}\pfrac{\varphi}{x^\beta}\,\dete,
\end{equation*}
the momentum can be eliminated from the equation of motion to yield:
\begin{equation*}
\pfrac{\tilde{\pi}^\alpha}{x^\alpha}=\pfrac{}{x^\alpha}\left(g^{\alpha\beta}\pfrac{\varphi}{x^\beta}\,\dete\right)
-\dete\,m^2\,\varphi.
\end{equation*}
By executing the partial derivative of the metric and its determinant, we can bring the Klein-Gordon equation in dynamic spacetime in the form of a tensor equation:
\begin{align} \label{eq:KG;}
g^{\alpha\beta}\varphi_{;\alpha;\beta}-2g^{\alpha\beta}S^\xi{}_{\alpha\xi}\varphi_{;\beta}+m^2\varphi=0,
\end{align}
where we defined the covariant derivative of a general tensor in analogy to equation (\ref{def:vierbeinpostulate}), meaning that every index is once contracted with either the affine or the spin connection.

\subsection{Maxwell-Proca equation in curved spacetime}
In order to obtain the field equations for the Maxwell-Proca, we again need to compute the canonical Equations (\ref{eq:a_deri}) and (\ref{eq:p_div}). The first one is identical to Eq.~(\ref{eq:pro_mom}), so we do not need to compute it again. The second one yields
\begin{align}
\pfrac{\tilde{p}^{\mu\alpha}}{x^\alpha}=-\pfrac{\tilde{\HCd_\text{P}}}{a_\mu}=m^2 \eta^{ij}e_i{}^\alpha e_j{}^\mu a_\alpha\varepsilon=m^2 g^{\mu\alpha}a_\alpha \varepsilon.
\end{align}
Since we demand metric compatibility, we can express the partial derivative of the metric with respect to the connection
\begin{align*}
\pfrac{g^{\mu\nu}}{x^\alpha}=-\gamma^{\mu}{}_{\xi\alpha}g^{\xi\nu}-\gamma\indices{^\nu_{\xi\alpha}}g^{\mu\xi},
\end{align*}
where the affine connection is connected to the spin connection via Eq.~(\ref{eq:def-affconn}). The derivative of the determinant of the metric is given by
\begin{align*}
\pfrac{\sqrt{-g}}{x^\alpha}=\sqrt{-g}\gamma\indices{^\xi_{\xi\alpha}}.
\end{align*}
For the vector field equation, we thus receive
\begin{equation} \label{eq:MP;}
f\indices{^{\mu\alpha}_{;\alpha}}-S\indices{^\mu_{\xi\alpha}}f^{\xi\alpha}-2S\indices{^\xi_{\alpha\xi}}f^{\mu\alpha}-m^2 a^\mu=0,
\end{equation}
where we identified the torsion tensor according to Eq.~(\ref{eq:e-deri}) and defined the covariant derivative in analogy to Eq.~(\ref{def:vierbeinpostulate}).

\subsection{Dirac equation in curved spacetime}\label{sec:gen-dirac1}
In order to obtain the field equations for the spinor field, we just need to compute the canonical equations (\ref{eq:psi-deri2}) to (\ref{eq:kappa-div2}). They result in
\begin{subequations}
\begin{align}
\Dderr_\mu \psi=\hphantom{-}\pfrac{\tilde{\HCd}_\text{D}}{\tilde{\bar{\kappa}}^\mu}=-\frac{\rmi M}{2}\left(\gamma_k\psi+\frac{6\tau_{kj}}{\varepsilon}e^j{}_\alpha \tilde{\kappa}^\alpha\right)e^k{}_\mu\label{eq:can-dirac1b}\\
\Dderr_\alpha\tilde{\kappa}^\alpha=-\pfrac{\tilde{\HCd}_\text{D}}{\bar{\psi}}=-\frac{\rmi M}{2}\gamma_ke^k{}_\alpha\tilde{\kappa}^\alpha-(m-M)\bar{\psi}\varepsilon\label{eq:can-dirac2b}\\
\bar{\psi}\Dderl_\mu=\hphantom{-}\pfrac{\tilde{\HCd}_D}{\tilde{\kappa}^\mu}=\hphantom{-}\frac{\rmi M}{2}\left(\bar{\psi}\gamma_j-\tilde{\bar{\kappa}}^\alpha\frac{6\tau_{kj}}{\varepsilon}e^k{}_\alpha\right)e^j{}_\mu\label{eq:can-dirac3b}\\
\bar{\tilde{\kappa}}^\alpha \Dderl_\alpha=-\pfrac{\tilde{\HCd}_\text{D}}{\psi}=\hphantom{-}\frac{\rmi M}{2}e^k{}_\alpha \tilde{\bar{\kappa}}^\alpha\gamma_k-(m-M)\bar{\psi}\varepsilon\label{eq:can-dirac4b}
\end{align}
\end{subequations}
As we can see, due to the appearance of the gauge field $\omega_{ij\nu}$, these field equations are now properly covariant.

\medskip
To express the coupled set of canonical equations as a field equation for the spinor $\psi$,
we solve Eqs.~(\ref{eq:can-dirac1b}) and~(\ref{eq:can-dirac3b}) for the spinor momentum fields $\tilde{\kappa}^\alpha$
and $\tilde{\kappabar}^\alpha$, respectively,
\begin{subequations}\label{eq:can-momenta-cov}
\begin{align}
\tilde{\kappa}^\alpha&=\left[-\frac{\rmi}{2}\gamma^j\,\psi+\frac{\rmi}{3M}\,\sigma^{ji}\,e\indices{_i^\beta}
\left(\pfrac{\psi}{x^\beta}-\frac{\rmi}{4}\ho_{nm\beta}\,\sigma^{nm}\,\psi\right)\right]e\indices{_j^\alpha}\dete\label{eq:can-dirac1c}\\
\tilde{\kappabar}^\alpha&=\left[\hphantom{-}\frac{\rmi}{2}\psibar\,\gamma^j-\frac{\rmi}{3M}
\left(\pfrac{\psibar}{x^\beta}+\frac{\rmi}{4}\ho_{nm\beta}\,\psibar\,\sigma^{nm}\right)\sigma^{ji}\,e\indices{_i^\beta}\right]e\indices{_j^\alpha}\dete,\label{eq:can-dirac3c}
\end{align}
\end{subequations}
and insert~(\ref{eq:can-dirac1c}) into Eq.~(\ref{eq:can-dirac2b}):
\begin{align}
&\quad\pfrac{}{x^\alpha}\left\{\left[-\ihalf e\indices{_j^\alpha}\gamma^j\psi+\frac{\rmi}{3M}e\indices{_j^\alpha}\,\sigma^{ji}\,e\indices{_i^\beta}
\left(\pfrac{\psi}{x^\beta}-\iquarter\ho_{nm\beta}\,\sigma^{nm}\,\psi\right)\right]\dete\right\}\nonumber\\
&=\left(\ihalf M\gamma_ke\indices{^k_\alpha}\!-\!\iquarter\ho_{kl\alpha}\sigma^{kl}\right)\!
\left[\ihalf e\indices{_j^\alpha}\gamma^j\psi\!-\!\frac{\rmi}{3M}e\indices{_j^\alpha}\sigma^{ji}\,e\indices{_i^\beta}\!
\left(\pfrac{\psi}{x^\beta}\!-\!\iquarter\ho_{nm\beta}\sigma^{nm}\psi\right)\!\right]\!\dete\nonumber\\
&\quad-\left(m-M\right)\psi\,\dete.
\label{eq:can-dirac1d}
\end{align}
The final version of the generalized Dirac equation~(\ref{eq:can-dirac1d}) is worked out in~\aref{sec:gen-dirac}. The result is
\begin{align} \label{eq:Diracfinall}
&\rmi\gamma^\beta \left( \Dderr_\beta - S^\alpha{}_{\beta\alpha} \right) \psi-m\psi=\nonumber\\
&\quad\,-\frac{\rmi}{3M}\Bigg[\!\left(2\sigma^{\xi\beta}\,S\indices{^\alpha_{\xi\alpha}}
+\sigma^{\alpha\xi}\,S\indices{^\beta_{\xi\alpha}}\right)
\Dderr_\beta\psi+\iquarter \sigma^{\alpha\beta}\,\sigma^{\mu\nu}
R_{\mu\nu\alpha\beta}\psi\Bigg],
\end{align}
which shows, that the spin coupling of the Dirac particle causes an effective mass correction term proportional to the Riemann tensor $R_{\mu\nu\alpha\beta}$. Neglecting torsion, this equation further simplifies to:
\begin{equation*}
\rmi\,\gamma^\beta\Dderr_\beta \psi
-\left(m+\frac{1}{24M}\sigma^{\alpha\beta}\,\sigma^{\mu\nu}\,\bar{R}_{\mu\nu\alpha\beta}\right)\psi=0.
\end{equation*}
For the case of torsion-free spacetime the contraction of the Riemann-Cartan tensor with the $\sigma$-matrices is shown in Ref.~\cite{struckmeier21a}
to reduce to twice the Ricci scalar with respect to the Levi-Civita connection $\bar{R}$, times the unit matrix in the spinor indices: 
\begin{equation*}
\rmi\,\gamma^\beta\Dderr_\beta \psi-\left(m+
\frac{\bar{R}}{12M}\right)\psi=0.
\end{equation*}
One thus encounters in the Dirac equation an additional mass term due to a direct interaction of $\psi$ with the gravitational field.

\subsection{Consistency equation revisited\label{sec:consistency}}
Rearranging the terms of Eq.~(\ref{eq:consistency0}) on page~\pageref{eq:consistency0}, this equation takes on the form:
\begin{align*}
0&=\iquarter\pfrac{}{x^\beta}\left(\psibar\,\sigma\indices{_i^j}\tilde{\kappa}^\beta-\tilde{\kappabar}^\beta\sigma\indices{_i^j}\,\psi\right)
+\iquarter\left(\psibar\,\sigma\indices{_i^n}\tilde{\kappa}^\beta-\tilde{\kappabar}^\beta\sigma\indices{_i^n}\,\psi\right)\ho\indices{^j_{n\beta}}\\
&\quad-\iquarter\left(\psibar\,\sigma\indices{_n^j}\tilde{\kappa}^\beta-\tilde{\kappabar}^\beta\sigma\indices{_n^j}\,\psi\right)\ho\indices{^n_{i\beta}}
+\left(\pfrac{\tilde{k}\indices{_i^{[\alpha\beta]}}}{x^\beta}-\tilde{k}\indices{_n^{[\alpha\beta]}}\ho\indices{^n_{i\beta}}\right)e\indices{^j_\alpha}\\
&\quad+\hoc\indices{_i^{n[\alpha\beta]}}\left(\pfrac{\ho\indices{^j_{n\alpha}}}{x^\beta}+\ho\indices{^j_{m\beta}}\ho\indices{^m_{n\alpha}}\right)
-\left(\pfrac{\ho\indices{^n_{i\alpha}}}{x^\beta}-\ho\indices{^n_{m\alpha}}\ho\indices{^m_{i\beta}}\right)\hoc\indices{_n^{\,j[\alpha\beta]}}\\
&\quad+\tilde{k}\indices{_i^{[\alpha\beta]}}\left(\pfrac{e\indices{^j_\alpha}}{x^\beta}+\ho\indices{^j_{n\beta}}\,e\indices{^n_\alpha}\right).
\end{align*}
Inserting the abbreviations~(\ref{eq:T-def2}) and~(\ref{eq:R-def}) for the last three terms yields:
\begin{align*}
0&=\left(\pfrac{\tilde{k}\indices{_i^{[\mu\alpha]}}}{x^\alpha}-\tilde{k}\indices{_n^{[\mu\alpha]}}\ho\indices{^n_{i\alpha}}\right)e\indices{^j_\mu}\\
&+\iquarter\pfrac{}{x^\beta}\left(\psibar\,\sigma\indices{_i^j}\tilde{\kappa}^\beta-\tilde{\kappabar}^\beta\sigma\indices{_i^j}\,\psi\right)
+\iquarter\left(\psibar\,\sigma\indices{_i^n}\tilde{\kappa}^\beta-\tilde{\kappabar}^\beta\sigma\indices{_i^n}\,\psi\right)\ho\indices{^j_{n\beta}}\\
&-\iquarter\left(\psibar\sigma\indices{_n^j}\tilde{\kappa}^\beta-\tilde{\kappabar}^\beta\sigma\indices{_n^j}\psi\right)\ho\indices{^n_{i\beta}}
-\onehalf\hoc\indices{_i^{n\alpha\beta}}R\indices{^j_{n\alpha\beta}}+\onehalf R\indices{^n_{i\alpha\beta}}\hoc\indices{_n^{j\alpha\beta}}
+\tilde{k}\indices{_i^{[\alpha\beta]}}S\indices{^j_{\alpha\beta}}.
\end{align*}
The first term can now be replaced by the metric energy-momentum tensors according to the field equation~(\ref{eq:k-div3}):
\begin{align}
0&=\iquarter\pfrac{}{x^\beta}\left(\psibar\,\sigma\indices{_i^j}\tilde{\kappa}^\beta-\tilde{\kappabar}^\beta\sigma\indices{_i^j}\,\psi\right)
+\iquarter\left(\psibar\,\sigma\indices{_i^n}\tilde{\kappa}^\beta-\tilde{\kappabar}^\beta\sigma\indices{_i^n}\,\psi\right)\ho\indices{^j_{n\beta}}\nonumber\\
&\quad-\iquarter\left(\psibar\,\sigma\indices{_n^j}\tilde{\kappa}^\beta-\tilde{\kappabar}^\beta\sigma\indices{_n^j}\,\psi\right)\ho\indices{^n_{i\beta}}
-\pfrac{\tilde{\HCd}_0}{e\indices{^i_\alpha}}e\indices{^j_\alpha}-\pfrac{\tilde{\HCd}_{\mathrm{Gr}}}{e\indices{^i_\alpha}}e\indices{^j_\alpha}\nonumber\\
&\quad+\onehalf\left(R\indices{^n_{i\alpha\beta}}\,\hoc\indices{_n^{\,j\alpha\beta}}-\hoc\indices{_i^{n\alpha\beta}}\,R\indices{^j_{n\alpha\beta}}\right)
+\tilde{k}\indices{_i^{\alpha\beta}}\,S\indices{^j_{\alpha\beta}}.
\label{eq:cons-equation}
\end{align}
The terms involving the derivatives of the Hamiltonian with respect to the vierbeins define the metric energy-momentum tensors
according to Eqs.~(\ref{eq:emt-KG-local}), (\ref{eq:emt-Proca-local}), (\ref{eq:emt-dirac-local}), and~(\ref{eq:emt-HGr}).
The terms related to $\tilde{q}$ simplify, inserting its correlation to the Riemann tensor from Eq.~(\ref{eq:R-q})
\begin{equation}
\quad\;\hphantom{-}\onehalf\left(R\indices{^n_{i\alpha\beta}}\,\hoc\indices{_n^{\,j\alpha\beta}}
-\hoc\indices{_i^{n\alpha\beta}}\,R\indices{^j_{n\alpha\beta}}\right)\nonumber\\
=-g_1g_2\,\dete\left(R\indices{_i^j}-R\indices{^j_i}\right). 
\label{eq:Rq-comm}
\end{equation}
It is, in addition, convenient and physically justified to introduce, in analogy to the energy-momentum tensor of matter built from the derivative of the Hamiltonian or Lagrangian with respect to the vierbein,
the \emph{spin-momentum tensor} built from the derivative of the matter Lagrangian \eqref{def:matterlagrangian} with respect to the spin connection:
\begin{equation}
\tilde{\Sigma}\indices{_i^j^\beta} :=  \pfrac{\tilde{\LCd}_{\mathrm{matter}}}{\omega\indices{^i_j_\beta}}
 = - \pfrac{\tilde{\HCd}_{\mathrm{Gau}_2}}{\omega\indices{^i_j_\beta}}
\end{equation}
Since only the Dirac Lagrangian, $\tilde{\LCd}_{\mathrm{D}}$, depends on the spin connection, we find
\begin{equation}\label{eq:spin-density2}
\tilde{\Sigma}\indices{^i^j^\beta}=\iquarter\left(\psibar\,\sigma\indices{^i^j}\tilde{\kappa}^\beta-\tilde{\kappabar}^\beta\sigma\indices{^i^j}\,\psi\right),
\end{equation}
which is skew-symmetric in $i$ and $j$. The explicit formula for $\tilde{\Sigma}\indices{_i^j^\beta}$ is calculated in the \aref{app:spintensor}, \eref{eq:spindensitytensor-expl}.

Raising the index $i$ in Eq.~(\ref{eq:cons-equation}), and inserting \eref{eq:eqm-k} gives then:
\begin{equation}\label{eq:cons-eq}
\pfrac{\tilde{\Sigma}\indices{^i^j^\beta}}{x^\beta}
+\tilde{\Sigma}\indices{^i^n^\beta}\ho\indices{^j_{n\beta}}-\tilde{\Sigma}\indices{^j^n^\beta}\ho\indices{^i_n_\beta}
-2g_1g_2\dete R^{[ij]}+g_3\dete S\indices{^i^{\alpha\beta}}S\indices{^j_{\alpha\beta}}
=\tilde{T}_0^{ij}+\tilde{T}_\text{Gr}^{ij}.
\end{equation}
This equation relates the spin-momentum tensor $\tilde{\Sigma}$ of the spinor fields and the metric energy-momentum tensor
$\tilde{T}_0^{ij}$ of all source fields with the energy-momentum tensor of $\tilde{\HCd}_{\mathrm{Gr}}$.
Equation~(\ref{eq:cons-eq}) can now be split into skew-symmetric and symmetric portions in $i$ and $j$.
The skew-symmetric portion follows as:
\begin{equation}\label{eq:cons-eq-skew}
\pfrac{\tilde{\Sigma}\indices{^i^j^\beta}}{x^\beta}
+\tilde{\Sigma}\indices{^i^n^\beta}\ho\indices{^j_{n\beta}}-\tilde{\Sigma}\indices{^j^n^\beta}\ho\indices{^i_n_\beta}
-\tilde{T}_0^{[ij]}=\tilde{T}_\text{Gr}^{[ij]}+2g_1g_2\,\dete\,R^{[ij]}.
\end{equation}
For the particular gravitational Hamiltonian, defined in Eq.~(\ref{eq:ham-free-grav}), the Riemann tensor terms on the right-hand
side of equation~(\ref{eq:cons-eq-skew}) cancel identically by virtue of Eq.~(\ref{eq:HGr-skewpart}).
As the energy-momentum tensors of the Klein-Gordon and Proca systems are symmetric, as can easily be seen in the Equations \eqref{eq:emt-KG-local} and \eqref{eq:emt-Proca-local},
only the spinor fields enter into the left-hand side.
These terms cancel by virtue of Eq.~(\ref{eq:spin-tensor-deri}) from \aref{app:spintensor}.
Hence, we are left with the identity
\begin{equation}\label{eq:spin-tensor-deri-0}
\pfrac{\tilde{\Sigma}\indices{^i^j^\beta}}{x^\beta}
+\tilde{\Sigma}\indices{^i^n^\beta}\ho\indices{^j_{n\beta}}-\tilde{\Sigma}\indices{^j^n^\beta}\ho\indices{^i_n_\beta}
\equiv \dete\left({\Sigma}\indices{^i^j^\beta_{;\beta}} -2{S}\indices{^\mu_\beta_\mu}\,{\Sigma}\indices{^i^j^\beta}\right)=\tilde{T}_{\mathrm{D}}^{[ij]}.
\end{equation}
such that \eref{eq:cons-eq-skew} is identically satisfied for the given Hamiltonians and does not provide any information on the field dynamics.

\medskip
Writing now the symmetric portion of Eq.~(\ref{eq:cons-eq}) in metric coordinates gives:
\begin{equation}\label{eq:cons-eq-symm}
\tilde{T}_0^{(\mu\nu)}=g_3\,\dete\,S\indices{^\mu^{\alpha\beta}}\,S\indices{^\nu_{\alpha\beta}}-\tilde{T}_\text{Gr}^{(\mu\nu)}.
\end{equation}
The torsion term $\sim g_3$ obviously contributes to the total energy-momentum density of spacetime.
With \eref{eq:TGr-symmpart} for $\tilde{T}_\text{Gr}^{(\mu\nu)}$, and setting
$R:= R_\alpha{}^\alpha$, this gives in metric coordinates:
\begin{align} \label{eq:HGr-symmpart}
&g_1\,\left(R\indices{_\alpha_\beta_\gamma^\nu}\,R\indices{^\alpha^\beta^\gamma^\mu}
-\quarter g^{\mu\nu} R\indices{_\alpha_\beta_\gamma_\delta}\,R\indices{^\alpha^\beta^\gamma^\delta}\right) 
-2g_1g_2\left(R^{(\mu\nu)}-\onehalf g^{\mu\nu}\,R+3g_2\,g^{\mu\nu}\right)\nonumber\\
&\quad-g_3\left(S\indices{_\alpha_\beta^\mu}\,S\indices{^\alpha^\beta^\nu}\
-\onehalf g^{\mu\nu}\,S\indices{_\alpha_\beta_\gamma}\,S\indices{^\alpha^\beta^\gamma}\right) = \tilde{T}_0^{(\mu\nu)}.
\end{align}
This is the so called CCGG field equation extending the Einstein field equation by the quadratic Riemann-Cartan tensor concomitant
(quadratic traceless Kretsch\-mann term) and torsion.

In the weak-field limit the term qudratic in curvature is relatively small and can be neglected.
If also the torsion is discarded,
Eq.~(\ref{eq:HGr-symmpart}) reduces
to the classical Einstein equation with the cosmological constant term, 
$\Lambda:=-3g_2$, and $-2 g_1 g_2=(8\pi G)^{-1} \equiv M_\mathrm{p}^2$ with the Einstein 
gravitational constant $G$ and the reduced Planck mass, $M_\mathrm{p}$, respectively:
\begin{equation*}
R\indices{^{(\nu\mu)}}- g^{\nu\mu}\left(\onehalf R+\Lambda \right)=8\pi G 
\,{T}_0^{(\mu\nu)}.
\end{equation*}
This identification implies then the final form of the CCGG equation:
\begin{align} \label{eq:CCGG2}
&g_1\,\left(R\indices{_\alpha_\beta_\gamma^\nu}\,R\indices{^\alpha^\beta^\gamma^\mu}
-\quarter g^{\mu\nu} R\indices{_\alpha_\beta_\gamma_\delta}\,R\indices{^\alpha^\beta^\gamma^\delta}\right) 
+\frac{1}{8\pi G}\left(R^{(\mu\nu)}-\onehalf g^{\mu\nu}\,R-\Lambda g^{\mu\nu}\right)\nonumber\\
&\quad-g_3\left(S\indices{_\alpha_\beta^\mu}\,S\indices{^\alpha^\beta^\nu}\
-\onehalf g^{\mu\nu}\,S\indices{_\alpha_\beta_\gamma}\,S\indices{^\alpha^\beta^\gamma}\right) = {T}_0^{(\mu\nu)}.
\end{align}
Note that the cosmological constant term $\Lambda$ is not introduced ``by hand'' but actually
emerges here as a geometrical term from the gauge procedure~\cite{struckmeier21a}, the vacuum energy of spacetime.
It is independent of the the vacuum energy of matter that will, though, counteract the observed cosmological constant~\cite{Vasak:2022gps}.

\medskip
The CCGG equation can be interpreted as a local energy cancellation
if the right-hand side of \eref{eq:cons-eq-symm} is interpreted as the total (symmetric) energy-momentum of spacetime, similar to the \emph{strain-energy tensor} in elastic media.
Then we can write for the left-hand side of \eref{eq:CCGG2}:
\begin{align} \label{def:strain-energy-tensor}
-{\Theta}_{\mathrm{Gr}}^{(\mu\nu)} := &g_1\,\left(R\indices{_\alpha_\beta_\gamma^\nu}\,R\indices{^\alpha^\beta^\gamma^\mu}
-\quarter g^{\mu\nu} R\indices{_\alpha_\beta_\gamma_\delta}\,R\indices{^\alpha^\beta^\gamma^\delta}\right)
+\frac{1}{8\pi G}\left(R^{(\mu\nu)}-\onehalf g^{\mu\nu}\,R-\Lambda g^{\mu\nu}\right)\nonumber\\
&\quad-g_3\left(S\indices{_\alpha_\beta^\mu}\,S\indices{^\alpha^\beta^\nu}\
-\onehalf g^{\mu\nu}\,S\indices{_\alpha_\beta_\gamma}\,S\indices{^\alpha^\beta^\gamma}\right)\, .
\end{align}
Consequently, ${T}_0^{(\mu\nu)}$ corresponds to the \emph{stress-energy tensor} and with
\begin{equation}
{\Theta}_{\mathrm{Gr}}^{(\mu\nu)}  +  {T}_0^{(\mu\nu)} = 0,
\end{equation}
we encounter a Universe with everywhere locally vanishing total energy.
That recovers the conjecture of \emph{Zero-Energy Universe}
which has been discussed long ago by Lorentz~\cite{lorentz1916}, Levi-Civita~\cite{levi-civita1917}, Jordan~\cite{jordan39}, and,
independently in each case, by Sciama~\cite{sciama53}, Feynman~\cite{feynman62}, Tryon~\cite{Tryon:1973xi},
Rosen~\cite{Rosen:1994vj}, Cooperstock~\cite{cooperstock95}, Hamada~\cite{Hamada:2022fko}, Melia~\cite{Melia:2022ifj}and Hawking~\cite{hawking03}, based on various physical reasonings.

\medskip
By switching on and off the different terms in \eref{eq:CCGG2} the Einstein-Cartan limit ($g_1 =0$ while $g_1\,g_2 = const$) and the Einstein-Hilbert limit ($g_1 = g_3 = 0$) are obtained.
Since in the torsion-free limit ($g_3 = 0$) the Schwarzschild metric has been shown~\cite{kehm17} to be a solution of \eref{eq:CCGG2}, it is consistent with solar-scale observations.

\subsection{Spin-curvature tensor coupling equation\label{sec:spin-torsion-1}}
We replace in the canonical field equation~(\ref{eq:t-div-2})
the momentum field $\tilde{k}\indices{_i^{\alpha\mu}}$ by the result of
the canonical equation~(\ref{eq:eqm-k}) for the ``free gravity'' Hamiltonian~(\ref{eq:ham-free-grav}), 
and the $\hoc\indices{_i^{j\mu\alpha}}$-dependent terms according to Eq.~(\ref{eq:R-q}).
Eq.~(\ref{eq:t-div-2}) is then equivalently expressed as
\begin{align*}
&g_1\!\left[\left( R\indices{_i^{j\mu\alpha}}\!\!-\bar{R}\indices{_i^{j\mu\alpha}}\right)_{;\alpha}
-\!\left(R\indices{_i^{j\xi\alpha}}-\bar{R}\indices{_i^{j\xi\alpha}}\right)S\indices{^\mu_\xi_\alpha}
\!-2\!\left(R\indices{_i^{j\mu\xi}}-\bar{R}\indices{_i^{j\mu\xi}}\right)S\indices{^\alpha_\xi_\alpha}\right]\\
&\quad= g_3\,\,S\indices{_i^\alpha^\mu}e\indices{^j_\alpha}+ {\Sigma}\indices{_i^j^\mu}.
\end{align*}
Inserting $\bar{R}\indices{_i^{j\mu\alpha}}$ gives finally:
\begin{align}
&\;g_1\left(R\indices{_i^{j\mu\alpha}_{;\alpha}}-R\indices{_i^{j\xi\alpha}}S\indices{^\mu_\xi_\alpha}
-2R\indices{_i^{j\mu\xi}}S\indices{^\alpha_\xi_\alpha}\right)-g_3\,S\indices{_i^\alpha^\mu}e\indices{^j_\alpha}\nonumber\\
&+\;2g_1g_2\left[e\indices{_k^\alpha}\,e\indices{_i^\xi}\,S\indices{^\mu_\xi_\alpha}
+\left(e\indices{_k^\xi}\,e\indices{_i^\mu}-e\indices{_k^\mu}\,e\indices{_i^\xi}\right)S\indices{^\alpha_\xi_\alpha}\right]\eta^{kj} = \Sigma\indices{_i^j^\mu}.
\label{eq:poisson-type}
\end{align}
This is a source equation, where the spinor fields act as the source for the curvature and torsion of spacetime.
It simplifies considerably neglecting all torsion terms:
\begin{equation*}
g_1\,R\indices{_i^{j\mu\alpha}_{;\alpha}}=\Sigma\indices{_i^j^\mu}.
\end{equation*}
To obtain the coordinate space representation of Eq.~(\ref{eq:poisson-type}), we contract it with $e\indices{^i_\nu}\,e\indices{_j^\beta}$:
\begin{align*}
&g_1\left(R\indices{_\nu^{\beta\mu\alpha}_{;\alpha}}-R\indices{_\nu^{\beta\xi\alpha}}S\indices{^\mu_\xi_\alpha}
-2R\indices{_\nu^{\beta\mu\xi}}S\indices{^\alpha_\xi_\alpha}\right)\nonumber\\
&-M_\mathrm{p}^2 \left(S\indices{^\mu_\nu^\beta}
+\delta{_\nu^\mu}\,S\indices{^\alpha^\beta_\alpha}-g^{\beta\mu}\,S\indices{^\alpha_\nu_\alpha}\right)-g_3\,S\indices{_\nu^\beta^\mu}
=\Sigma\indices{_\nu^\beta^\mu}.
\end{align*}
Raising finally the index $\nu$, we observe that merely the skew-symmetric portion of the term
proportional to $g_3$ contributes as all other terms are already skew-symmetric in $\nu$ and $\beta$:
\begin{align}
&g_1\left(R\indices{^\nu^{\beta\mu\alpha}_{;\alpha}}-R\indices{^\nu^{\beta\xi\alpha}}S\indices{^\mu_\xi_\alpha}
-2R\indices{^\nu^{\beta\mu\xi}}S\indices{^\alpha_\xi_\alpha}\right) \nonumber \\
&-M_\mathrm{p}^2 \left(S\indices{^\mu^\nu^\beta}
+g^{\nu\mu}\,S\indices{^\alpha^\beta_\alpha}-g^{\beta\mu}\,S\indices{^\alpha^\nu_\alpha}\right)-g_3\,S\indices{^{[\nu\beta]}^\mu}
=\Sigma\indices{^\nu^\beta^\mu}.
\label{eq:poisson-type-coord}
 \end{align}
On the other hand, this proves that with this ansatz for gauge gravity the symmetric term proportional to $g_3$ has to vanish identically, which implies that torsion must be restricted to a completely anti-symmetric form (aka pseudo-vector torsion):
\begin{align}\boxed{
S^{(\alpha\beta)\mu}=0.}
\label{eq:totallyantisyymetric torsion}
\end{align}
This dynamic constraint resolves many discussions on the structure of torsion, cf. e.g.~\cite{Capozziello:2001mq}, and also ensures that the autoparallel and geodetic transport are identical.
We shall in the following take this constraint into account and be led to considerably simplified formulae.
For \eref{eq:poisson-type-coord} this gives
\begin{align}
g_1\left(R\indices{^\nu^{\beta\mu\alpha}_{;\alpha}}-R\indices{^\nu^{\beta\xi\alpha}}S\indices{^\mu_\xi_\alpha}\right)-(M_\mathrm{p}^2 +g_3)\,S\indices{^{\nu\beta}^\mu}=\Sigma\indices{^\nu^\beta^\mu}.
\end{align}
Now for metric compatible geometries the affine connection satisfies the identity
\begin{equation}\label{eq:gammaing}
\gamma\indices{^{\lambda}_{\mu\nu}} =
\genfrac{\lbrace}{\rbrace}{0pt}{0}{\lambda}{\mu\nu}+
K\indices{^{\lambda}_{\mu\nu}}. 
\end{equation}
The first term on the r.h.s. is the symmetric Christoffel symbol that in the Einstein-Hilbert theory is identified with the affine connection and also called the Levi-Civita connection. It is determined entirely by the metric or vierbein via the Levi-Civita relation
\begin{equation} \label{def:LCgamma}
\genfrac{\lbrace}{\rbrace}{0pt}{0}{\lambda}{\mu\nu} := \onehalf g^{\lambda\alpha}
\left( \pfrac{g_{\nu\alpha}}{x^{\mu}} \, + \,
       \pfrac{g_{\alpha\mu}}{x^{\nu}} \, - \,
       \pfrac{g_{\mu\nu}}   {x^{\alpha}} \right)
\end{equation}
For the second term, the contortion $K\indices{^{\lambda}_{\mu\nu}}$, the identity
\begin{equation*}
 K\indices{^{\lambda}_{\mu\nu}} \equiv S\indices{^{\lambda}_{\mu\nu}}
\end{equation*}
holds for totally anti-symmetric torsion tensors.
Hence here we have
\begin{align*}
\gamma\indices{^{\lambda}_{\mu\nu}} =
\genfrac{\lbrace}{\rbrace}{0pt}{0}{\lambda}{\mu\nu} + S\indices{^{\lambda}_{\mu\nu}} 
\end{align*}
Denoting the covariant derivative with respect to the Levi-Civita connection by $\bar{\nabla}$, we finally arrive at
\begin{align}\boxed{
g_1\left( \bar{\nabla}_\alpha R\indices{^\nu^{\beta\mu\alpha}}
+R\indices{^\xi^{\beta\mu\alpha}}S\indices{^\nu_\xi_\alpha}
-R\indices{^\xi^{\nu\mu\alpha}}S\indices{^\beta_\xi_\alpha}
\right)-(M_\mathrm{p}^2 +g_3)\,S\indices{^{\nu\beta}^\mu}=\Sigma\indices{^\nu^\beta^\mu}.}
\end{align}
Notice that with $g_1 = g_3 = 0$ this equation reduces to the so called Cartan equation, an algebraic equation between spin momentum and (non-propagating) torsion.
Here torsion is a propagating field as will be shown below in \sref{sec:spin-torsion-2}.
Moreover, since the Riemann-Cartan curvature can be split up into a Riemann and a torsion-dependent ``Cartan'' portions,
\begin{equation*}
 R\indices{^\xi^{\beta\mu\alpha}} = \bar{R}\indices{^\xi^{\beta\mu\alpha}} + P\indices{^\xi^{\beta\mu\alpha}},
\end{equation*}
we can for the purpose of finding a solution isolate the metric (vierbein) and torsion dependent terms in that equation.

\subsection{Spin-torsion tensor coupling equation\label{sec:spin-torsion-2}}
Inserting in Eq.~(\ref{eq:k-div2}) the relation \eref{eq:eqm-k} gives
\begin{equation}\label{eq:spin-torsion}
g_3\left(S\indices{^{\nu\mu\alpha}_{;\alpha}}-S\indices{^\nu^{\beta\alpha}}\,S\indices{^\mu_\beta_\alpha}
\right)+T_\text{Gr}^{\nu\mu}=-T_0^{\nu\mu}.
\end{equation}
The symmetric portion in $\nu$, $\mu$ of this equation is then
\begin{equation}
T_0^{\nu\mu}=
g_3S\indices{^\nu^\beta^\alpha}S\indices{^\mu_\beta_\alpha}-T_\text{Gr}^{(\nu\mu)},
\end{equation}
since the symmetric part of $S$ vanishes. This coincides with the consistency equation \eref{eq:cons-eq-symm}, and gives in its explicit form \eref{eq:HGr-symmpart}.

\medskip
The skew-symmetric portion of Eq.~(\ref{eq:spin-torsion}),
\begin{equation*}
g_3\,\dete S\indices{^{\nu\mu\alpha}_{;\alpha}}
+\tilde{T}_\text{Gr}^{[\nu\mu]}=-\tilde{T}_\mathrm{D}^{[\nu\mu]},
\end{equation*}
simplifies with Eq.~(\ref{eq:HGr-skewpart}) to:
\begin{equation}\label{eq:spin-torsion-skewpart-metr}
M_{\mathrm{p}}^2 \,R_{[\nu\mu]}
+g_3S\indices{_{\nu\mu}^\alpha_{;\alpha}}
=-T\indices{_{{\mathrm{D}}\,[\nu\mu]}}.
\end{equation}
Using the identity
\begin{equation}\label{eq:ricci-identity}
R_{[\nu\mu]}\equiv S\indices{^\alpha_{\nu\mu;\alpha}}-2S\indices{^\beta_\nu_\mu}S\indices{^\alpha_\beta_\alpha}
+S\indices{^\alpha_{\nu\alpha;\mu}}-S\indices{^\alpha_{\mu\alpha;\nu}},
\end{equation}
gives for the now totally skew-symmetric torsion tensor
\begin{align}
&\quad M_{\mathrm{p}}^2 S\indices{^\alpha_{\nu\mu;\alpha}}+g_3S\indices{_{\nu\mu}^\alpha_{;\alpha}}=-T\indices{_{\mathrm{D}\,[\nu\mu]}}.
\label{eq:spin-torsion-skewpart-metr2}
\end{align}
This is a proper source equation for the torsion tensor.
With~\eref{eq:spin-tensor-deri-0} the right-hand side can be alternatively substituted by the divergence of the spin-momentum density:
\begin{equation}\label{eq:spin-torsion-skewpart-metr3}
\left( M_{\mathrm{p}}^2 +g_3 \right) \,S\indices{_{\nu\mu}^\alpha_{;\alpha}}= -T\indices{_{\mathrm{D}\,[\nu\mu]}}
=  -{\Sigma}\indices{_{\nu\mu}^\alpha_{;\alpha}}.
\end{equation}
which turns out to be a conservation law for torsion-spin:\index{conservation law for spin-torsion}
\begin{equation}\label{eq:spin-torsion-skewpart-metr4}\boxed{
\left[ \left( M_{\mathrm{p}}^2 +g_3 \right) S\indices{^{\nu\mu\alpha}}+ {\Sigma}\indices{^{\nu\mu}^\alpha}\right]_{;\alpha} = 0.}
\end{equation}
The expression in brackets, an algebraic relation between torsion and spin density, vanishes in the Einstein-Cartan theory~\cite{Hehl:1971qi}. 
Here, in contrast, it is a propagating field. 

\medskip
In terms of the Levi-Civita derivative,~\eref{eq:spin-torsion-skewpart-metr4} becomes
\begin{equation}\label{eq:spin-torsion-M}
\bar{\nabla}_\alpha \,\left[ \left(M_{\mathrm{p}}^2 +g_3\right) S\indices{^{\nu\mu}^\alpha}+ {\Sigma}\indices{^{\nu\mu}^\alpha} \right] =
S\indices{^\nu_{\beta\alpha}}\,{\Sigma}\indices{^\mu^{\beta\alpha}}
-S\indices{^\mu_{\beta\alpha}}\,{\Sigma}\indices{^\nu^{\beta\alpha}}.
\end{equation}

\subsection{Field equations of matter with anti-symmetric torsion}
Replacing the covariant derivative in the Klein-Gordon equation \eqref{eq:KG;} by the Levi-Civita derivative and the torsion term gives
\begin{align}\boxed{
g^{\alpha\beta}\bar{\nabla}_\alpha \bar{\nabla}_\beta \varphi+m^2\varphi=0,}
\end{align}
We see, that torsion drops out of the field equation for the scalar field.

\medskip
Similarly, the Maxwell-Proca equation \eqref{eq:MP;} becomes
\begin{align}\boxed{
\bar{\nabla}_\alpha f^{\mu\alpha}-m^2 a^\mu =0.}
\end{align}
Since all torsion dependent terms cancel out, the Proca-Maxwell equation is equal to that known from the standard Einsteinian gravity.
This means that while gauge fields influence curvature via their energy-momentum content, they do not ``feel'' torsion in their own propagation.
It is straightforward to extend this conclusion also to non-abelian gauge fields.
However, for fields not subject to $\mathrm{SU}(N)$ symmetry, the option to chose the Maxwell-type exterior derivative for the free Lagrangian is not unique. With having instead the quadratic covariant derivative in the Proca Lagrangian in section \ref{sec:a-contribution}, vector-torsion coupling would arise.

\medskip
Also the Dirac equation simplifies for a totally anti-symmetric torsion:
\begin{align}\boxed{
\rmi\left(\gamma^\beta-\frac{1}{3M}\sigma^{\xi\alpha}S^\beta{}_{\xi\alpha}\right)\Dderr_\beta \psi-\left(m+\frac{1}{24M}\sigma^{\alpha\beta}\sigma^{nm}R_{nm\alpha\beta}\right)\psi=0.}
\end{align}
Compared to the standard Dirac equation in curved spacetime, two ``anomalous'' corrections related to the mass parameter $M$ arise.
Firstly, a term proportional to torsion modifies the dynamic affine connection $\gamma^\beta = \gamma^i\,e\indices{_i^\beta}$ that is projected via the vierbein field to the base manifold. 
Secondly, there is a mass correction proportional to the Riemann tensor.
A further, regular correction arising from torsion is encountered once we realize that the spin connection can be, according to the vierbein postulate \eqref{def:vierbeinpostulate}, split up into a Levi-Civita portion
$\bar{\omega}\indices{^i_j_\mu}$ (Ricci rotation coefficients) and a torsion term:
\begin{equation*}
 \omega\indices{^i_j_\mu} = \bar{\omega}\indices{^i_j_\mu} + S\indices{^i_j_\mu}\,.
\end{equation*}
Then the covariant derivative \eqref{Def:Spincovderivative1} too has a Levi-Civita and a torsional portion:
\begin{equation}
 \Dderr_\mu =\pfracr{}{x^\mu}-\iquarter\sigma^{lj}\,\omega\indices{_{lj\mu}}
 = \overrightarrow{\bar{\mathcal{D}}}_\mu-\iquarter\sigma^{lj}\,S\indices{_{lj\mu}}.
\end{equation}

\section{Conclusion\label{sec:conclusion}}

\subsection{Summary\label{sec:summary}}
The Covariant Canonical Gauge Gravity (CCGG) is a semi-classical relativistic field theory in the De Donder-Weyl formulation in the Hamiltonian picture.
For implementing local symmetries, the framework of the canonical transformation theory is deployed, providing a rigorous mathematical blueprint for implementing gauge fields for any kind of symmetries, starting off from a limited set of fundamental assumptions.

\medskip
After proving in the past that the Yang-Mills theories can be reproduced in this way, the approach is applied here to the external Lorentz and diffeomorphism symmetry on principal frame bundles.
The key initial postulates are the Hamiltonian principle of least action, and the principles of General Relativity and Equivalence.
The sole input is then the transformation behaviour of the involved matter fields, here the real Klein-Gordon, Maxwell-Proca, and the Dirac fields.
In order to establish a $1:1$ correspondence between the Lagranian and Hamiltonian pictures, we request the  Hamiltonians to be non-degenerate, i.e. to contain quadratic terms in the momentum fields.
The framework enables in the first place to \emph{derive} the field equations of motion for any matter fields in any kind of spacetime geometry.

\medskip
For the $\mathrm{SO}(1,3)\times\mathrm{Diff}(M)$ symmetry addressed here in the vierbein formalism, we identify the spin connection to be the gauge field for any theory of gravity.
The specification to a particular form of matter dynamics, and to a reasonable phenomenologically sound gravity ansatz, enables then the generic dynamic equations to be specified for that concrete dynamic system of selected particles and geometry.
The results are a set of field equations for matter fields in curved geometry that, for scalar and vector fields, confirm the so called comma-to-semicolon rule with the Levi-Civita connection.
For the Dirac field, in contrast, we apply the Gasiorowicz quadratic formulation, that leads to anomalous couplings to torsion and a to a curvature dependent mass correction.
This generalizes past work going back to Sciama, Kibble et al. that was based on the linear Einstein-Cartan ansatz, and sheds a new light on the so called Poincare Gauge Theory in its quadratic realization developed by Hehl et al.

\medskip
However, the strength of the covariant Hamiltonian approach is manifested in further findings. For the selected linear-quadratic gravity ansatz these are for example:
\begin{itemize}
\item Einstein's field equation is extended by a quadratic traceless Kretschmann term and torsion
\item The source term on the right-hand side of that equation turns out to be the total symmetrized metric (Hilbert) energy-momentum tensor of matter.
\item The Einstein-Cartan or Einstein-Hilbert theories are obtained as limiting cases by setting respectively $g_1=0$ or $g_1 =g_3 = 0$
\item The extended theory does not contradict solar-scale observations as the usual Schwarzschild metric is its valid solution
\item The ground state of geometry is the maximally symmetric, de Sitter or Anti-de Sitter, configuration
\item The quadratic term in the gravity Hamiltonian lends inertia to spacetime w.r.t. deformations vs. de Sitter or Anti-de Sitter geometry
\item The dynamics of curvature and torsion is given by covariant source equations \`{a} la Gauss law of electrodynamics, with source terms based on fermion spin only
\item Torsion is a totally anti-symmetric (pseudo-vector only), propagating field sourced by fermion spin
\item Torsional energy density acts as dark energy and contributes tot the total energy density of spacetime
 \item The total energy-momentum of matter and spacetime vanishes locally everywhere (``zero-energy universe``)
 \item The cosmological constant is of geometrical origin, independent of the vacuum energy of matter, and can be interpreted as the vacuum energy of spacetime
 \item After embedding in curved geometry, an emerging length (energy) parameter arises in the Dirac theory
 \item Fermions interact with torsion and acquire a curvature-dependent mass correction.
\end{itemize}
Below,  for the sake of convenience, the key dynamic equations for both, matter and spacetime, based on the CCGG ansatz, are collected:

\begin{itemize}
 \item Klein-Gordon equation

 $g^{\alpha\beta}\bar{\nabla}_\alpha \bar{\nabla}_\beta \varphi+m^2\varphi=0 $

 \item Proca equation

 $\bar{\nabla}_\alpha f^{\mu\alpha}-m^2 a^\mu =0$

  \item Dirac equation

  $\rmi\left(\gamma^\beta-\frac{1}{3M}\sigma^{\xi\alpha}S^\beta{}_{\xi\alpha}\right)
  \left(\overrightarrow{\bar{\mathcal{D}}}_\mu-\iquarter\sigma^{lj}\,S\indices{_{lj\mu}}\right) \\
  -\left(m+\frac{1}{24M}\sigma^{\alpha\beta}\sigma^{nm}R_{nm\alpha\beta}\right)\psi=0 $

  \item Curvature equation

 $g_1\left( \bar{\nabla}_\alpha R\indices{^\nu^{\beta\mu\alpha}}
+R\indices{^\xi^{\beta\mu\alpha}}S\indices{^\nu_\xi_\alpha}
-R\indices{^\xi^{\nu\mu\alpha}}S\indices{^\beta_\xi_\alpha}
\right)-(M_{\mathrm{p}}^2 +g_3)\,S\indices{^{\nu\beta}^\mu}=\Sigma\indices{^\nu^\beta^\mu}$

\item Spin-torsion equation

$S^{\alpha\beta\mu} \equiv S^{[\alpha\beta\mu]} \\
 { \nabla}_\alpha\, S\indices{^\alpha_{\nu\mu}}=-\frac{1}{M_{\mathrm{p}}^2 +g_3}T\indices{_{\mathrm{D}\,[\nu\mu]}} \\
 \,\, \Leftrightarrow \,\,
\bar{\nabla}_\alpha \left[ \left(M_{\mathrm{p}}^2 +g_3\right) S\indices{^{\nu\mu}^\alpha}+ {\Sigma}\indices{^{\nu\mu}^\alpha} \right] =
 S\indices{^\nu_{\beta\alpha}}\,{\Sigma}\indices{^\mu^{\beta\alpha}}
 -S\indices{^\mu_{\beta\alpha}}\,{\Sigma}\indices{^\nu^{\beta\alpha}}$

 \item CCGG field equation

 $ g_1\,\left(R\indices{_\alpha_\beta_\gamma^\mu}\,R\indices{^\alpha^\beta^\gamma^\nu}
-\quarter g^{\mu\nu} R\indices{_\alpha_\beta_\gamma_\delta}\,R\indices{^\alpha^\beta^\gamma^\delta}\right) 
+M_{\mathrm{p}}^2 \left(R^{(\mu\nu)}-\onehalf g^{\mu\nu}\,R-\Lambda g^{\mu\nu}\right) \nonumber\\
\quad-g_3\left(S\indices{_\alpha_\beta^\mu}\,S\indices{^\alpha^\beta^\nu}\
-\onehalf g^{\mu\nu}\,S\indices{_\alpha_\beta_\gamma}\,S\indices{^\alpha^\beta^\gamma}\right) =
{T}_{\mathrm{KG}}^{(\mu\nu)} + {T}_{\mathrm{P}}^{(\mu\nu)} + {T}_{\mathrm{D}}^{(\mu\nu)}$
\end{itemize}

The formulas for the energy-momentum tensors are discussed in \sref{sec:EMT}, the source terms spin-momentum $\Sigma\indices{^\nu^\beta^\mu}$ and skew-symmetric energy-momentum $T_{\mathrm{D}}^{[\nu\mu]}$ are defined
in Eqs.~\eqref{eq:spindensitytensor-expl} and \eqref{eq:emt-dirac-local-skew}, respectively.

\subsection{Discussion \label{sec:discussion}}
The formulation presented here has a generic and a specific aspect.
Its specific features are incorporated in the choice of the matter and gravity Hamiltonians, and for both some ambiguities exist.
For the Klein-Gordon field we might for example include a Higgs-type potential term replacing the mass term.

\medskip
The question whether a massive vector field $a_\mu$ acts as a source for torsion of spacetime was found to depend on the
model for the free vector field dynamics, as discussed briefly in Sect.~\ref{sec:a-contribution}.
Here we defined the field tensor in the Lagrangian as the \emph{exterior derivative} of the vector field $a_\mu$, aka Maxwell-Proca field,
and hence the momentum field tensor in the corresponding Hamiltonian was defined to be \emph{skew-symmetric}.
Then the respective term in the action~(\ref{action-integral5}) did \emph{not} couple directly to the gauge field $\ho\indices{^i_j_\nu}$.
Consequently, vector field terms do not turn up in the gauge Hamiltonian~(\ref{eq:HGgammaomega}) and thus neither
in the skew-symmetric portion~(\ref{eq:cons-eq-skew}) of the consistency equation, nor in the spin-curvature tensor equation~(\ref{eq:poisson-type-coord})
and in the spin-torsion tensor equation~(\ref{eq:spin-torsion-skewpart-metr}).
The other option for the Proca field was discussed earlier in Refs.~\cite{struckmeier17a,struckmeier18a},
where the diffeomorphism-invariance of the final action was achieved by converting the partial derivative of the vector field
in the initial action~(\ref{action-integral3}) into a \emph{covariant derivative} in the course of the gauge procedure.
In that case, the vector field was found to act as a source of torsion of spacetime.

\medskip
Space dynamics is formulated in terms of three equations. In addition to an extended Einstein field equations we get Maxwell-like equations for curvature and (propagating) torsion sourced by the fermion spin,
that are extensions of the equations derived earlier within the Poincare Gauge Theory, see e.g. \cite{kibble61, hehl2023lectures}.
Those extensions replace Kibble's ''conservation equations`` by contribution from the terms in the Hamiltonian with the coupling constants $g_1$ and $g_3$.

\medskip
\!\!Further novel features derive from the quadratic Gasiorowicz-Dirac Lagrangian. While it reproduces the standard Dirac equation in the free case, anomalous terms arise once interactions are turn on.
A Fermi-like interaction term arises in the electromagnetic case, and for gravity novel interactions with torsion and curvature are found.
Both are connected by the emerging length parameter, and so could, in principle, be subject of independent measurements.
The spin-curvature interaction should have immediate consequences in scenarios with $R\gg0$, e.g.\ inflation \cite{Benisty:2019jqz} and neutron star mergers.
On the other hand, depending on the size of the parameter $M$, even the ambient curvature might affect massless ferminons, i.e. influence the neutrino mass.
An independent estimate of the parameter $M$ is needed, e.g. from the electron and the muon $g-2$ experiments \cite{Sastry:1999is}.
However, there is a priori no reason to expect the value of $M$ to be independent of the particle flavour, and since neutrino experiments are notoriously difficult,
it does not seem likely that there will be estimates any time soon.

\medskip
Last not least, the so far omitted interactions between matter fields has to be taken into account in a combined way following the process sketched above.
The presented gauge procedure leading to the final action~(\ref{action-integral5}) can in particular be extended to include also internal symmetries.

\medskip
Our anticipation based on previous work~\cite{struckmeier17} is that the action of a $\mathrm{SO}(1,3)\times\mathrm{SU}(N)\times\mathrm{Diff}(M)$ gauge theory becomes by a straightforward analogy:
\begin{align}
S=\int_{V}&\d^4x\Bigg\{\tilde{\pi}^\nu\pfrac{\varphi}{x^\nu}
+\tilde{\kappabar}_{cI}^\nu\left(\pfrac{\psi^{cI}}{x^\nu}-\rmi g\,a\indices{^I_J_\nu}\,\psi^{cJ}
-\iquarter\ho\indices{^i_{j\nu}}\,\sigma\indices{^c_b_i^j}\,\psi^{bI}\right)\nonumber\\
&+\left(\pfrac{\psibar_{cI}}{x^\nu}+\rmi g\,\psibar_{cJ}\,a\indices{^J_I_\nu}
+\iquarter\psibar_{bI}\,\ho\indices{^i_{j\nu}}\,\sigma\indices{^b_c_i^j}\right)\tilde{\kappa}^{cI\nu}\nonumber\\
&+\onehalf\tilde{p}\indices{^J_I^{\mu\nu}}\left(\pfrac{a\indices{^{I}_{J\mu}}}{x^{\nu}}
-\pfrac{a\indices{^{I}_{J\nu}}}{x^{\mu}}-\rmi g\left(a\indices{^{I}_{N\nu}}\,a\indices{^{N}_{J\mu}}
-a\indices{^{I}_{N\mu}}\,a\indices{^{N}_{J\nu}}\right)\right)\nonumber\\
&+\onehalf\tilde{k}\indices{_i^{\mu\nu}}\left(\pfrac{e\indices{^i_\mu}}{x^\nu}-\pfrac{e\indices{^i_\nu}}{x^\mu}
+\ho\indices{^i_{j\nu}}\,e\indices{^j_\mu}-\ho\indices{^i_{j\mu}}\,e\indices{^j_\nu}\right)
\nonumber\\
&+\onehalf\hoc\indices{_i^{j\mu\nu}}\left(\pfrac{\ho\indices{^i_{j\mu}}}{x^\nu}
-\pfrac{\ho\indices{^i_{j\nu}}}{x^\mu}+\ho\indices{^i_{n\nu}}\,\ho\indices{^n_{j\mu}}
-\ho\indices{^i_{n\mu}}\,\ho\indices{^n_{j\nu}}\right)-\tilde{\HCd}_0-\tilde{\HCd}_{\mathrm{Gr}}\Bigg\}.
\label{action-integral7}
\end{align}
Herein, the spinor indices are written explicitly using lower case Latin letters $b$ and $c$,
whereas the $\mathrm{SU}(N)$ indices are denoted by upper case Latin letters.
As before, the lower case Latin letters $i$, $j$, and $n$ stand for the Lorentz indices and Greek letters for the coordinate space indices.
The Hamiltonian $\tilde{\HCd}_0$ is supposed here to also describe the dynamics of the Hermitian matrix $a\indices{^I_J_\mu}$ of \emph{massless} vector fields.
An example is the strict (unbroken) color symmetry $\mathrm{SU}(3)_c$ of the quark-gluon plasma.
By analogy with Yang-Mills theories, the theory~(\ref{action-integral7}) being a quadratic Palatini theory
might be void of ghosts~\cite{koivisto08} and also renormalizable~\cite{lasenby05}.

\section*{Acknowledgements}
The authors are indebted to the ``Walter Greiner-Gesell\-schaft zur F\"{o}rderung
der physi\-ka\-lischen Grundlagenforschung e.V.'' (WGG) in Frankfurt for their support.\,
DV especially thanks the Fueck Stiftung for support. \\
The authors also wish to thank Vladimir Denk for carefully reading the manuscript and suggesting improvements to the text,
and Johannes Kirsch, Armin van de Venn, David Benisty, Eduardo Guendelman, Peter Hess and Friedrich Hehl for valuable discussions.


\begin{thebibliography}{10}

\bibitem{weyl19}
Hermann Weyl.
\newblock {Eine neue Erweiterung der Relativit{\"a}tstheorie}.
\newblock {\em Annalen der Physik IV Folge}, 59:101, 1919.

\bibitem{einstein55}
Albert Einstein.
\newblock {\em The Meaning of Relativity}.
\newblock Princeton University Press, Princeton, 1955.

\bibitem{YM54}
Yang and Mills.
\newblock {Conservation of Isotopic Spin and Isotopic Gauge Invariance}.
\newblock {\em Phys. Rev.}, 96:191, 1954.

\bibitem{sciama62}
D.~W. Sciama.
\newblock The analogy between charge and spin in general relativity.
\newblock In {\em Recent Developments in General Relativity}, pages 415--439.
  Pergamon Press, Oxford; PWN, Warsaw, 1962.
\newblock {F}estschrift for {I}nfeld.

\bibitem{kibble61}
T.~W.~B. Kibble.
\newblock Lorentz invariance and the gravitational field.
\newblock {\em J. Math. Phys.}, 2:212--221, 3 1961.

\bibitem{utiyama56}
Ryoyu Utiyama.
\newblock {Invariant {T}heoretical {I}nterpretation of {I}nteraction}.
\newblock {\em Phys. Rev.}, 101(5):1597, 3 1956.

\bibitem{hehl2023lectures}
Friedrich~W. Hehl.
\newblock {\em {Four Lectures on Poincar{\'e} Gauge Field Theory}}, pages
  5--61.
\newblock Springer US, Boston, MA, 1980.

\bibitem{hayashi80}
K.~Hayashi and T.~Shirafuji.
\newblock Gravity from the {P}oincaré gauge theory of fundamental
  interactions.
\newblock {\em Prog. Theor. Phys.}, 64(3):866, 883, 1435, 2222, 9 1980.

\bibitem{hayashi80II}
K.~Hayashi and T.~Shirafuji.
\newblock {Gravity from Poincar\'e Gauge Theory of the Fundamental Particles
  II}.
\newblock {\em Prog. Theor. Phys.}, 64(3):883, September 1980.

\bibitem{hayashi80III}
K.~Hayashi and T.~Shirafuji.
\newblock {Gravity from Poincar\'e Gauge Theory of the Fundamental Particles
  III}.
\newblock {\em Prog. Theor. Phys.}, 64(3):866, September 1980.

\bibitem{hayashi80IV}
K.~Hayashi and T.~Shirafuji.
\newblock {Gravity from Poincar\'e Gauge Theory of the Fundamental Particles
  IV}.
\newblock {\em Prog. Theor. Phys.}, 64(3):2222, September 1980.

\bibitem{hayashi81V}
K.~Hayashi and T.~Shirafuji.
\newblock {Gravity from Poincar\'e Gauge Theory of the Fundamental Particles
  V}.
\newblock {\em Prog. Theor. Phys.}, 65(3):525, September 1981.

\bibitem{struckmeier08}
J.~Struckmeier and A.~Redelbach.
\newblock {Covariant {H}amiltonian Field Theory}.
\newblock {\em Int. J. Mod. Phys. E}, 17:435--491, 2008.

\bibitem{Chen:2015vya}
Chiang-Mei Chen, James~M. Nester, and Roh-Suan Tung.
\newblock {Gravitational energy for GR and Poincar\'e gauge theories: A
  covariant Hamiltonian approach}.
\newblock {\em Int. J. Mod. Phys. D}, 24(11):1530026, 2015.

\bibitem{StrRei12}
J.~Struckmeier and H.~Reichau.
\newblock General {U$(N)$} gauge transformations in the realm of covariant
  {H}amiltonian field theory.
\newblock In Walter Greiner, editor, {\em Exciting Interdisciplinary Physics},
  {FIAS} {I}nterdisciplinary {S}cience {S}eries, page 367. Springer
  International Publishing Switzerland, 2013.

\bibitem{struckmeier17}
J.~Struckmeier, D.~Vasak, and H.~Stoecker.
\newblock {C}ovariant {H}amiltonian {R}epresentation of {N}oether's {T}herorem
  and its {A}pplication to {SU(N)} {G}auge {T}heories.
\newblock In S.~Schramm and M.~Schaefer, editors, {\em {N}ew {H}orizons in
  {F}undamental {P}hysics}, FIAS Interdisciplinary Science Series. Springer
  International Publishing Switzerland, 2017.

\bibitem{struckmeier17a}
J.~Struckmeier, J.~Muench, D.~Vasak, J.~Kirsch, M.~Hanauske, and H.~Stoecker.
\newblock {Canonical transformation path to gauge theories of gravity}.
\newblock {\em Phys. Rev. D}, 95:124048, 6 2017.

\bibitem{dedonder30}
Th. {De Donder}.
\newblock {\em {Th{\'e}orie Invariantive Du Calcul des Variations}}.
\newblock Gaulthier-Villars \& Cie., Paris, 1930.

\bibitem{weyl35}
Hermann Weyl.
\newblock {Geodesic Fields in the Calculus of Variation for Multiple
  Integrals}.
\newblock {\em Ann. Math.}, 36(3):607--629, 1935.

\bibitem{hehl76}
F.~W. Hehl, P.~von~der Heyde, G.~D. Kerlick, and J.~M. Nester.
\newblock {General relativity with spin and torsion: {F}oundations and
  prospects}.
\newblock {\em Rev. Mod. Phys.}, 48(3):393, 1976.

\bibitem{Tryon:1973xi}
Edward~P. Tryon.
\newblock {Is the universe a vacuum fluctuation}.
\newblock {\em Nature}, 246:396, 1973.

\bibitem{struckvasak18c}
J.~Struckmeier, D.~Vasak, A.~Redelbach, and H.~Stoecker.
\newblock Pauli-type coupling between spinors and curved spacetime.
\newblock 2020.
\newblock arXiv:1812.09669.

\bibitem{Benisty:2019jqz}
David Benisty, Eduardo~I. Guendelman, Emmanuel~N. Saridakis, Horst Stoecker,
  Jurgen Struckmeier, and David Vasak.
\newblock {Inflation from fermions with curvature-dependent mass}.
\newblock {\em Phys. Rev. D}, D100(4):043523, 2019.

\bibitem{vasak19a}
D.~Vasak, J.~Kirsch, D.~Kehm, and J.~Struckmeier.
\newblock Covariant canonical gauge gravitation and cosmology.
\newblock In {\em J. Phys. Conf. Ser.}, volume 1194, page 012108, 2019.

\bibitem{vasak19b}
D.~Vasak, J.~Kirsch, and J.~Struckmeier.
\newblock {Locally contorted space-time invokes inflation, dark energy, and a
  non-singular Big Bang}.
\newblock 2019.

\bibitem{misner}
Ch.~W. Misner, K.~S. Thorne, and J.~A. Wheeler.
\newblock {\em {Gravitation}}.
\newblock W. H. Freeman and Company, New York, 1973.

\bibitem{gasiorowicz66}
S.~Gasiorowicz.
\newblock {\em {Elementary particle physics}}.
\newblock Wiley, New York, 1966.

\bibitem{Benisty:2018ufz}
D.~Benisty, E.~I. Guendelman, D.~Vasak, J.~Struckmeier, and H.~Stoecker.
\newblock Quadratic curvature theories formulated as covariant canonical gauge
  theories of gravity.
\newblock {\em Phys. Rev. D}, 98:106021, 2018.

\bibitem{Vasak:2022gps}
D.~Vasak, J.~Kirsch, J.~Struckmeier, and H.~Stoecker.
\newblock {On the cosmological constant in the deformed Einstein-Cartan gauge
  gravity in De Donder-Weyl Hamiltonian formulation}.
\newblock {\em Astron. Nachr./AN}, e20220069, 2022.

\bibitem{carroll13}
Sean Carroll.
\newblock {\em Spacetime and Geometry}.
\newblock Prentice Hall, 2013.

\bibitem{einstein18}
Albert Einstein.
\newblock {Private letter to {Hermann} {W}eyl}.
\newblock ETH Z{\"u}rich Library, Archives and Estates, 3 1918.

\bibitem{struckmeier21a}
J.~Struckmeier and D.~Vasak.
\newblock Covariant canonical gauge theory of gravitation for fermions.
\newblock {\em Astron. Nachr.}, (e20220069), 2022.

\bibitem{lorentz1916}
H.~Lorentz.
\newblock {Over Einstein’s theorie der zwaartekracht (iii)}.
\newblock {\em Koninklikje Akademie van Wetenschappen the Amsterdam. Verslangen
  van de Gewone Vergaderingen der Wisen Natuurkundige Afdeeling}, 25:468--486,
  1916.

\bibitem{levi-civita1917}
T.~Levi-Civita.
\newblock {On the analytic expression that must be given to the gravitational
  tensor in Einstein’s theory}.
\newblock {\em Atti della Accademia Nazionale dei Lincei, Rendiconti Lincei,
  Scienze Fisiche e Naturali.}, 26(4), 1917.

\bibitem{jordan39}
P.~Jordan.
\newblock Bemerkungen zur kosmologie.
\newblock {\em Annalen der Physik}, 428(1):64--70, 1939.

\bibitem{sciama53}
D.W. Sciama.
\newblock On the origin of inertia.
\newblock {\em Monthly Notices of the Royal Astronomical Society}, 113:34,
  1953.

\bibitem{feynman62}
R.~Feynman, W.~Morinigo, and W.~Wagner.
\newblock {\em Feynman Lectures On Gravitation (Frontiers in Physics)}.
\newblock Westview Press, Boulder, Colorado, 2002.

\bibitem{Rosen:1994vj}
N.~Rosen.
\newblock {The energy of the universe}.
\newblock {\em Gen. Rel. Grav.}, 26:319--321, 1994.

\bibitem{cooperstock95}
F.~I. Cooperstock and M.~Israelit.
\newblock {The Energy of the Universe}.
\newblock {\em Foundations of Physics, Vol. 25, No. 4, 1995}, (4), 1995.

\bibitem{Hamada:2022fko}
{Ken-ji} Hamada.
\newblock {Revealing A Trans-Planckian World Solves The Cosmological Constant
  Problem}.

\bibitem{Melia:2022ifj}
Fulvio Melia.
\newblock {Initial energy of a spatially flat universe: A hint of its possible
  origin}.
\newblock {\em Astron. Nachr.}, 343(3):e224010, 2022.

\bibitem{hawking03}
S.~Hawking.
\newblock {\em The Theory of Everything}.
\newblock New Millenium Press, 2003.

\bibitem{kehm17}
D.~Kehm, J.~Kirsch, J.~Struckmeier, D.~Vasak, and M.~Hanauske.
\newblock {Violation of Birkhoff's theorem for pure quadratic gravity action}.
\newblock {\em Astron. Nachr./AN}, 338(9-10):1015--1018, 2017.

\bibitem{Capozziello:2001mq}
S.~Capozziello, G.~Lambiase, and C.~Stornaiolo.
\newblock {Geometric classification of the torsion tensor in space-time}.
\newblock {\em Annalen Phys.}, 10:713--727, 2001.

\bibitem{Hehl:1971qi}
F.~W. Hehl and B.~K. Datta.
\newblock {Nonlinear spinor equation and asymmetric connection in general
  relativity}.
\newblock {\em J. Math. Phys.}, 12:1334--1339, 1971.

\bibitem{struckmeier18a}
J\"urgen Struckmeier, David Vasak, and Johannes Kirsch.
\newblock {Generic Theory of Geometrodynamics from Noether’s Theorem for the
  $\mathrm {Diff}(M)$ Symmetry Group}.
\newblock In Johannes Kirsch, Jan Steinheimer-Froschauer, Stefan Schramm, and
  Horst Stöcker, editors, {\em Discoveries at the Frontiers of Science: From
  Nuclear Astrophysics to Relativistic Heavy Ion Collisions}, pages 143--181.
  Springer Nature Switzerland AG, 2020.

\bibitem{Sastry:1999is}
Ramchander~R. Sastry.
\newblock {Quantum electrodynamics with the Pauli term}.
\newblock 3 1999.

\bibitem{koivisto08}
T.~Koivisto.
\newblock Covariant conservation of energy-momentum in modified gravities.
\newblock {\em Class. Quant. Grav.}, 23:4289, 2006.

\bibitem{lasenby05}
A.~N. Lasenby, C.~J.~L. Doran, and R.~Heineke.
\newblock {Analytic solutions to Riemann-squared gravity with background
  isotropic torsion}.
\newblock {\em arXiv:gr-qc/0509014}, 2005.

\bibitem{peskin95}
M.E. Peskin and D.V. Schroeder.
\newblock {\em Quantum Field Theory}.
\newblock Westview Press, Boulder, Colorado, 1995.

\end{thebibliography}



\newpage
\appendix
\counterwithin{figure}{subsection}
\counterwithin{equation}{subsection}
\section*{Appendices}
\addcontentsline{toc}{section}{Appendices}
\renewcommand{\thesubsection}{\Alph{subsection}}

\subsection{Transformation rules for a generating function of type $\tilde{\FCd}_3^\nu$ for the action~(\ref{action-integral3})\label{app1}}
The integrand condition similar to Eq.~(\ref{F3derivative2}) for the action~(\ref{action-integral3}) becomes
\begin{align}
&\qquad\tilde{\pi}^\nu\,\pfrac{\varphi}{x^\nu}+\tilde{p}^{\mu\nu}\,\pfrac{a_\mu}{x^\nu}
+\pfrac{\psibar}{x^\nu}\,\tilde{\kappa}^{\nu}+\tilde{\kappabar}^\nu\,\pfrac{\psi}{x^\nu}
+\tilde{k}\indices{_i^{\mu\nu}}\,\pfrac{e\indices{^i_\mu}}{x^\nu}
+\hoc\indices{_i^{j\mu\nu}}\,\pfrac{\ho\indices{^i_{j\mu}}}{x^\nu}\nonumber\\
&\qquad-\tilde{\HCd}\left(\varphi,\tilde{\pi}^\nu,a_{\mu},\tilde{p}^{\mu\nu},\psi,\tilde{\kappabar}\indices{^\nu},
\psibar,\tilde{\kappa}^{\nu},e\indices{^i_\mu},\tilde{k}\indices{_i^{\mu\nu}},\ho\indices{^i_{j\mu}},
\hoc\indices{_i^{j\mu\nu}},x\right)\nonumber\\
&=\left(\tilde{\Pi}^\nu\pfrac{\Phi}{X^\nu}\!+\!\tilde{P}^{\mu\nu}\pfrac{A_\mu}{X^\nu}
+\!\pfrac{\Psibar}{X^\nu}\tilde{\KCd}^{\nu}\!+\!\tilde{\KCdbar}\indices{^\nu}\pfrac{\Psi}{X^\nu}
\!+\!\tilde{K}\indices{_I^{\mu\nu}}\pfrac{E\indices{^I_\mu}}{X^\nu}
+\HOc\indices{_I^{J\mu\nu}}\pfrac{\HO\indices{^I_{J\mu}}}{X^\nu}\right)\nonumber\\
&\qquad\times\detpartial{X}{x}\nonumber\\
&\qquad-\tilde{\HCd}^{\prime}\left(\Phi,\tilde{\Pi}^\nu,A_{\mu},\tilde{P}^{\mu\nu},\Psi,\tilde{\KCdbar}\indices{^\nu},
\Psibar,\tilde{\KCd}^{\nu},E\indices{^I_\mu},\tilde{K}\indices{_I^{\mu\nu}},\HO\indices{^I_{J\mu}},
\HOc\indices{_I^{J\mu\nu}},X\right)\detpartial{X}{x}\nonumber\\
&\qquad+\pfrac{}{x^\nu}\left(\tilde{\FCd}_{3}^\nu+\tilde{\pi}^\nu\varphi+\tilde{p}^{\mu\nu}a_{\mu}
+\psibar\,\tilde{\kappa}^{\nu}+\tilde{\kappabar}\indices{^\nu}\psi+\tilde{k}\indices{_i^{\mu\nu}}e\indices{^i_\mu}
+\hoc\indices{_i^{j\mu\nu}}\ho\indices{^i_{j\mu}}\right).
\label{forminvarianceH3}
\end{align}
Expanding the last line of Eq.~(\ref{forminvarianceH3}) yields:
\begin{align}
&-\pfrac{\tilde{\pi}^\beta}{x^\nu}\delta_\beta^\nu\varphi
-\pfrac{\tilde{p}^{\mu\beta}}{x^\nu}\delta_\beta^\nu a_{\mu}
-\pfrac{\tilde{\kappabar}\indices{^\beta}}{x^\nu}\delta_\beta^\nu\psi
-\delta_\beta^\nu\psibar\pfrac{\tilde{\kappa}^{\beta}}{x^\nu}
-\pfrac{\tilde{k}\indices{_i^{\mu\beta}}}{x^\nu}\delta_\beta^\nu e\indices{^i_\mu}
\!-\pfrac{\hoc\indices{_i^{j\mu\beta}}}{x^\nu}\delta_\beta^\nu\ho\indices{^i_{j\mu}}\nonumber\\
&\qquad-\tilde{\HCd}\nonumber\\
&=\left(\!\tilde{\Pi}^\nu\pfrac{\Phi}{X^\nu}+\!\tilde{P}^{\mu\nu}\pfrac{A_{\mu}}{X^\nu}
+\pfrac{\Psibar}{X^\nu}\tilde{\KCd}^{\nu}\!+\tilde{\KCdbar}\indices{^\nu}\pfrac{\Psi}{X^\nu}
+\tilde{K}\indices{_I^{\mu\nu}}\pfrac{E\indices{^I_\mu}}{X^\nu}
+\HOc\indices{_I^{J\mu\nu}}\pfrac{\HO\indices{^I_{J\mu}}}{X^\nu}\!\right)\nonumber\\
&\qquad\times\detpartial{X}{x}\nonumber\\
&\qquad+\,\pfrac{\tilde{\FCd}_{3}^\lambda}{\Phi}\,\pfrac{X^\nu}{x^\lambda}\,\pfrac{\Phi}{X^\nu}
\,+\,\pfrac{\tilde{\FCd}_{3}^\nu}{\tilde{\pi}^\beta}\,\pfrac{\tilde{\pi}^\beta}{x^\nu}
\,+\,\pfrac{\tilde{\FCd}_{3}^\lambda}{A_{\mu}}\,\pfrac{X^\nu}{x^\lambda}\,\pfrac{A_{\mu}}{X^\nu}
\,+\,\pfrac{\tilde{\FCd}_{3}^\nu}{\tilde{p}^{\mu\beta}}\,\pfrac{\tilde{p}^{\mu\beta}}{x^\nu}\nonumber\\
&\qquad+\,\pfrac{\Psibar}{X^\nu}\,\pfrac{\tilde{\FCd}_{3}^\lambda}{\Psibar}\,\pfrac{X^\nu}{x^\lambda}
\,+\,\pfrac{\tilde{\FCd}_{3}^\nu}{\tilde{\kappa}^{\beta}}\,\pfrac{\tilde{\kappa}^{\beta}}{x^\nu}
\,+\,\pfrac{\tilde{\FCd}_{3}^\lambda}{\Psi}\,\pfrac{X^\nu}{x^\lambda}\,\pfrac{\Psi}{X^\nu}
\,+\,\pfrac{\tilde{\kappabar}\indices{^\beta}}{x^\nu}\,\pfrac{\tilde{\FCd}_{3}^\nu}{\tilde{\kappabar}\indices{^\beta}}\nonumber\\
&\qquad+\,\pfrac{\tilde{\FCd}_{3}^\lambda}{E\indices{^I_\mu}}\,\pfrac{X^\nu}{x^\lambda}\,\pfrac{E\indices{^I_\mu}}{X^\nu}
\,+\,\pfrac{\tilde{\FCd}_{3}^\nu}{\tilde{k}\indices{_i^{\mu\beta}}}\,\pfrac{\tilde{k}\indices{_i^{\mu\beta}}}{x^\nu}
\,+\,\pfrac{\tilde{\FCd}_{3}^\lambda}{\HO\indices{^I_{J\mu}}}\,\pfrac{X^\nu}{x^\lambda}\,\pfrac{\HO\indices{^I_{J\mu}}}{X^\nu}\nonumber\\
&\qquad+\,\pfrac{\tilde{\FCd}_{3}^\nu}{\hoc\indices{_i^{j\mu\beta}}}\,\pfrac{\hoc\indices{_i^{j\mu\beta}}}{x^\nu}
\,+\,\left.\pfrac{\tilde{\FCd}_{3}^\nu}{x^\nu}\right|_{\text{expl}}-\tilde{\HCd}^{\prime}\detpartial{X}{x}.
\label{F21derivative1}
\end{align}
We again derive the canonical transformation rules for the fields by comparing the coefficients of Eq.~(\ref{F21derivative1}):
\begin{subequations} \label{canonicalrules}
\begin{align}
\tilde{\Pi}^\nu&\equiv-\pfrac{\tilde{\FCd}_{3}^\lambda}{\Phi}\,\pfrac{X^\nu}{x^\lambda}\detpartial{x}{X}&
\delta_\beta^\nu\,\varphi&\equiv-\pfrac{\tilde{\FCd}_{3}^\nu}{\tilde{\pi}^\beta}\\
\tilde{P}^{\mu\nu}&\equiv-\pfrac{\tilde{\FCd}_{3}^\lambda}{A_{\mu}}\,\pfrac{X^\nu}{x^\lambda}\detpartial{x}{X}&
\delta_\beta^\nu\,a_{\mu}&\equiv-\pfrac{\tilde{\FCd}_{3}^\nu}{\tilde{p}^{\mu\beta}}\\
\tilde{\KCd}^{\nu}&\equiv-\pfrac{\tilde{\FCd}_{3}^\lambda}{\Psibar}\,\pfrac{X^\nu}{x^\lambda}\detpartial{x}{X}&
\delta_\beta^\nu\psibar&\equiv-\pfrac{\tilde{\FCd}_{3}^\nu}{\tilde{\kappa}^{\beta}}\\
\tilde{\KCdbar}\indices{^\nu}&\equiv-\pfrac{\tilde{\FCd}_{3}^\lambda}{\Psi}\,\pfrac{X^\nu}{x^\lambda}\detpartial{x}{X}&
\delta_\beta^\nu\psi&\equiv-\pfrac{\tilde{\FCd}_{3}^\nu}{\tilde{\kappabar}\indices{^\beta}}\\
\tilde{K}\indices{_I^{\mu\nu}}&\equiv-\pfrac{\tilde{\FCd}_{3}^\lambda}{E\indices{^I_\mu}}\,\pfrac{X^\nu}{x^\lambda}\detpartial{x}{X}&
\delta_\beta^\nu\,e\indices{^i_\mu}&\equiv-\pfrac{\tilde{\FCd}_{3}^\nu}{\tilde{k}\indices{_i^{\mu\beta}}}\\
\HOc\indices{_I^{J\mu\nu}}&\equiv-\pfrac{\tilde{\FCd}_{3}^\lambda}{\HO\indices{^{I}_{J\mu}}}\,\pfrac{X^\nu}{x^\lambda}\detpartial{x}{X}&
\delta_\beta^\nu\,\ho\indices{^i_{j\mu}}&\equiv-\pfrac{\tilde{\FCd}_{3}^\nu}{\hoc\indices{_i^{j\mu\beta}}}
\end{align}
and, as usual, for the Hamiltonian
\begin{equation}\label{eq:ham-rule}
\tilde{\HCd}^{\prime}=\left(\tilde{\HCd}+\pfrac{\tilde{\FCd}_{3}^\nu}{x^\nu}\bigg|_{\text{expl}}\right)\detpartial{x}{X}.
\end{equation}
\end{subequations}

\subsection{Spinor representation of the gauge field transformation rule~(\ref{eq:conn-trans})} \label{appS}

The parameters of the transformation given by the spinor transformation matrix $S$ such that
$\Psi=S\psi$ depend on those of the Lorentz transformation $\Lambda\indices{_I^i}$.
With the Dirac matrices $\Gamma^I$ and $\gamma^i$ in the local Lorentz frames,
the spinor representation of the Lorentz transformation follows from the requirement of
form-invariance of the Dirac equation under Lorentz transformations:
\begin{equation*}
\rmi\Gamma^I\pfrac{\Psi}{X^I}-m\Psi=0\qquad\Leftrightarrow\qquad\rmi\gamma^i\pfrac{\psi}{x^i}-m\psi=0
\end{equation*}
exactly if
\begin{equation}\label{eq:gamma-S}
S^{-1}\Gamma^I\,S\,\Lambda\indices{_I^i}=\gamma^i,\qquad
S^{-1}\Gamma_I\,S=\gamma_i\,\Lambda\indices{^i_I}.
\end{equation}
For a \emph{local} Lorentz transformation, the matrices $S$ and $\Lambda\indices{^i_I}$ are $x$-dependent.
The derivative of Eq.~(\ref{eq:gamma-S}) follows as
\begin{equation}\label{eq:gamma-S-deri}
\Gamma_I\,\pfrac{S}{x^\mu}=\pfrac{S}{x^\mu}\,\gamma_i\,\Lambda\indices{^i_I}
+S\,\gamma_i\,\pfrac{\Lambda\indices{^i_I}}{x^\mu}.
\end{equation}
Contracting Eq.~(\ref{eq:gamma-S-deri}) with $S^{-1}\Gamma^I$ from the left yields
\begin{equation*}
S^{-1}\Gamma^I\Gamma_I\,\pfrac{S}{x^\mu}=S^{-1}\Gamma^I\,\pfrac{S}{x^\mu}\,\gamma_i\,\Lambda\indices{^i_I}
+S^{-1}\Gamma^I\,S\,\gamma_i\,\pfrac{\Lambda\indices{^i_I}}{x^\mu},
\end{equation*}
hence
\begin{equation}\label{eq:gamma-S-deri-2}
4S^{-1}\,\pfrac{S}{x^\mu}=\gamma^j\,S^{-1}\pfrac{S}{x^\mu}\,\gamma_i\,\Lambda\indices{^i_I}\,\Lambda\indices{^I_j}
+\Lambda\indices{^I_j}\,\pfrac{\Lambda\indices{^i_I}}{x^\mu}\,\gamma^j\,\gamma_i.
\end{equation}
Contracting Eq.~(\ref{eq:gamma-S-deri}) with $S^{-1}$ from the left and with $\Lambda\indices{^I_j}\,\gamma^j$ from the right gives, on the other hand,
\begin{equation*}
S^{-1}\,\Gamma_I\,\pfrac{S}{x^\mu}\,\Lambda\indices{^I_j}\,\gamma^j
=S^{-1}\,\pfrac{S}{x^\mu}\,\gamma_i\,\Lambda\indices{^i_I}\,\Lambda\indices{^I_j}\,\gamma^j
+\gamma_i\,\pfrac{\Lambda\indices{^i_I}}{x^\mu}\,\Lambda\indices{^I_j}\,\gamma^j,
\end{equation*}
hence
\begin{equation}\label{eq:gamma-S-deri-3}
4S^{-1}\,\pfrac{S}{x^\mu}=\gamma_i\,S^{-1}\,\pfrac{S}{x^\mu}\,\Lambda\indices{^i_I}\,\Lambda\indices{^I_j}\,\gamma^j
-\Lambda\indices{^I_j}\,\pfrac{\Lambda\indices{^i_I}}{x^\mu}\,\gamma_i\,\gamma^j,
\end{equation}
Equations~(\ref{eq:gamma-S-deri-2}) and~(\ref{eq:gamma-S-deri-3}) are now added to give
\begin{align*}
8S^{-1}\,\pfrac{S}{x^\mu}&=\Lambda\indices{^I_j}\pfrac{\Lambda\indices{^i_I}}{x^\mu}\left(\gamma^j\,\gamma_i-\gamma_i\,\gamma^j\right)
+S^{-1}\left(\Gamma^I\,\pfrac{S}{x^\mu}\gamma_j\,\Lambda\indices{^j_I}+\Gamma_I\,\pfrac{S}{x^\mu}\gamma^j\,\Lambda\indices{^I_j}\right)\\
&=-2\rmi\Lambda\indices{^i_I}\,\pfrac{\Lambda\indices{^I_j}}{x^\mu}\,\sigma\indices{_i^j}
+2\gamma^i\,S^{-1}\,\pfrac{S}{x^\mu}\,\gamma_i.
\end{align*}
The last term vanishes by virtue of the Dirac algebra if $S$ stands for the spinor representation $S$ of the (infinitesimal) local Lorentz transformation
$\Lambda\indices{^I_i}\!=\!\delta^I_i+\epsilon\indices{^I_i}$ with infinitesimal parameters \mbox{$\epsilon_{ij}(x)=-\epsilon_{ji}(x)$} \cite{peskin95}:
\begin{equation*}
S(x)=\exp\left(-\iquarter\epsilon_{ij}(x)\sigma^{ij}\right)\qquad\Rightarrow\qquad\gamma^k S^{-1}\pfrac{S}{x^\mu}\gamma_k
=-\iquarter\pfrac{\epsilon_{ij}}{x^\mu}\underbrace{\gamma^k\sigma^{ij}\gamma_k}_{\equiv0}\equiv0.
\end{equation*}
We conclude that for \emph{local} Lorentz transformations
\begin{equation}\label{eq:conn-trans-spinor0}
\iquarter\Lambda\indices{^i_I}\,\pfrac{\Lambda\indices{^I_j}\,}{x^\mu}\sigma\indices{_i^j}=-S^{-1}\pfrac{S}{x^\mu}.
\end{equation}
With the commutators
\begin{equation}\label{eq:sigma-def}
\sigma^{ij}\equiv\ihalf\left(\gamma^i\,\gamma^j-\gamma^j\,\gamma^i\right),\qquad
\Sigma^{IJ}\equiv\ihalf\left(\Gamma^I\,\Gamma^J-\Gamma^J\,\Gamma^I\right),
\end{equation}
we finally encounter the spinor representation of the transformation rule~(\ref{eq:conn-trans}) for the gauge field $\ho\indices{^i_{j\mu}}$:
\begin{equation}\label{eq:conn-trans-spinor}
\iquarter\HO\indices{_I_{J\nu}}\Sigma^{IJ}=\left(\iquarter\,S\,\ho\indices{_i_{j\mu}}\,\sigma^{ij}\,S^{-1}+\pfrac{S}{x^\mu}S^{-1}\right)\pfrac{x^\mu}{X^\nu}.
\end{equation}

\subsection{Explicit calculation of the contributions of the vierbein field $E\indices{^I_\alpha}$ and the
gauge field $\HO\indices{^I_{J\alpha}}$ to Eq.~(\ref{F21derivative2})\label{app2}}
The coefficients related to the spacetime transformation in the transformation rule (\ref{F21derivative2})
for the Hamiltonian are to be converted into dependencies of the physical fields.
This means for the term proportional to the momentum field $\tilde{k}\indices{_i^{\mu\nu}}$:
\begin{align*}
-\tilde{k}\indices{_i^{\mu\nu}}\,\pfrac{}{x^\nu}\left(\Lambda\indices{^i_J}
\pfrac{X^\alpha}{x^\mu}\right)E\indices{^J_\alpha}
=-\tilde{k}\indices{_i^{\mu\nu}}\left(\pfrac{\Lambda\indices{^i_J}}{x^\nu}\pfrac{X^\alpha}{x^\mu}+
\Lambda\indices{^i_J}\ppfrac{X^\alpha}{x^\mu}{x^\nu}\right)E\indices{^J_\alpha}
\end{align*}
The derivative of $\Lambda\indices{^i_J}$ is expressed in terms of
the transformation rule~(\ref{eq:conn-trans}) for the gauge field $\ho\indices{^i_{j\nu}}$
\begin{equation*}
\pfrac{\Lambda\indices{^i_J}}{x^\nu}=\Lambda\indices{^i_I}\,\HO\indices{^I_{J\xi}}\,\pfrac{X^\xi}{x^\nu}-
\ho\indices{^i_{j\nu}}\,\Lambda\indices{^j_J},
\end{equation*}
whereas the second derivative of $X^\alpha(x)$ from is obtained from
the canonical transformation rule~(\ref{eq:tet-trans}), written in the equivalent form
\begin{equation*}
\pfrac{X^\alpha}{x^\mu}=E\indices{_I^\alpha}\,\Lambda\indices{^I_i}\,e\indices{^i_\mu}
\end{equation*}
as
\begin{align}
\ppfrac{X^\alpha}{x^\mu}{x^\nu}&=
\pfrac{E\indices{_I^\alpha}}{x^\nu}\Lambda\indices{^I_i}e\indices{^i_\mu}
+E\indices{_I^\alpha}\Lambda\indices{^I_i}\pfrac{e\indices{^i_\mu}}{x^\nu}
+E\indices{_I^\alpha}\pfrac{\Lambda\indices{^I_j}}{x^\nu}\,e\indices{^j_\mu}\nonumber\\
&=\pfrac{E\indices{_I^\alpha}}{x^\nu}E\indices{^I_\eta}\pfrac{X^\eta}{x^\mu}
\!+\!\pfrac{X^\alpha}{x^\xi}e\indices{_i^\xi}\pfrac{e\indices{^i_\mu}}{x^\nu}
\!+\!E\indices{_I^\alpha}\!\!\left(\!\Lambda\indices{^I_i}\ho\indices{^i_{j\nu}}\!
-\Omega\indices{^I_{J\xi}}\Lambda\indices{^J_j}\pfrac{X^\xi}{x^\nu}\right)\!e\indices{^j_\mu}\nonumber\\
&=-E\indices{_I^\alpha}\pfrac{E\indices{^I_\eta}}{X^\xi}\pfrac{X^\xi}{x^\nu}\pfrac{X^\eta}{x^\mu}
+e\indices{_i^\xi}\,\pfrac{e\indices{^i_\mu}}{x^\nu}\,\pfrac{X^\alpha}{x^\xi}\nonumber\\
&\quad+\pfrac{X^\alpha}{x^\xi}\,e\indices{_i^\xi}\,\ho\indices{^i_{j\nu}}\,e\indices{^j_\mu}-
E\indices{_I^\alpha}\,\Omega\indices{^I_{J\xi}}\,E\indices{^J_\eta}\,\pfrac{X^\eta}{x^\mu}\pfrac{X^\xi}{x^\nu}\nonumber\\
&=e\indices{_i^\xi}\!\left(\!\pfrac{e\indices{^i_\mu}}{x^\nu}+\ho\indices{^i_{j\nu}}e\indices{^j_\mu}\!\right)\!\pfrac{X^\alpha}{x^\xi}-
E\indices{_I^\alpha}\!\left(\!\pfrac{E\indices{^I_\eta}}{X^\xi}+\Omega\indices{^I_{J\xi}}E\indices{^J_\eta}\!\right)\!
\pfrac{X^\eta}{x^\mu}\pfrac{X^\xi}{x^\nu}.
\label{eq:X2-trans}
\end{align}
This yields
\begin{align*}
&\quad-\tilde{k}\indices{_i^{\mu\nu}}\,\pfrac{}{x^\nu}\left(\Lambda\indices{^i_J}
\pfrac{X^\alpha}{x^\mu}\right)E\indices{^J_\alpha}\\
&=-\tilde{k}\indices{_i^{\mu\nu}}\pfrac{\Lambda\indices{^i_J}}{x^\nu}\pfrac{X^\alpha}{x^\mu}E\indices{^J_\alpha}-
\tilde{k}\indices{_i^{(\mu\nu)}}\Lambda\indices{^i_J}\ppfrac{X^\alpha}{x^\mu}{x^\nu}E\indices{^J_\alpha}\\
&=-\tilde{k}\indices{_i^{\mu\nu}}\left(\Lambda\indices{^i_I}\HO\indices{^I_{J\xi}}\pfrac{X^\xi}{x^\nu}-
\ho\indices{^i_{j\nu}}\Lambda\indices{^j_J}\right)E\indices{^J_\alpha}\pfrac{X^\alpha}{x^\mu}
-\tilde{k}\indices{_i^{(\mu\nu)}}\Lambda\indices{^i_J}E\indices{^J_\alpha}\ppfrac{X^\alpha}{x^\mu}{x^\nu}\\
&=-\tilde{k}\indices{_i^{\mu\nu}}\Lambda\indices{^i_I}\HO\indices{^I_{J\xi}}E\indices{^J_\alpha}
\pfrac{X^\alpha}{x^\mu}\pfrac{X^\xi}{x^\nu}+\tilde{k}\indices{_i^{\mu\nu}}
\ho\indices{^i_{j\nu}}\Lambda\indices{^j_J}E\indices{^J_\alpha}\pfrac{X^\alpha}{x^\mu}\\
&\quad+\onehalf\tilde{k}\indices{_i^{\mu\nu}}\Lambda\indices{^i_I}\pfrac{X^\xi}{x^\nu}\pfrac{X^\eta}{x^\mu}
\left(\pfrac{E\indices{^I_\eta}}{X^\xi}+\pfrac{E\indices{^I_\xi}}{X^\eta}+\HO\indices{^I_{J\xi}}
E\indices{^J_\eta}+\HO\indices{^I_{J\eta}}E\indices{^J_\xi}\right)\nonumber\\
&\quad-\onehalf \tilde{k}\indices{_n^{\mu\nu}}
\Lambda\indices{^n_J}E\indices{^J_\alpha}\pfrac{X^\alpha}{x^\xi}e\indices{_i^\xi}\left(\pfrac{e\indices{^i_\mu}}{x^\nu}+
\pfrac{e\indices{^i_\nu}}{x^\mu}+\ho\indices{^i_{j\nu}}e\indices{^j_\mu}+\ho\indices{^i_{j\mu}}e\indices{^j_\nu}\right)\\
&=\tilde{k}\indices{_i^{\mu\nu}}\ho\indices{^i_{j\nu}}e\indices{^j_\mu}
-\tilde{K}\indices{_I^{\eta\xi}}\HO\indices{^I_{J\xi}}E\indices{^J_\eta}\detpartial{X}{x}\\
&\quad-\onehalf \tilde{k}\indices{_i^{\mu\nu}}\left(\pfrac{e\indices{^i_\mu}}{x^\nu}+\pfrac{e\indices{^i_\nu}}{x^\mu}
+\ho\indices{^i_{j\nu}}e\indices{^j_\mu}+\ho\indices{^i_{j\mu}}e\indices{^j_\nu}\right)\\
&\quad+\onehalf \tilde{K}\indices{_I^{\eta\xi}}\left(\pfrac{E\indices{^I_\eta}}{X^\xi}+\pfrac{E\indices{^I_\xi}}{X^\eta}
+\HO\indices{^I_{J\xi}}E\indices{^J_\eta}+\HO\indices{^I_{J\eta}}E\indices{^J_\xi}\right)\detpartial{X}{x},
\end{align*}
hence finally
\begin{align*}
-\tilde{k}\indices{_i^{\mu\nu}}\,\pfrac{}{x^\nu}&\left(\Lambda\indices{^i_J}\pfrac{X^\alpha}{x^\mu}\right)E\indices{^J_\alpha}
=-\onehalf\tilde{k}\indices{_i^{\mu\nu}}\left(\pfrac{e\indices{^i_\mu}}{x^\nu}+\pfrac{e\indices{^i_\nu}}{x^\mu}-
\ho\indices{^i_{j\nu}}e\indices{^j_\mu}+\ho\indices{^i_{j\mu}}e\indices{^j_\nu}\right)\nonumber\\
&\qquad+\onehalf\tilde{K}\indices{_I^{\mu\nu}}\left(\pfrac{E\indices{^I_\mu}}{X^\nu}+\pfrac{E\indices{^I_\nu}}{X^\mu}-
\Omega\indices{^I_{J\nu}}E\indices{^J_\mu}+\Omega\indices{^I_{J\mu}}E\indices{^J_\nu}\right)\detpartial{X}{x}.
\end{align*}
In the same way, all coefficients in the term proportional to $\hoc\indices{_i^{j\mu\nu}}$
of Eq.~(\ref{F21derivative2}) are now converted into dependencies on the physical fields.
To this end, we first write this term in expanded form:
\begin{align*}
&\quad-\hoc\indices{_i^{j\mu\nu}}\left[\HO\indices{^I_{J\alpha}}\,\pfrac{}{x^\nu}
\left(\Lambda\indices{^i_I}\,\Lambda\indices{^J_j}\,\pfrac{X^\alpha}{x^\mu}\right)
+\pfrac{}{x^\nu}\left(\Lambda\indices{^i_I}\,\pfrac{\Lambda\indices{^I_j}}{x^\mu}\right)\right]\\
&=-\hoc\indices{_i^{j\mu\nu}}\HO\indices{^I_{J\alpha}}
\left(\pfrac{\Lambda\indices{^i_I}}{x^\nu}\,\Lambda\indices{^J_j}\,\pfrac{X^\alpha}{x^\mu}
+\Lambda\indices{^i_I}\,\pfrac{\Lambda\indices{^J_j}}{x^\nu}\,\pfrac{X^\alpha}{x^\mu}
+\Lambda\indices{^i_I}\,\Lambda\indices{^J_j}\,\ppfrac{X^\alpha}{x^\mu}{x^\nu}\right)\\
&\quad-\hoc\indices{_i^{j\mu\nu}}\left(\pfrac{\Lambda\indices{^i_I}}{x^\nu}\,\pfrac{\Lambda\indices{^I_j}}{x^\mu}
+\Lambda\indices{^i_I}\ppfrac{\Lambda\indices{^I_j}}{x^\mu}{x^\nu}\right).
\end{align*}
With the transformation rule~(\ref{omegatransform1}) solved for $\HO\indices{^I_{J\alpha}}$
\begin{equation*}
\HO\indices{^I_{J\alpha}}=\left(\Lambda\indices{^I_n}\,\ho\indices{^n_{m\xi}}
-\pfrac{\Lambda\indices{^I_m}}{x^\xi}\right)\Lambda\indices{^m_J}\pfrac{x^\xi}{X^\alpha},
\end{equation*}
Eq.~(\ref{F21derivative3}) is expressed equivalently in terms of the original fields as
\begin{align}
&-\hoc\indices{_i^{j\mu\nu}}\left[\left(\Lambda\indices{^I_n}\ho\indices{^n_{m\xi}}
-\pfrac{\Lambda\indices{^I_m}}{x^\xi}\right)\right.\nonumber\\
&\qquad\qquad\times\left(\pfrac{\Lambda\indices{^i_I}}{x^\nu}\delta^m_j\delta^\xi_\mu+
\Lambda\indices{^i_I}\pfrac{\Lambda\indices{^J_j}}{x^\nu}\Lambda\indices{^m_J}\delta^\xi_\mu+
\Lambda\indices{^i_I}\delta^m_j\ppfrac{X^\alpha}{x^\mu}{x^\nu}\pfrac{x^\xi}{X^\alpha}\right)\nonumber\\
&\left.\qquad\qquad+\,\pfrac{\Lambda\indices{^i_I}}{x^\nu}\,\pfrac{\Lambda\indices{^I_j}}{x^\mu}+
\Lambda\indices{^i_I}\ppfrac{\Lambda\indices{^I_j}}{x^\mu}{x^\nu}\right]\nonumber\\
&=-\hoc\indices{_i^{j\mu\nu}}\!\left(\ho\indices{^n_{j\mu}}\Lambda\indices{^I_n}\pfrac{\Lambda\indices{^i_I}}{x^\nu}
+\ho\indices{^i_{n\mu}}\Lambda\indices{^n_J}\pfrac{\Lambda\indices{^J_j}}{x^\nu}
+\ho\indices{^i_{j\xi}}\ppfrac{X^\alpha}{x^\mu}{x^\nu}\pfrac{x^\xi}{X^\alpha}
-\cancel{\pfrac{\Lambda\indices{^I_j}}{x^\mu}\pfrac{\Lambda\indices{^i_I}}{x^\nu}}\right.\nonumber\\
&\left.\;-\,\Lambda\indices{^i_I}\pfrac{\Lambda\indices{^I_n}}{x^\mu}\Lambda\indices{^n_J}\pfrac{\Lambda\indices{^J_j}}{x^\nu}
-\Lambda\indices{^i_I}\pfrac{\Lambda\indices{^I_j}}{x^\xi}\ppfrac{X^\alpha}{x^\mu}{x^\nu}\pfrac{x^\xi}{X^\alpha}
+\cancel{\pfrac{\Lambda\indices{^i_I}}{x^\nu}\pfrac{\Lambda\indices{^I_j}}{x^\mu}}
+\!\Lambda\indices{^i_I}\ppfrac{\Lambda\indices{^I_j}}{x^\mu}{x^\nu}\right)\nonumber\\
&=-\hoc\indices{_i^{j\mu\nu}}\!\left[\ho\indices{^i_{n\mu}}\Lambda\indices{^n_J}\pfrac{\Lambda\indices{^J_j}}{x^\nu}
-\ho\indices{^n_{j\mu}}\Lambda\indices{^i_I}\pfrac{\Lambda\indices{^I_n}}{x^\nu}
+\!\left(\ho\indices{^i_{j\xi}}\!-\Lambda\indices{^i_I}\pfrac{\Lambda\indices{^I_j}}{x^\xi}\right)
\ppfrac{X^\alpha}{x^\mu}{x^\nu}\!\pfrac{x^\xi}{X^\alpha}\right.\nonumber\\
&\left.\qquad\qquad\quad-\,\Lambda\indices{^i_I}\pfrac{\Lambda\indices{^I_n}}{x^\mu}
\Lambda\indices{^n_J}\pfrac{\Lambda\indices{^J_j}}{x^\nu}
+\Lambda\indices{^i_I}\ppfrac{\Lambda\indices{^I_j}}{x^\mu}{x^\nu}\right].\label{F21derivative4}
\end{align}
Now, the transformation rule~(\ref{omegatransform1}) is inserted in the form
\begin{equation*}
\Lambda\indices{^i_I}\,\pfrac{\Lambda\indices{^I_j}}{x^\nu}=
\ho\indices{^i_{j\nu}}-\Lambda\indices{^i_I}\,\HO\indices{^I_{J\alpha}}\Lambda\indices{^J_j}\pfrac{X^\alpha}{x^\nu}
\end{equation*}
and its derivative
\begin{align*}
\Lambda\indices{^i_I}\,&\ppfrac{\Lambda\indices{^I_j}}{x^\mu}{x^\nu}=
\onehalf\Lambda\indices{^i_I}\left(\pfrac{\Lambda\indices{^I_n}}{x^\nu}\ho\indices{^n_{j\mu}}+
\pfrac{\Lambda\indices{^I_n}}{x^\mu}\ho\indices{^n_{j\nu}}\right)+
\onehalf\left(\pfrac{\ho\indices{^i_{j\mu}}}{x^\nu}+\pfrac{\ho\indices{^i_{j\nu}}}{x^\mu}\right)\\
&\quad-\onehalf\Lambda\indices{^i_I}\left(\pfrac{\HO\indices{^I_{J\xi}}}{X^\alpha}+\pfrac{\HO\indices{^I_{J\alpha}}}{X^\xi}\right)
\Lambda\indices{^J_j}\pfrac{X^\xi}{x^\mu}\pfrac{X^\alpha}{x^\nu}\\
&\quad-\onehalf\Lambda\indices{^i_I}\HO\indices{^I_{J\alpha}}\left(
\pfrac{\Lambda\indices{^J_j}}{x^\nu}\pfrac{X^\alpha}{x^\mu}+\pfrac{\Lambda\indices{^J_j}}{x^\mu}\pfrac{X^\alpha}{x^\nu}\right)
-\Lambda\indices{^i_I}\,\HO\indices{^I_{J\alpha}}\Lambda\indices{^J_j}\ppfrac{X^\alpha}{x^\mu}{x^\nu}.
\end{align*}
Putting it together
\begin{align*}
&-\hoc\indices{_i^{j\mu\nu}}\Bigg[
\onehalf\left(\pfrac{\ho\indices{^i_{j\mu}}}{x^\nu}+\pfrac{\ho\indices{^i_{j\nu}}}{x^\mu}\right)
-\onehalf\left(\pfrac{\HO\indices{^I_{K\beta}}}{X^\alpha}+\pfrac{\HO\indices{^I_{K\alpha}}}{X^\beta}\right)
\Lambda\indices{^i_I}\Lambda\indices{^K_j}\pfrac{X^\beta}{x^\mu}\pfrac{X^\alpha}{x^\nu}\\
&\qquad\quad+\ho\indices{^i_{n\mu}}
\left(\ho\indices{^n_{j\nu}}-\Lambda\indices{^n_I}\,\HO\indices{^I_{J\alpha}}\Lambda\indices{^J_j}\pfrac{X^\alpha}{x^\nu}\right)\\
&\qquad\quad-\onehalf\ho\indices{^n_{j\mu}}\left(\ho\indices{^i_{n\nu}}-\Lambda\indices{^i_I}\,\HO\indices{^I_{J\alpha}}
\Lambda\indices{^J_n}\pfrac{X^\alpha}{x^\nu}\right)\nonumber\\
&\qquad\quad+\onehalf\ho\indices{^n_{j\nu}}
\left(\ho\indices{^i_{n\mu}}-\Lambda\indices{^i_I}\,\HO\indices{^I_{J\alpha}}\Lambda\indices{^J_n}\pfrac{X^\alpha}{x^\mu}\right)
+\cancel{\Lambda\indices{^i_I}\,\HO\indices{^I_{J\alpha}}\Lambda\indices{^J_j}\ppfrac{X^\alpha}{x^\mu}{x^\nu}}\\
&\qquad\quad-\,
\left(\ho\indices{^i_{n\mu}}-\Lambda\indices{^i_I}\,\HO\indices{^I_{J\alpha}}\Lambda\indices{^J_n}\pfrac{X^\alpha}{x^\mu}\right)
\left(\ho\indices{^n_{j\nu}}-\Lambda\indices{^n_K}\,\HO\indices{^K_{L\beta}}\Lambda\indices{^L_j}\pfrac{X^\beta}{x^\nu}\right)\\
&\qquad\quad-\onehalf\Lambda\indices{^i_I}\HO\indices{^I_{J\alpha}}\left(
\Lambda\indices{^J_n}\,\ho\indices{^n_{j\nu}}
-\HO\indices{^J_{K\beta}}\Lambda\indices{^K_j}\pfrac{X^\beta}{x^\nu}\right)\pfrac{X^\alpha}{x^\mu}\\
&\qquad\quad-\onehalf\Lambda\indices{^i_I}\HO\indices{^I_{J\alpha}}\left(
\Lambda\indices{^J_n}\,\ho\indices{^n_{j\mu}}
-\HO\indices{^J_{K\beta}}\Lambda\indices{^K_j}\pfrac{X^\beta}{x^\mu}\right)\pfrac{X^\alpha}{x^\nu}\\
&\qquad\quad-\cancel{\Lambda\indices{^i_I}\,\HO\indices{^I_{J\alpha}}\Lambda\indices{^J_j}\ppfrac{X^\alpha}{x^\mu}{x^\nu}}\,\Bigg],
\end{align*}
which simplifies, after expanding
\begin{align*}
&-\hoc\indices{_i^{j\mu\nu}}\!\left[
\onehalf\left(\pfrac{\ho\indices{^i_{j\mu}}}{x^\nu}+\pfrac{\ho\indices{^i_{j\nu}}}{x^\mu}\right)
-\onehalf\left(\pfrac{\HO\indices{^I_{K\beta}}}{X^\alpha}+\pfrac{\HO\indices{^I_{K\alpha}}}{X^\beta}\right)
\Lambda\indices{^i_I}\Lambda\indices{^K_j}\pfrac{X^\alpha}{x^\nu}\pfrac{X^\beta}{x^\mu}\right.\\
&\qquad\quad+\ho\indices{^i_{n\mu}}
\ho\indices{^n_{j\nu}}-\ho\indices{^i_{n\mu}}\Lambda\indices{^n_I}\,\HO\indices{^I_{J\alpha}}\Lambda\indices{^J_j}\pfrac{X^\alpha}{x^\nu}
-\onehalf\ho\indices{^n_{j\mu}}\ho\indices{^i_{n\nu}}+\onehalf\ho\indices{^n_{j\mu}}\Lambda\indices{^i_I}\,\HO\indices{^I_{J\alpha}}
\Lambda\indices{^J_n}\pfrac{X^\alpha}{x^\nu}\nonumber\\
&\qquad\quad+\onehalf\ho\indices{^n_{j\nu}}\ho\indices{^i_{n\mu}}-\onehalf\ho\indices{^n_{j\nu}}\Lambda\indices{^i_I}\,
\HO\indices{^I_{J\alpha}}\Lambda\indices{^J_n}\pfrac{X^\alpha}{x^\mu}
-\ho\indices{^i_{n\mu}}\,\ho\indices{^n_{j\nu}}
+\ho\indices{^i_{n\mu}}\Lambda\indices{^n_I}\,\HO\indices{^I_{J\alpha}}\Lambda\indices{^J_j}\pfrac{X^\alpha}{x^\nu}\\
&\qquad\quad+\ho\indices{^n_{j\nu}}\Lambda\indices{^i_I}\,\HO\indices{^I_{J\alpha}}\Lambda\indices{^J_n}\pfrac{X^\alpha}{x^\mu}
-\Lambda\indices{^i_I}\,\HO\indices{^I_{J\alpha}}
\,\HO\indices{^J_{K\beta}}\Lambda\indices{^K_j}\pfrac{X^\alpha}{x^\mu}\pfrac{X^\beta}{x^\nu}\\
&\qquad\quad-\onehalf\Lambda\indices{^i_I}\HO\indices{^I_{J\alpha}}
\Lambda\indices{^J_n}\,\ho\indices{^n_{j\nu}}\pfrac{X^\alpha}{x^\mu}+\onehalf\Lambda\indices{^i_I}\HO\indices{^I_{J\alpha}}
\HO\indices{^J_{K\beta}}\Lambda\indices{^K_j}\pfrac{X^\alpha}{x^\mu}\pfrac{X^\beta}{x^\nu}\\
&\qquad\quad-\onehalf\Lambda\indices{^i_I}\HO\indices{^I_{J\alpha}}
\Lambda\indices{^J_n}\,\ho\indices{^n_{j\mu}}\pfrac{X^\alpha}{x^\nu}+\onehalf\Lambda\indices{^i_I}\HO\indices{^I_{J\alpha}}
\HO\indices{^J_{K\beta}}\Lambda\indices{^K_j}\pfrac{X^\alpha}{x^\nu}\pfrac{X^\beta}{x^\mu}\Bigg]\\
&=-\onehalf\hoc\indices{_i^{j\mu\nu}}\left[
\pfrac{\ho\indices{^i_{j\mu}}}{x^\nu}+\pfrac{\ho\indices{^i_{j\nu}}}{x^\mu}
+\ho\indices{^i_{n\mu}}\,\ho\indices{^n_{j\nu}}-\ho\indices{^i_{n\nu}}\,\ho\indices{^n_{j\mu}}\right.\\
&\qquad\qquad\quad-\left(\pfrac{\HO\indices{^I_{K\alpha}}}{X^\beta}+\pfrac{\HO\indices{^I_{K\beta}}}{X^\alpha}
+\HO\indices{^I_{J\alpha}}\HO\indices{^J_{K\beta}}-\HO\indices{^I_{J\beta}}\HO\indices{^J_{K\alpha}}\right)
\Lambda\indices{^i_I}\Lambda\indices{^K_j}\pfrac{X^\alpha}{x^\mu}\pfrac{X^\beta}{x^\nu}\Bigg]\\
&=-\onehalf\hoc\indices{_i^{j\mu\nu}}\left(
\pfrac{\ho\indices{^i_{j\mu}}}{x^\nu}+\pfrac{\ho\indices{^i_{j\nu}}}{x^\mu}
+\ho\indices{^i_{n\mu}}\,\ho\indices{^n_{j\nu}}-\ho\indices{^i_{n\nu}}\,\ho\indices{^n_{j\mu}}\right)\\
&\quad\,+\onehalf\HOc\indices{_I^{J\mu\nu}}\left(\pfrac{\HO\indices{^I_{J\mu}}}{X^\nu}+\pfrac{\HO\indices{^I_{J\nu}}}{X^\mu}
+\HO\indices{^I_{K\mu}}\,\HO\indices{^K_{J\nu}}-\HO\indices{^I_{K\nu}}\,\HO\indices{^K_{J\mu}}\right)\detpartial{X}{x}.
\end{align*}

\subsection{Explicit derivation of the generic Consistency equation \label{sec:gen-consistapp}}
The eight terms in \eref{eq:consistency0} involving the spinor fields add up to:
\begin{align}
&\quad\iquarter\bigg(\pfrac{\tilde{\kappabar}^\alpha}{x^\alpha}\sigma\indices{^i^j}\psi+\tilde{\kappabar}^\alpha\sigma\indices{^i^j}\pfrac{\psi}{x^\alpha}
-\pfrac{\psibar}{x^\alpha}\,\sigma\indices{^i^j}\tilde{\kappa}^\alpha-\psibar\,\sigma\indices{^i^j}\pfrac{\tilde{\kappa}^\alpha}{x^\alpha}\nonumber\\
&\qquad-\tilde{\kappabar}^\alpha\,\sigma\indices{^j^n}\,\psi\,\ho\indices{^i_{n\alpha}}
+\psibar\,\sigma\indices{^j^n}\,\tilde{\kappa}^\alpha\ho\indices{^i_{n\alpha}}
+\tilde{\kappabar}^\alpha\,\sigma\indices{^i^n}\,\psi\,\ho\indices{^j_{n\alpha}}
-\psibar\,\sigma\indices{^i^n}\,\tilde{\kappa}^\alpha\ho\indices{^j_{n\alpha}}\bigg)\nonumber\\
&=\iquarter\left(-\pfrac{\tilde{\HCd}_0}{\psi}-\iquarter\,\tilde{\kappabar}^\alpha\,\ho_{nm\alpha}\,\sigma^{nm}\right)\sigma\indices{^i^j}\psi
+\iquarter\tilde{\kappabar}^\alpha\sigma\indices{^i^j}\left(\pfrac{\tilde{\HCd}_0}{\tilde{\kappabar}^\alpha}+\iquarter\,\ho_{nm\alpha}\,\sigma^{nm}\,\psi\right)\nonumber\\
&\quad+\iquarter\left(-\pfrac{\tilde{\HCd}_0}{\tilde{\kappa}^\alpha}+\iquarter\,\psibar\,\ho_{nm\alpha}\,\sigma^{nm}\right)\sigma\indices{^i^j}\tilde{\kappa}^\alpha
+\iquarter\psibar\,\sigma\indices{^i^j}\left(\pfrac{\tilde{\HCd}_0}{\psibar}-\iquarter\,\ho_{nm\alpha}\,\sigma^{nm}\,\tilde{\kappa}^\alpha\right)\nonumber\\
&\quad+\iquarter\bigg[\psibar\,\sigma\indices{^j^n}\,\tilde{\kappa}^\alpha\ho\indices{^i_{n\alpha}}
-\tilde{\kappabar}^\alpha\,\sigma\indices{^j^n}\,\psi\,\ho\indices{^i_{n\alpha}}
+\tilde{\kappabar}^\alpha\,\sigma\indices{^i^n}\,\psi\,\ho\indices{^j_{n\alpha}}-\psibar\,\sigma\indices{^i^n}\,\tilde{\kappa}^\alpha\ho\indices{^j_{n\alpha}}\bigg]\nonumber\\
&=-\iquarter\pfrac{\tilde{\HCd}_0}{\psi}\sigma\indices{^i^j}\psi+\iquarter\tilde{\kappabar}^\alpha\sigma\indices{^i^j}\pfrac{\tilde{\HCd}_0}{\tilde{\kappabar}^\alpha}
-\iquarter\pfrac{\tilde{\HCd}_0}{\tilde{\kappa}^\alpha}\sigma\indices{^i^j}\tilde{\kappa}^\alpha
+\iquarter\psibar\,\sigma\indices{^i^j}\pfrac{\tilde{\HCd}_0}{\psibar}\nonumber\\
&\quad+\iquarter\bigg[
\iquarter\tilde{\kappabar}^\alpha\left(\sigma\indices{^i^j}\sigma^{nm}-\sigma^{nm}\sigma\indices{^i^j}\right)\psi
-\iquarter\psibar\left(\sigma\indices{^i^j}\sigma^{nm}-\sigma^{nm}\sigma\indices{^i^j}\right)\tilde{\kappa}^\alpha\bigg]\ho_{nm\alpha}\nonumber\\
&\quad+\iquarter\bigg[\tilde{\kappabar}^\alpha\left(\sigma\indices{^i^n}\,\ho\indices{^j_{n\alpha}}
-\sigma\indices{^j^n}\,\ho\indices{^i_{n\alpha}}\right)\psi-\psibar\left(\sigma\indices{^i^n}\,\ho\indices{^j_{n\alpha}}
-\sigma\indices{^j^n}\,\ho\indices{^i_{n\alpha}}\right)\tilde{\kappa}^\alpha\bigg].
\label{eq:dirac-terms}
\end{align}
By virtue of the algebra of the Dirac matrices, the commutator of $\sigma$-matrices amounts to
\begin{equation}\label{eq:comm-sigma-square}
\sigma\indices{^i^j}\,\sigma\indices{^n^m}-\sigma\indices{^n^m}\,\sigma\indices{^i^j}
\equiv -2\rmi\left(\eta^{im}\,\sigma\indices{^n^j}-\eta^{nj}\,\sigma\indices{^i^m}+\eta^{mj}\,\sigma^{in}-\eta^{in}\,\sigma^{mj}\right),
\end{equation}
hence, contracted with $\iquarter\ho\indices{_n_{m\alpha}}$:
\begin{align*}
\iquarter\ho\indices{_n_{m\alpha}}\!\left(\sigma\indices{^i^j}\sigma\indices{^n^m}-\sigma\indices{^n^m}\sigma\indices{^i^j}\right)
&=\onehalf\!\left(-\ho\indices{^i_{n\alpha}}\sigma\indices{^n^j}\!-\ho\indices{^j_{m\alpha}}\sigma\indices{^i^m}\!-
\ho\indices{^j_n_\alpha}\sigma\indices{^i^n}\!-\ho\indices{^i_{m\alpha}}\sigma\indices{^m^j}\right)\\
&=-\sigma\indices{^i^n}\,\ho\indices{^j_{n\alpha}}+\sigma\indices{^j^n}\,\ho\indices{^i_{n\alpha}}.
\end{align*}
The last two lines of Eq.~(\ref{eq:dirac-terms}) thus cancel.
The consistency equation~(\ref{eq:consistency1}) thus reduces to:
\begin{align*}
0&=\frac{\rmi}{4}\left(\pfrac{\tilde{\HCd}_0}{\psi}\sigma\indices{_i^j}\psi
-\tilde{\kappabar}^\alpha\sigma\indices{_i^j}\pfrac{\tilde{\HCd}_0}{\tilde{\kappabar}^\alpha}
+\pfrac{\tilde{\HCd}_0}{\tilde{\kappa}^\alpha}\sigma\indices{_i^j}\tilde{\kappa}^\alpha
-\psibar\,\sigma\indices{_i^j}\pfrac{\tilde{\HCd}_0}{\psibar}\right)\\
&\quad+\left(\pfrac{\tilde{k}\indices{_i^{[\alpha\beta]}}}{x^\beta}
-\ho\indices{^n_{i\beta}}\tilde{k}\indices{_n^{[\alpha\beta]}}\right)e\indices{^j_\alpha}
+\tilde{k}\indices{_i^{[\alpha\beta]}}\left(\pfrac{e\indices{^j_\alpha}}{x^\beta}+\ho\indices{^j_{n\beta}}e\indices{^n_\alpha}\right)\\
&\quad+\hoc\indices{_i^{m[\alpha\beta]}}\left(\pfrac{\ho\indices{^j_{m\alpha}}}{x^\beta}+\ho\indices{^j_{n\beta}}\ho\indices{^n_{m\alpha}}\right)
-\hoc\indices{_m^{\,j[\alpha\beta]}}\left(\pfrac{\ho\indices{^m_{i\alpha}}}{x^\beta}-\ho\indices{^m_{n\alpha}}\ho\indices{^n_{i\beta}}\right).
\end{align*}
Finally, the canonical equation~(\ref{eq:e-deri2}), (\ref{eq:omega-deri2}), and~(\ref{eq:k-div2}) are inserted,
\begin{align*}
0&=\frac{\rmi}{4}\left(\pfrac{\tilde{\HCd}_0}{\psi}\sigma\indices{_i^j}\psi
-\tilde{\kappabar}^\alpha\sigma\indices{_i^j}\pfrac{\tilde{\HCd}_0}{\tilde{\kappabar}^\alpha}
+\pfrac{\tilde{\HCd}_0}{\tilde{\kappa}^\alpha}\sigma\indices{_i^j}\tilde{\kappa}^\alpha
-\psibar\,\sigma\indices{_i^j}\pfrac{\tilde{\HCd}_0}{\psibar}\right)
-\pfrac{\tilde{\HCd}_0}{e\indices{^i_\alpha}}e\indices{^j_\alpha}\nonumber\\
&\quad-\pfrac{\tilde{\HCd}_{\mathrm{Gr}}}{e\indices{^i_\alpha}}e\indices{^j_\alpha}
+\tilde{k}\indices{_i^{[\alpha\beta]}}\pfrac{\tilde{\HCd}_\mathrm{Gr}}{\tilde{k}\indices{_j^{\alpha\beta}}}
+\hoc\indices{_i^{m[\alpha\beta]}}\pfrac{\tilde{\HCd}_\mathrm{Gr}}{\hoc\indices{_j^{m\alpha\beta}}}
-\hoc\indices{_m^{j[\alpha\beta]}}\pfrac{\tilde{\HCd}_\mathrm{Gr}}{\hoc\indices{_m^{i\alpha\beta}}},
\end{align*}
to give \eref{eq:consistency2}.

\subsection{Explicit derivation of the generalized Dirac equation\label{sec:gen-dirac}}
Equation~(\ref{eq:can-dirac1d}) writes in expanded form:
\begin{align*}
&\qquad e\indices{^k_\xi}\pfrac{e\indices{_k^\xi}}{x^\alpha}
\left[\ihalf e\indices{_j^\alpha}\gamma^j\psi-\frac{\rmi}{3M}e\indices{_j^\alpha}\,\sigma^{ji}\,e\indices{_i^\beta}
\left(\pfrac{\psi}{x^\beta}-\iquarter\ho_{nm\beta}\,\sigma^{nm}\,\psi\right)\right]\dete\\
&\quad-\ihalf\pfrac{e\indices{_j^\alpha}}{x^\alpha}\gamma^j\psi\,\dete-\ihalf e\indices{_j^\alpha}\gamma^j\pfrac{\psi}{x^\alpha}\dete\\
&\quad+\frac{\rmi}{3M}e\indices{_j^\alpha}\,\sigma^{ji}\,e\indices{_i^\beta}\left(
\cancel{\ppfrac{\psi}{x^\beta}{x^\alpha}}-\iquarter\pfrac{\ho_{nm\beta}}{x^\alpha}\,\sigma^{nm}\,\psi
-\iquarter\ho_{nm\beta}\,\sigma^{nm}\pfrac{\psi}{x^\alpha}\right)\dete\\
&\quad+\frac{\rmi}{3M}\left(\pfrac{e\indices{_j^\alpha}}{x^\alpha}\,\sigma^{ji}\,e\indices{_i^\beta}+
e\indices{_j^\alpha}\,\sigma^{ji}\,\pfrac{e\indices{_i^\beta}}{x^\alpha}\right)
\left(\pfrac{\psi}{x^\beta}-\iquarter\ho_{nm\beta}\,\sigma^{nm}\,\psi\right)\dete\\
&=\ihalf M\gamma_k\,e\indices{^k_\alpha}\left[\bcancel{\ihalf e\indices{_j^\alpha}\gamma^j\psi}
-\frac{\rmi}{3M}e\indices{_j^\alpha}\,\sigma^{ji}\,e\indices{_i^\beta}
\left(\pfrac{\psi}{x^\beta}-\iquarter\ho_{nm\beta}\,\sigma^{nm}\,\psi\right)\right]\dete\\
&\quad-\iquarter\,\ho_{kl\alpha}\,\sigma^{kl}\left[\ihalf e\indices{_j^\alpha}\gamma^j\psi
-\frac{\rmi}{3M}e\indices{_j^\alpha}\,\sigma^{ji}\,e\indices{_i^\beta}
\left(\pfrac{\psi}{x^\beta}-\iquarter\ho_{nm\beta}\,\sigma^{nm}\,\psi\right)\right]\dete\\
&\quad-\left(m-\bcancel{M}\right)\psi\,\dete,
\end{align*}
hence
\begin{align*}
&\quad\frac{\rmi}{3M}\Bigg[-e\indices{_j^\alpha}\,\sigma^{ji}\,e\indices{_i^\beta}\,e\indices{^k_\xi}\pfrac{e\indices{_k^\xi}}{x^\alpha}
\left(\pfrac{\psi}{x^\beta}-\iquarter\ho_{nm\beta}\,\sigma^{nm}\,\psi\right)\\
&\qquad\quad-\iquarter e\indices{_j^\alpha}\,\sigma^{ji}\,e\indices{_i^\beta}\left(
\pfrac{\ho_{nm\beta}}{x^\alpha}\,\sigma^{nm}\,\psi
+\ho_{nm\beta}\,\sigma^{nm}\pfrac{\psi}{x^\alpha}\right)\\
&\qquad\quad+\left(\pfrac{e\indices{_j^\alpha}}{x^\alpha}\,\sigma^{ji}\,e\indices{_i^\beta}+
e\indices{_j^\alpha}\,\sigma^{ji}\,\pfrac{e\indices{_i^\beta}}{x^\alpha}\right)
\left(\pfrac{\psi}{x^\beta}-\iquarter\ho_{nm\beta}\,\sigma^{nm}\,\psi\right)\\
&\qquad\quad+\iquarter\,\ho_{kl\alpha}\,\sigma^{kl}e\indices{_j^\alpha}\,\sigma^{ji}\,e\indices{_i^\beta}
\left(\pfrac{\psi}{x^\beta}-\iquarter\ho_{nm\beta}\,\sigma^{nm}\,\psi\right)\Bigg]\\
&=\rmi\,e\indices{_j^\alpha}\,\gamma^j\left(\pfrac{\psi}{x^\alpha}-\iquarter\ho_{nm\alpha}\,\sigma^{nm}\,\psi\right)-m\,\psi
+\ihalf\gamma^j\left(\pfrac{e\indices{_j^\alpha}}{x^\alpha}
-e\indices{_j^\alpha}\,e\indices{^k_\xi}\pfrac{e\indices{_k^\xi}}{x^\alpha}\right)\psi\\
&\quad-\oneeighth\ho_{nm\alpha}\,e\indices{_j^\alpha}\left(\gamma^j\,\sigma^{nm}-\sigma^{nm}\,\gamma^j\right)\psi.
\end{align*}
The last line is converted according to:
\begin{align*}
\oneeighth\,\ho_{nm\alpha}\,e\indices{_j^\alpha}\left(\gamma^j\sigma^{nm}-\sigma^{nm}\gamma^j\right)
&=\iquarter\,\ho_{nm\alpha}\,e\indices{_j^\alpha}\left(\eta^{jn}\gamma^m-\eta^{mj}\gamma^n\right)\\
&=\ihalf\gamma^j\,e\indices{_k^\alpha}\,\ho\indices{^k_{j\alpha}},
\end{align*}
which yields
\begin{align*}
&\frac{\rmi}{3M}\Bigg[-\iquarter e\indices{_j^\alpha}\,e\indices{_i^\beta}\left(\sigma^{ji}\pfrac{\ho_{nm\beta}}{x^\alpha}
+\iquarter\,\ho_{kl\alpha}\,\sigma^{kl}\,\sigma^{ji}\,\ho_{nm\beta}\right)\sigma^{nm}\,\psi\\
&\qquad+\iquarter e\indices{_j^\alpha}\,e\indices{_i^\beta}\,\ho_{nm\alpha}\left(\sigma^{nm}\sigma^{ji}\,
-\sigma^{ji}\,\sigma^{nm}\right)\pfrac{\psi}{x^\beta}\\
&\qquad+\left(\pfrac{e\indices{_j^\alpha}}{x^\alpha}\,\sigma^{ji}\,e\indices{_i^\beta}+
e\indices{_j^\alpha}\,\sigma^{ji}\,\pfrac{e\indices{_i^\beta}}{x^\alpha}
-e\indices{_j^\alpha}\,\sigma^{ji}\,e\indices{_i^\beta}\,e\indices{^k_\xi}\pfrac{e\indices{_k^\xi}}{x^\alpha}\right)
\left(\pfrac{\psi}{x^\beta}-\iquarter\ho_{nm\beta}\,\sigma^{nm}\,\psi\right)\Bigg]\\
&=\rmi\,\gamma^j\,e\indices{_j^\alpha}\left(\pfrac{\psi}{x^\alpha}-\iquarter\ho_{nm\alpha}\,\sigma^{nm}\,\psi\right)-m\,\psi
+\ihalf\gamma^j\left(\pfrac{e\indices{_j^\alpha}}{x^\alpha}-e\indices{_k^\alpha}\,\ho\indices{^k_{j\alpha}}
-e\indices{_j^\alpha}e\indices{^k_\xi}\pfrac{e\indices{_k^\xi}}{x^\alpha}\right)\psi.
\end{align*}
The first two lines are equivalently expressed as:
\begin{align*}
&\frac{\rmi}{3M}\Bigg[-\iquarter e\indices{_j^\alpha}\,e\indices{_i^\beta}\left(\sigma^{ji}\pfrac{\ho_{nm\beta}}{x^\alpha}
+\iquarter\,\ho_{kl\alpha}\,\sigma^{ji}\,\sigma^{kl}\,\ho_{nm\beta}\right)\sigma^{nm}\,\psi\\
&\qquad+\iquarter e\indices{_j^\alpha}\,e\indices{_i^\beta}\,\ho_{kl\alpha}\left(\sigma^{kl}\sigma^{ji}\,
-\sigma^{ji}\,\sigma^{kl}\right)\left(\pfrac{\psi}{x^\beta}-\iquarter\ho_{nm\beta}\,\sigma^{nm}\,\psi\right)\\
&\qquad+\left(\pfrac{e\indices{_j^\alpha}}{x^\alpha}\,\sigma^{ji}\,e\indices{_i^\beta}+
e\indices{_j^\alpha}\,\sigma^{ji}\,\pfrac{e\indices{_i^\beta}}{x^\alpha}
-e\indices{_j^\alpha}\,\sigma^{ji}\,e\indices{_i^\beta}\,e\indices{^k_\xi}\pfrac{e\indices{_k^\xi}}{x^\alpha}\right)
\left(\pfrac{\psi}{x^\beta}-\iquarter\ho_{nm\beta}\,\sigma^{nm}\,\psi\right)\Bigg]\\
&=\rmi\,\gamma^j\,e\indices{_j^\alpha}\left(\pfrac{\psi}{x^\alpha}-\iquarter\ho_{nm\alpha}\,\sigma^{nm}\,\psi\right)-m\,\psi
+\ihalf\gamma^j\left(\pfrac{e\indices{_j^\alpha}}{x^\alpha}-e\indices{_k^\alpha}\,\ho\indices{^k_{j\alpha}}
-e\indices{_j^\alpha}e\indices{^k_\xi}\pfrac{e\indices{_k^\xi}}{x^\alpha}\right)\psi.
\end{align*}
The product of two $\sigma$ matrices in the first and second line is converted according to:
\begin{equation*}
\iquarter\ho_{kl\alpha}\left(\sigma^{kl}\sigma^{ji}-\sigma^{ji}\,\sigma^{kl}\right)
=\ho\indices{^i_{k\alpha}}\,\sigma^{kj}-\ho\indices{^j_{k\alpha}}\,\sigma^{ki}
\end{equation*}
which yields
\begin{align*}
&\quad\frac{\rmi}{3M}\Bigg[-\iquarter e\indices{_j^\alpha}\,e\indices{_i^\beta}\sigma^{ji}\left(\pfrac{\ho_{nm\beta}}{x^\alpha}
+\iquarter\,\sigma^{kl}\,\ho_{kl\alpha}\,\ho_{nm\beta}\right)\sigma^{nm}\,\psi\\
&\qquad+e\indices{_j^\alpha}\,e\indices{_i^\beta}\left(\ho\indices{^i_{k\alpha}}\,\sigma^{kj}
-\ho\indices{^j_{k\alpha}}\,\sigma^{ki}\right)\left(\pfrac{\psi}{x^\beta}-\iquarter\ho_{nm\beta}\,\sigma^{nm}\,\psi\right)\\
&\qquad+\left(\pfrac{e\indices{_j^\alpha}}{x^\alpha}\,\sigma^{ji}\,e\indices{_i^\beta}+
e\indices{_j^\alpha}\,\sigma^{ji}\,\pfrac{e\indices{_i^\beta}}{x^\alpha}
-e\indices{_j^\alpha}\,\sigma^{ji}\,e\indices{_i^\beta}\,e\indices{^k_\xi}\pfrac{e\indices{_k^\xi}}{x^\alpha}\right)
\left(\pfrac{\psi}{x^\beta}-\iquarter\ho_{nm\beta}\,\sigma^{nm}\,\psi\right)\Bigg]\\
&=\rmi\,\gamma^j\,e\indices{_j^\alpha}\left(\pfrac{\psi}{x^\alpha}-\iquarter\ho_{nm\alpha}\,\sigma^{nm}\,\psi\right)-m\,\psi
+\ihalf\gamma^j\left(\pfrac{e\indices{_j^\alpha}}{x^\alpha}-e\indices{_k^\alpha}\,\ho\indices{^k_{j\alpha}}
-e\indices{_j^\alpha}e\indices{^k_\xi}\pfrac{e\indices{_k^\xi}}{x^\alpha}\right)\psi
\end{align*}
By virtue of the identity
\begin{equation*}
e\indices{_j^\alpha}\,e\indices{_i^\beta}\,\sigma^{ji}\,\iquarter\,\ho_{kl\alpha}\,\sigma^{kl}\,\sigma^{nm}\,\ho_{nm\beta}
=e\indices{_j^\alpha}\,e\indices{_i^\beta}\,\sigma^{ji}\,\sigma^{nm}\,\ho_{nk\alpha}\,\ho\indices{^k_{m\beta}}
\end{equation*}
the generalized Dirac equation acquires its final form:
\begin{align*}
&\frac{\rmi}{3M}\Bigg[-\iquarter e\indices{_j^\alpha}\,e\indices{_i^\beta}\,\sigma^{ji}\,\sigma^{nm}
\left(\pfrac{\ho_{nm\beta}}{x^\alpha}+\ho_{nk\alpha}\,\ho\indices{^k_{m\beta}}\right)\psi\\
&\qquad+e\indices{_j^\alpha}\,e\indices{_i^\beta}\left(\ho\indices{^i_{k\alpha}}\,\sigma^{kj}-\ho\indices{^j_{k\alpha}}\,\sigma^{ki}\right)
\left(\pfrac{\psi}{x^\beta}-\iquarter\ho_{nm\beta}\,\sigma^{nm}\,\psi\right)\\
&\qquad+\left(\pfrac{e\indices{_j^\alpha}}{x^\alpha}\,\sigma^{ji}\,e\indices{_i^\beta}+
e\indices{_j^\alpha}\,\sigma^{ji}\,\pfrac{e\indices{_i^\beta}}{x^\alpha}
-e\indices{_j^\alpha}\,\sigma^{ji}\,e\indices{_i^\beta}\,e\indices{^k_\xi}\pfrac{e\indices{_k^\xi}}{x^\alpha}\right)
\left(\pfrac{\psi}{x^\beta}-\iquarter\ho_{nm\beta}\,\sigma^{nm}\,\psi\right)\Bigg]\\
&=\rmi\,\gamma^j\,e\indices{_j^\alpha}\left(\pfrac{\psi}{x^\alpha}-\iquarter\ho_{nm\alpha}\,\sigma^{nm}\,\psi\right)-m\,\psi
+\ihalf\gamma^j\left(\pfrac{e\indices{_j^\alpha}}{x^\alpha}-e\indices{_k^\alpha}\,\ho\indices{^k_{j\alpha}}
-e\indices{_j^\alpha}e\indices{^k_\xi}\pfrac{e\indices{_k^\xi}}{x^\alpha}\right)\psi.
\end{align*}

\begin{align}
&\frac{\rmi}{3M}\sigma^{ji}\Bigg[\!\left(\pfrac{e\indices{_j^\alpha}}{x^\alpha}\,e\indices{_i^\beta}\!
-e\indices{_k^\alpha}\,\ho\indices{^k_{j\alpha}}\,e\indices{_i^\beta}\!
+e\indices{_j^\alpha}\,\pfrac{e\indices{_i^\beta}}{x^\alpha}\!
-e\indices{_j^\alpha}\,e\indices{_k^\beta}\,\ho\indices{^k_{i\alpha}}\!
-e\indices{_j^\alpha}\,e\indices{_i^\beta}\,e\indices{^k_\xi}\pfrac{e\indices{_k^\xi}}{x^\alpha}\!\right)\nonumber\\
&\qquad\quad\times\left(\pfrac{\psi}{x^\beta}-\iquarter\ho_{nm\beta}\,\sigma^{nm}\psi\right)
-\iquarter e\indices{_j^\alpha}e\indices{_i^\beta}\,\sigma^{nm}
\left(\pfrac{\ho_{nm\beta}}{x^\alpha}+\ho_{nk\alpha}\,\ho\indices{^k_{m\beta}}\right)\psi\Bigg]\nonumber\\
&=\rmi\,\gamma^j\,e\indices{_j^\beta}\left(\pfrac{\psi}{x^\beta}-\iquarter\ho_{nm\beta}\,\sigma^{nm}\,\psi\right)-m\,\psi\nonumber\\
&\qquad+\ihalf\gamma^j\left(\pfrac{e\indices{_j^\alpha}}{x^\alpha}-e\indices{_k^\alpha}\,\ho\indices{^k_{j\alpha}}
-e\indices{_j^\alpha}\,e\indices{^k_\xi}\,\pfrac{e\indices{_k^\xi}}{x^\alpha}\right)\psi.
\label{eq:gen-dirac-final}
\end{align}
With the identity~(\ref{eq:partial-deri-tetr}), the last term of the generalized Dirac equation~(\ref{eq:gen-dirac-final}) can be re-written 
as:
\begin{align*}
\pfrac{e\indices{_j^\alpha}}{x^\alpha}-e\indices{_k^\alpha}\ho\indices{^k_{j\alpha}}
-e\indices{_j^\alpha}e\indices{^k_\xi}\pfrac{e\indices{_k^\xi}}{x^\alpha}
&=-e\indices{_j^\xi}\gamma\indices{^\alpha_{\xi\alpha}}-e\indices{_j^\alpha}e\indices{^k_\xi}
\left(e\indices{_m^\xi}\ho\indices{^m_{k\alpha}}-\gamma\indices{^\xi_{\beta\alpha}}e\indices{_k^\beta}\right)\\
&=-e\indices{_j^\xi}\,\gamma\indices{^\alpha_{\xi\alpha}}-\cancel{e\indices{_j^\alpha}\,\ho\indices{^m_{m\alpha}}}+e\indices{_j^\xi}\,\gamma\indices{^\alpha_{\alpha\xi}}\\
&=2e\indices{_j^\xi}\,S\indices{^\alpha_{\alpha\xi}}.
\end{align*}
The spin connection term vanishes due to the skew-symmetry in its first index pair.

Similarly, by virtue of the skew-symmetry of $\sigma^{ji}$, we get the tensor expression
\begin{align*}
&\quad\sigma^{ji}\left(\pfrac{e\indices{_j^\alpha}}{x^\alpha}\,e\indices{_i^\beta}
-e\indices{_k^\alpha}\,\ho\indices{^k_{j\alpha}}\,e\indices{_i^\beta}
+e\indices{_j^\alpha}\,\pfrac{e\indices{_i^\beta}}{x^\alpha}
-e\indices{_j^\alpha}\,e\indices{_k^\beta}\,\ho\indices{^k_{i\alpha}}
-e\indices{_j^\alpha}\,e\indices{_i^\beta}\,e\indices{^n_\xi}\pfrac{e\indices{_n^\xi}}{x^\alpha}\right)\\
&=\sigma^{ji}\left(-\gamma\indices{^\alpha_{\xi\alpha}}\,e\indices{_j^\xi}\,e\indices{_i^\beta}
-\gamma\indices{^\beta_{\xi\alpha}}\,e\indices{_i^\xi}\,e\indices{_j^\alpha}
+\gamma\indices{^\alpha_{\alpha\xi}}\,e\indices{_i^\beta}\,e\indices{_j^\alpha}\right)\\
&=\sigma^{ji}\left(2e\indices{_i^\beta}\,e\indices{_j^\xi}\,S\indices{^\alpha_{\alpha\xi}}
-e\indices{_i^\xi}\,e\indices{_j^\alpha}\,S\indices{^\beta_{\xi\alpha}}\right).
\end{align*}
The generalized Dirac equation~(\ref{eq:gen-dirac-final}) thus simplifies 
to
\begin{align}
&\rmi\gamma^\beta\Dderr_\beta \psi-m\psi=\nonumber\\
&\quad\,-\frac{\rmi}{3M}\sigma^{ji}\Bigg[\left(2e\indices{_i^\beta}\,e\indices{_j^\xi}\,S\indices{^\alpha_{\xi\alpha}}
+e\indices{_i^\xi}\,e\indices{_j^\alpha}\,S\indices{^\beta_{\xi\alpha}}\right)
\Dderr_\beta\psi\nonumber\\
&\qquad\quad+\iquarter e\indices{_j^\alpha}e\indices{_i^\beta}\,\sigma^{nm}
\left(\pfrac{\ho_{nm\beta}}{x^\alpha}+\ho_{nk\alpha}\,\ho\indices{^k_{m\beta}}\right)\psi\Bigg]+\rmi\gamma^\xi S^\alpha{}_{\xi\alpha}\psi\label{eq:gen-dirac-mc}
\end{align}
This equation may be further simplified by noting, that the term involving the derivative of the spin connection is just the Riemann tensor due to the antisymmetry of $\sigma^{ij}$.
The equation thus turns into
\begin{align*}
&\rmi\gamma^\beta \left( \Dderr_\beta - S^\alpha{}_{\beta\alpha} \right) \psi-m\psi=\nonumber\\
&\quad\,-\frac{\rmi}{3M}\Bigg[\!\left(2\sigma^{\xi\beta}\,S\indices{^\alpha_{\xi\alpha}}
+\sigma^{\alpha\xi}\,S\indices{^\beta_{\xi\alpha}}\right)
\Dderr_\beta\psi+\iquarter \sigma^{\alpha\beta}\,\sigma^{\mu\nu}
R_{\mu\nu\alpha\beta}\psi\Bigg],
\end{align*}
giving \eref{eq:Diracfinall}.
\subsection{Riemann-Cartan tensor from the spin connection\label{app:riemanntensor}}
By virtue of
\begin{equation*}
\pfrac{e\indices{^i_\mu}}{x^\alpha}=e\indices{^i_\xi}\gamma\indices{^\xi_{\mu\alpha}}-\ho\indices{^i_{j\alpha}}\,e\indices{^j_\mu},
\end{equation*}
the Riemann-Cartan tensor can be set up on the basis of the affine connection $\gamma\indices{^\xi_\mu_\nu}$ as follows.
The second derivatives of the vierbeins have the two equivalent representations
\begin{align*}
\ppfrac{e\indices{^i_\mu}}{x^\alpha}{x^\beta}&=\pfrac{e\indices{^i_\xi}}{x^\beta}\gamma\indices{^\xi_{\mu\alpha}}
+e\indices{^i_\xi}\pfrac{\gamma\indices{^\xi_{\mu\alpha}}}{x^\beta}
-\pfrac{\ho\indices{^i_{j\alpha}}}{x^\beta}\,e\indices{^j_\mu}-\ho\indices{^i_{j\alpha}}\,\pfrac{e\indices{^j_\mu}}{x^\beta}\\
\ppfrac{e\indices{^i_\mu}}{x^\beta}{x^\alpha}&=\pfrac{e\indices{^i_\xi}}{x^\alpha}\gamma\indices{^\xi_{\mu\beta}}
+e\indices{^i_\xi}\pfrac{\gamma\indices{^\xi_{\mu\beta}}}{x^\alpha}
-\pfrac{\ho\indices{^i_{j\beta}}}{x^\alpha}\,e\indices{^j_\mu}-\ho\indices{^i_{j\beta}}\,\pfrac{e\indices{^j_\mu}}{x^\alpha}.
\end{align*}
The difference of both representations follows as:
\begin{align*}
0&=\pfrac{e\indices{^i_\xi}}{x^\beta}\gamma\indices{^\xi_{\mu\alpha}}-\pfrac{e\indices{^i_\xi}}{x^\alpha}\gamma\indices{^\xi_{\mu\beta}}
+\ho\indices{^i_{j\beta}}\,\pfrac{e\indices{^j_\mu}}{x^\alpha}-\ho\indices{^i_{j\alpha}}\,\pfrac{e\indices{^j_\mu}}{x^\beta}\\
&\quad+e\indices{^i_\xi}\left(\pfrac{\gamma\indices{^\xi_{\mu\alpha}}}{x^\beta}-\pfrac{\gamma\indices{^\xi_{\mu\beta}}}{x^\alpha}\right)
-\left(\pfrac{\ho\indices{^i_{j\alpha}}}{x^\beta}-\pfrac{\ho\indices{^i_{j\beta}}}{x^\alpha}\right)e\indices{^j_\mu}
\end{align*}
The above expression for the derivatives of the vierbeins can now be inserted:
\begin{align*}
0&=\left(e\indices{^i_\xi}\gamma\indices{^\xi_{\eta\beta}}-\cancel{\ho\indices{^i_{j\beta}}\,e\indices{^j_\eta}}\,\right)\gamma\indices{^\eta_{\mu\alpha}}
-\left(e\indices{^i_\xi}\gamma\indices{^\xi_{\eta\alpha}}-\bcancel{\ho\indices{^i_{j\alpha}}\,e\indices{^j_\eta}}\right)\gamma\indices{^\eta_{\mu\beta}}\\
&\quad+\ho\indices{^i_{j\beta}}\left(\,\cancel{e\indices{^j_\eta}\gamma\indices{^\eta_{\mu\alpha}}}-\ho\indices{^j_{n\alpha}}\,e\indices{^n_\mu}\right)
-\ho\indices{^i_{j\alpha}}\left(\bcancel{e\indices{^j_\eta}\gamma\indices{^\eta_{\mu\beta}}}-\ho\indices{^j_{n\beta}}\,e\indices{^n_\mu}\right)\\
&\quad+e\indices{^i_\xi}\left(\pfrac{\gamma\indices{^\xi_{\mu\alpha}}}{x^\beta}-\pfrac{\gamma\indices{^\xi_{\mu\beta}}}{x^\alpha}\right)
-\left(\pfrac{\ho\indices{^i_{j\alpha}}}{x^\beta}-\pfrac{\ho\indices{^i_{j\beta}}}{x^\alpha}\right)e\indices{^j_\mu},
\end{align*}
which yields, after reordering:
\begin{align*}
0&=e\indices{^i_\xi}\left(\pfrac{\gamma\indices{^\xi_{\mu\alpha}}}{x^\beta}-\pfrac{\gamma\indices{^\xi_{\mu\beta}}}{x^\alpha}
+\gamma\indices{^\xi_{\eta\beta}}\,\gamma\indices{^\eta_{\mu\alpha}}-\gamma\indices{^\xi_{\eta\alpha}}\,\gamma\indices{^\eta_{\mu\beta}}\right)\\
&\quad-\left(\pfrac{\ho\indices{^i_{j\alpha}}}{x^\beta}-\pfrac{\ho\indices{^i_{j\beta}}}{x^\alpha}
+\ho\indices{^i_{n\beta}}\,\ho\indices{^n_{j\alpha}}-\ho\indices{^i_{n\alpha}}\,\ho\indices{^n_{j\beta}}\right)e\indices{^j_\mu}.
\end{align*}
The sums in parentheses are the affine and the spin connection representations of the Riemann-Cartan tensor:
\begin{align*}
R\indices{^\xi_{\mu\beta\alpha}}&=\pfrac{\gamma\indices{^\xi_{\mu\alpha}}}{x^\beta}-\pfrac{\gamma\indices{^\xi_{\mu\beta}}}{x^\alpha}
+\gamma\indices{^\xi_{\eta\beta}}\,\gamma\indices{^\eta_{\mu\alpha}}-\gamma\indices{^\xi_{\eta\alpha}}\,\gamma\indices{^\eta_{\mu\beta}}\\
R\indices{^i_{j\beta\alpha}}&=\pfrac{\ho\indices{^i_{j\alpha}}}{x^\beta}-\pfrac{\ho\indices{^i_{j\beta}}}{x^\alpha}
+\ho\indices{^i_{n\beta}}\,\ho\indices{^n_{j\alpha}}-\ho\indices{^i_{n\alpha}}\,\ho\indices{^n_{j\beta}},
\end{align*}
hence
\begin{equation}\label{eq:Riem-correlation}
R\indices{^i_{j\beta\alpha}}=R\indices{^\xi_{\mu\beta\alpha}}\,e\indices{^i_\xi}\,e\indices{_j^\mu}\quad\Leftrightarrow\quad
R\indices{^\xi_{\mu\beta\alpha}}=R\indices{^i_{j\beta\alpha}}\,e\indices{_i^\xi}\,e\indices{^j_\mu}.
\end{equation}
\subsection{Discussion of the spin-momentum tensor $\tilde{\Sigma}\indices{^i^j^\beta}$\label{app:spintensor}}
In order to express the sum of the spin-momentum tensor terms in Eq.~(\ref{eq:cons-eq-skew}) by the spinor fields and their canonical momenta,
we deduce from Eqs.~(\ref{eq:can-dirac2b}) and~(\ref{eq:can-dirac4b}):
\begin{align*}
\psibar\,\sigma\indices{_i^j}\,\pfrac{\tilde{\kappa}^\beta}{x^\beta}-\pfrac{\tilde{\kappabar}^\beta}{x^\beta}\,\sigma\indices{_i^j}\,\psi
&=\left(-\frac{\rmi M}{2}\psibar\,\sigma\indices{_i^j}\gamma_n\,e\indices{^n_\beta}
+\iquarter\psibar\,\sigma\indices{_i^j}\,\sigma\indices{_n^m}\,\ho\indices{^n_{m\beta}}\right)\tilde{\kappa}^\beta\\
&\quad-\tilde{\kappabar}^\beta\left(\frac{\rmi M}{2}e\indices{^n_\beta}\,\gamma_n\,\sigma\indices{_i^j}\,\psi
-\iquarter\ho\indices{^n_{m\beta}}\,\sigma\indices{_n^m}\,\sigma\indices{_i^j}\,\psi\right),
\end{align*}
and from Eqs.~(\ref{eq:can-dirac1b}) and~(\ref{eq:can-dirac3b}):
\begin{align*}
\pfrac{\psibar}{x^\beta}\sigma\indices{_i^j}\tilde{\kappa}^\beta&-\tilde{\kappabar}^\beta\sigma\indices{_i^j}\,\pfrac{\psi}{x^\beta}\\
&=\left(\frac{\rmi M}{2}\psibar\,\gamma_n\,e\indices{^n_\beta}
-\tilde{\kappabar}^\alpha\,e\indices{^n_\alpha}\,\frac{3\rmi M\tau_{nm}}{\dete}\,e\indices{^m_\beta}
-\iquarter\,\psibar\,\ho_{nm\beta}\,\sigma^{nm}\right)\sigma\indices{_i^j}\tilde{\kappa}^\beta\\
&\quad+\tilde{\kappabar}^\beta\sigma\indices{_i^j}\!\left(\frac{\rmi M}{2}\gamma_n\,e\indices{^n_\beta}\,\psi
+e\indices{^n_\beta}\,\frac{3\rmi M\tau_{nm}}{\dete}\,e\indices{^m_\alpha}\,\tilde{\kappa}^\alpha
-\iquarter\,\ho_{nm\beta}\,\sigma^{nm}\,\psi\right),
\end{align*}
thus
\begin{align*}
&\quad\,\pfrac{\tilde{\Sigma}\indices{_i^j^\beta}}{x^\beta}
+\tilde{\Sigma}\indices{_i^n^\beta}\ho\indices{^j_{n\beta}}-\tilde{\Sigma}\indices{_n^j^\beta}\ho\indices{^n_{i\beta}}\\
&=\frac{\rmi}{4}\psibar\left(\sigma\indices{_i^n}\,\ho\indices{^j_{n\beta}}
-\ho\indices{^n_{i\beta}}\,\sigma\indices{_n^j}-\frac{\rmi M}{2}\sigma\indices{_i^j}\gamma_n\,e\indices{^n_\beta}
+\iquarter\ho\indices{_n_{m\beta}}\,\sigma\indices{_i^j}\,\sigma\indices{^n^m}\right)\tilde{\kappa}^\beta\nonumber\\
&\quad-\frac{\rmi}{4}\tilde{\kappabar}^\beta\left(\frac{\rmi M}{2}e\indices{^n_\beta}\,\gamma_n\,\sigma\indices{_i^j}
-\iquarter\ho\indices{_n_{m\beta}}\,\sigma\indices{^n^m}\,\sigma\indices{_i^j}
+\sigma\indices{_i^n}\,\ho\indices{^j_{n\beta}}
-\ho\indices{^n_{i\beta}}\,\sigma\indices{_n^j}\right)\psi\nonumber\\
&\quad+\frac{\rmi}{4}\left(\frac{\rmi M}{2}\psibar\,\gamma_n\,e\indices{^n_\beta}
-\tilde{\kappabar}^\alpha\,e\indices{^n_\alpha}\,\frac{3\rmi M\tau_{nm}}{\dete}\,e\indices{^m_\beta}
-\iquarter\,\psibar\,\ho_{nm\beta}\,\sigma^{nm}\right)\sigma\indices{_i^j}\tilde{\kappa}^\beta\\
&\quad-\frac{\rmi}{4}\tilde{\kappabar}^\beta\sigma\indices{_i^j}\left(-\frac{\rmi M}{2}\gamma_n\,e\indices{^n_\beta}\,\psi
-e\indices{^n_\beta}\,\frac{3\rmi M\tau_{nm}}{\dete}\,e\indices{^m_\alpha}\,\tilde{\kappa}^\alpha
+\iquarter\,\ho_{nm\beta}\,\sigma^{nm}\,\psi\right),
\end{align*}
hence
\begin{align*}
&\quad\,\pfrac{\tilde{\Sigma}\indices{_i^j^\beta}}{x^\beta}
+\tilde{\Sigma}\indices{_i^n^\beta}\ho\indices{^j_{n\beta}}-\tilde{\Sigma}\indices{_n^j^\beta}\ho\indices{^n_{i\beta}}\\
&=\frac{\rmi}{4}\psibar\left[\sigma\indices{_i^n}\ho\indices{^j_{n\beta}}-\ho\indices{^n_{i\beta}}\sigma\indices{_n^j}
+\iquarter\ho\indices{^n_{m\beta}}\left(\sigma\indices{_i^j}\sigma\indices{_n^m}-\sigma\indices{_n^m}\sigma\indices{_i^j}\right)
-\frac{\rmi M}{2}e\indices{^n_\beta}\!\left(\sigma\indices{_i^j}\gamma_n-\gamma_n\sigma\indices{_i^j}\right)\right]\!\tilde{\kappa}^\beta\nonumber\\
&\quad-\frac{\rmi}{4}\tilde{\kappabar}^\beta\!\left[\sigma\indices{_i^n}\ho\indices{^j_{n\beta}}-\ho\indices{^n_{i\beta}}\sigma\indices{_n^j}
+\iquarter\ho\indices{^n_{m\beta}}\left(\sigma\indices{_i^j}\,\sigma\indices{_n^m}-\sigma\indices{_n^m}\,\sigma\indices{_i^j}\right)
-\frac{\rmi M}{2}e\indices{^n_\beta}\!\left(\sigma\indices{_i^j}\gamma_n-\gamma_n\sigma\indices{_i^j}\right)\right]\!\psi\nonumber\\
&\quad-\frac{3M}{4\dete}\tilde{\kappabar}^\alpha\,e\indices{^n_\alpha}\left(\sigma\indices{_i^j}\,\tau_{nm}-\tau_{nm}\,\sigma\indices{_i^j}\right)
e\indices{^m_\beta}\,\tilde{\kappa}^\beta.
\end{align*}
By virtue of the algebra of the Dirac matrices, the commutator of $\sigma$-matrices amounts to
\begin{equation}\label{eq:comm-sigma-square-0}
\sigma\indices{_i^j}\,\sigma\indices{_n^m}-\sigma\indices{_n^m}\,\sigma\indices{_i^j}
\equiv 2\rmi\left(\delta_n^j\,\sigma\indices{_i^m}-\eta_{in}\,\sigma\indices{^j^m}-\delta_i^m\,\sigma\indices{_n^j}+\eta^{mj}\,\sigma_{ni}\right),
\end{equation}
hence, contracted with $\iquarter\ho\indices{^n_{m\beta}}$:
\begin{align*}
\iquarter\left(\sigma\indices{_i^j}\sigma\indices{_n^m}\!-\sigma\indices{_n^m}\sigma\indices{_i^j}\right)\ho\indices{^n_{m\beta}}
&=\onehalf\!\left(\sigma\indices{^j^m}\ho\indices{_i_{m\beta}}-\sigma\indices{_i^m}\ho\indices{^j_{m\beta}}
\!-\sigma_{ni}\,\ho\indices{^n^j_\beta}\!+\sigma\indices{_n^j}\ho\indices{^n_{i\beta}}\right)\\
&=\ho\indices{^n_{i\beta}}\,\sigma\indices{_n^j}-\sigma\indices{_i^n}\,\ho\indices{^j_{n\beta}}.
\end{align*}
The spin tensor terms thus simplify to:
\begin{align}
&\quad\,\pfrac{\tilde{\Sigma}\indices{_i^j^\beta}}{x^\beta}
+\tilde{\Sigma}\indices{_i^n^\beta}\ho\indices{^j_{n\beta}}-\tilde{\Sigma}\indices{_n^j^\beta}\ho\indices{^n_{i\beta}}\nonumber\\
&=\frac{M}{8}\left[\psibar\,e\indices{^n_\beta}\left(\sigma\indices{_i^j}\gamma_n-\gamma_n\sigma\indices{_i^j}\right)\tilde{\kappa}^\beta
-\tilde{\kappabar}^\beta\,e\indices{^n_\beta}\!\left(\sigma\indices{_i^j}\gamma_n-\gamma_n\sigma\indices{_i^j}\right)\psi\right]\nonumber\\
&\quad-\frac{3M}{4\dete}\tilde{\kappabar}^\alpha\,e\indices{^n_\alpha}\left(\sigma\indices{_i^j}\,\tau_{nm}-\tau_{nm}\,\sigma\indices{_i^j}\right)
e\indices{^m_\beta}\,\tilde{\kappa}^\beta.
\label{eq:consistency-explicit-2}
\end{align}
Remarkably, all couplings of the spinor fields to the gauge field, i.e.\ the spin connection $\ho\indices{^j_{n\beta}}$, cancel.
Equation~(\ref{eq:consistency-explicit-2}) can be further simplifies by means of the Dirac algebra identities
\begin{align*}
\sigma\indices{_i^j}\,\gamma_n-\gamma_n\,\sigma\indices{_i^j}&\equiv\hphantom{-}2\rmi\left(\delta_n^j\,\gamma_i-\gamma^j\,\eta_{ni}\right)\\
\sigma\indices{_i^j}\,\tau_{nm}-\tau_{nm}\,\sigma\indices{_i^j}
&\equiv\hphantom{-}2\rmi\left(\eta_{ni}\,\tau\indices{_m^j}-\delta_n^j\,\tau\indices{_m_i}
+\eta_{mi}\,\tau\indices{^j_n}-\delta_m^j\,\tau\indices{_i_n}\right)\\
&\equiv-2\rmi\left(\eta_{ni}\,\tau\indices{^j_m}-\delta_n^j\,\tau\indices{_i_m}
+\eta_{mi}\,\tau\indices{_n^j}-\delta_m^j\,\tau\indices{_n_i}\right)\\
\tau\indices{_n^j}+\tau\indices{^j_n}&\equiv\hphantom{-}\frac{4\rmi}{3}\delta_n^j\,\Eins
\end{align*}
to yield
\begin{align*}
&\quad\,\pfrac{\tilde{\Sigma}\indices{_i^j^\beta}}{x^\beta}
+\tilde{\Sigma}\indices{_i^n^\beta}\ho\indices{^j_{n\beta}}-\tilde{\Sigma}\indices{_n^j^\beta}\ho\indices{^n_{i\beta}}\\
&=\frac{\rmi M}{4}\left[\psibar\,e\indices{^n_\beta}\left(\delta_n^j\,\gamma_i-\gamma^j\,\eta_{ni}\right)\tilde{\kappa}^\beta
-\tilde{\kappabar}^\beta\,e\indices{^n_\beta}\!\left(\delta_n^j\,\gamma_i-\gamma^j\,\eta_{ni}\right)\psi\right]\nonumber\\
&\quad+\frac{3\rmi M}{2\dete}\tilde{\kappabar}^\alpha\,e\indices{^n_\alpha}\left(\eta_{ni}\,\tau\indices{^j_m}-\delta_n^j\,\tau\indices{_i_m}
+\eta_{mi}\,\tau\indices{_n^j}-\delta_m^j\,\tau\indices{_n_i}\right)
e\indices{^m_\beta}\,\tilde{\kappa}^\beta,
\end{align*}
hence
\begin{align*}
&\quad\,\pfrac{\tilde{\Sigma}\indices{_i^j^\beta}}{x^\beta}
+\tilde{\Sigma}\indices{_i^n^\beta}\ho\indices{^j_{n\beta}}-\tilde{\Sigma}\indices{_n^j^\beta}\ho\indices{^n_{i\beta}}\\
&=\frac{\rmi M}{4}\left(\psibar\,\gamma_i\,\tilde{\kappa}^j-\tilde{\kappabar}^j\,\gamma_i\,\psi-\psibar\,\gamma^j\,\tilde{\kappa}_i
+\tilde{\kappabar}_i\,\gamma^j\,\psi\right)\\
&\quad-\frac{3\rmi M}{2\dete}\left(\tilde{\kappabar}^j\,\tau\indices{_i_n}\,\tilde{\kappa}^n+\tilde{\kappabar}^n\,\tau\indices{_n_i}\,\tilde{\kappa}^j
-\tilde{\kappabar}^n\,\tau\indices{_n^j}\,\tilde{\kappa}_i-\tilde{\kappabar}_i\,\tau\indices{^j_n}\,\tilde{\kappa}^n\right)\\
&=\frac{1}{2}\left(\pfrac{\tilde{\HCd}_\mathrm{D}}{e\indices{^i_\alpha}}e\indices{^j_\alpha}
-\pfrac{\tilde{\HCd}_\mathrm{D}}{e\indices{^n_\alpha}}e\indices{^m_\alpha}\eta^{nj}\eta_{mi}\right)
=\onehalf\left(\tilde{T}_\mathrm{D}{}\indices{_i^j}-\tilde{T}_\mathrm{D}{}\indices{^j_i}\right),
\end{align*}
where in the last line the fields are expressed in terms of the Hamiltonian form of the energy-momentum tensor of the Dirac system from Eq.~(\ref{eq:emt-dirac-local}).
Raising the index $i$ finally yields:
\begin{align}
&\quad\,\pfrac{\tilde{\Sigma}\indices{^i^j^\beta}}{x^\beta}
+\tilde{\Sigma}\indices{^i^n^\beta}\ho\indices{^j_{n\beta}}-\tilde{\Sigma}\indices{^j^n^\beta}\ho\indices{^i_n_\beta}\nonumber\\
&=\frac{1}{2}\pfrac{\tilde{\HCd}_\mathrm{D}}{e\indices{^n_\alpha}}\left(e\indices{^j_\alpha}\eta^{ni}-e\indices{^i_\alpha}\eta^{nj}\right)
=\onehalf\left(\tilde{T}_\mathrm{D}^{ij}-\tilde{T}_\mathrm{D}^{ji}\right)\nonumber\\
&=\tilde{T}_{\mathrm{D}}^{[ij]}.
\label{eq:spin-tensor-deri}
\end{align}
The spin tensor density terms of Eq.~(\ref{eq:spin-density2}) thus represent the skew-symmetric portion of the energy-momentum tensor of the free Dirac system
whose explicit representation is given in Eq.~(\ref{eq:emt-dirac-local-skew}).

The explicit representation of the spin tensor density $\Sigma^{ij\beta}$ in terms of the spinor fields on their derivatives is worked out in the following.
The spinor terms of the \emph{spin-momentum tensor} $\tilde{\Sigma}\indices{^i^j^\beta}$, defined in Eq.~(\ref{eq:spin-density2}), are
\begin{equation*}
\tilde{\Sigma}\indices{^i^j^\beta}=\iquarter\left(\psibar\,\sigma\indices{^i^j}\tilde{\kappa}^\beta-\tilde{\kappabar}^\beta\sigma\indices{^i^j}\,\psi\right).
\end{equation*}
Inserting the momentum fields according to Eqs.~(\ref{eq:can-dirac1c}) and~(\ref{eq:can-dirac3c}) yields:
\begin{align}
\Sigma\indices{^i^j^\beta}=&\;\psibar\,\sigma\indices{^i^j}\left[\frac{1}{8}\,\gamma^k\,\psi-\frac{1}{12M}\sigma^{kl}\,e\indices{_l^\alpha}
\left(\pfrac{\psi}{x^\alpha}-\frac{\rmi}{4}\ho_{nm\beta}\,\sigma^{nm}\,\psi\right)\right]e\indices{_k^\beta}\nonumber\\
&+\left[\frac{1}{8}\psibar\,\gamma^k-\frac{1}{12M}\left(\pfrac{\psibar}{x^\alpha}
+\frac{\rmi}{4}\ho_{nm\alpha}\,\psibar\,\sigma^{nm}\right)\sigma^{kl}\,e\indices{_l^\alpha}\right]e\indices{_k^\beta}\,\sigma\indices{^i^j}\,\psi\nonumber\\
=&\;\frac{1}{8}\psibar\left(\sigma\indices{^i^j}\,\gamma^k+\gamma^k\,\sigma\indices{^i^j}\right)e\indices{_k^\beta}\,\psi\nonumber\\
&-\frac{1}{12M}\left[\psibar\,\sigma\indices{^i^j}\,\sigma^{kl}
\pfrac{\psi}{x^\alpha}+\pfrac{\psibar}{x^\alpha}\,\sigma^{kl}\,\sigma\indices{^i^j}\,\psi\right.\nonumber\\
&\left.\qquad\qquad\mbox{}+\frac{\rmi}{4}\ho_{nm\alpha}\,\psibar\left(\sigma^{nm}\,\sigma^{kl}\,\sigma\indices{^i^j}
-\sigma\indices{^i^j}\,\sigma^{kl}\,\sigma^{nm}\right)\psi\right]e\indices{_l^\alpha}\,e\indices{_k^\beta}.
\label{eq:spindensitytensor-expl}
\end{align}

\end{document}